# Comprehensive Survey of QML: From Data Analysis to Algorithmic Advancements

Sahil Tomar[1], Rajeshwar Tripathi[2], and *Sandeep Kumar[3]

[1, 2, 3]Central Research Laboratory, BEL, Ghaziabad, India

*Corresponding author mail Id: sann.kaushik@gmail.com

**Abstract**— Quantum Machine Learning represents a paradigm shift at the intersection of Quantum Computing and Machine Learning, leveraging quantum phenomena such as superposition, entanglement, and quantum parallelism to address the limitations of classical approaches in processing high-dimensional and large-scale datasets. This survey provides a comprehensive analysis of Quantum Machine Learning, detailing foundational concepts, algorithmic advancements, and their applications across domains such as healthcare, finance, and quantum chemistry. Key techniques, including Quantum Support Vector Machine, Quantum Neural Network, Quantum Decision Trees, and hybrid quantum-classical models, are explored with a focus on their theoretical foundations, computational benefits, and comparative performance against classical counterparts. While the potential for exponential speedups and enhanced efficiency is evident, the field faces significant challenges, including hardware constraints, noise, and limited qubit coherence in the current era of Noisy Intermediate-Scale Quantum devices. Emerging solutions, such as error mitigation techniques, hybrid frameworks, and advancements in quantum hardware, are discussed as critical enablers for scalable and fault-tolerant Quantum Machine Learning systems. By synthesizing state-of-the-art developments and identifying research gaps, this survey aims to provide a foundational resource for advancing Quantum Machine Learning toward practical, real-world applications in tackling computationally intensive problems.

## ACRONYMS

| | |
|---|---|
| AI | Artificial Intelligence |
| CNN | Convolutional Neural Networks |
| DL | Deep Learning |
| EQDA | Exploratory Quantum Data Analysis |
| GAN | Generative Adversarial Networks |
| GMM | Gaussian Mixture Models |
| HPC | High Performance Computing |
| KNN | K-Nearest Neighbour |
| LSTM | Long Short-Term Memory |
| ML | Machine Learning |
| NISQ | Noisy Intermediate-Scale Quantum |
| NLP | Natural Language Processing |
| PCA | Principal Component Analysis |
| PQC | Post-Quantum Cryptography |
| QAOA | Quantum Approximate Optimization Algorithm |
| QC | Quantum Computing |
| QDA | Quantum Data Analysis |
| QEC | Quantum Error Correction |
| QFL | Quantum Federated Learning |
| QGAN | Quantum Generative Adversarial Networks |
| QIML | Quantum Inspired Machine Learning |
| QIP | Quantum Image Processing |
| QKD | Quantum Key Distribution |
| QKNN | Quantum K-Nearest Neighbour |

| | | |
|---|---|---|
| QLSTM | Quantum Long Short-Term Memory | |
| QML | Quantum Machine Learning | |
| QNLP | Quantum Natural Language Processing | |
| QNB | Quantum Naive Bayes | |
| QNN | Quantum Neural Networks | |
| QPCA | Quantum Principal Component Analysis | |
| QRNN | Quantum Recurrent Neural Networks | |
| QRL | Quantum Reinforcement Learning | |
| QSVM | Quantum Support Vector Machine | |
| QTSA | Quantum Time-Series Analysis | |
| QTL | Quantum Transfer Learning | |
| RL | Reinforcement Learning | |
| RNN | Recurrent Neural Networks | |
| SVM | Support Vector Machine | |
| TFQ | TensorFlow Quantum | |
| VQA | Variational Quantum Algorithm | |
| VQC | Variational Quantum Classifiers | |
| VQE | Variational Quantum Eigen | |

1. INTRODUCTION

The evolution of classical computing has been dominated by Moore's Law, which predicts the doubling of transistors in integrated circuits approximately every two years. However, the pace of this growth has started to slow, as we approach fundamental physical limits in semiconductor technology [1]. Classical computers, while effective in many areas, are increasingly unable to meet the demands of modern applications like Machine Learning (ML) and Deep Learning (DL), which require massive amounts of data processing and computational power. Traditional computing systems, which rely on binary data representation and sequential processing, struggle with the complexity of big data, posing limitations in fields like Artificial Intelligence (AI), where speed, scale, and flexibility are critical. This growing gap between computational needs and classical computing capabilities has sparked interest in QC. Quantum computers, built on the principles of quantum mechanics, offer the potential to revolutionize computational tasks. By leveraging quantum phenomena such as superposition and entanglement, quantum computers can, in theory, solve problems exponentially faster than classical computers, particularly in handling large-scale data sets and complex tasks that ML and DL models demand [2][3]. QC was conceptualized in the 2080s, with key contributions from pioneers like Richard Feynman and David Deutsch, who proposed that quantum mechanics could be harnessed for computational tasks beyond the reach of classical machines. The breakthrough came in the 2090s with Shor's algorithm, which demonstrated that quantum computers could factor large numbers exponentially faster than the best-known classical algorithms, a result that had profound implications for cryptography and computational complexity theory [3]. This and other quantum algorithms, such as Grover's search algorithm, laid the foundation for the development of QC as a field.

QC remained a theoretical endeavor for many years, with limited practical implementations. However, the advent of experimental techniques such as ion traps and superconducting qubits has brought QC closer to reality. In 2019, Google achieved a landmark achievement in QC, demonstrating "quantum supremacy" by performing a task that would have been infeasible for classical computers, marking a significant step forward in the development of quantum hardware [4]. The continued progress in quantum processors, combined with increased research funding and collaboration from leading tech companies, positions QC as a viable alternative for addressing the computational limits of classical systems [5]. ML, a subset of AI, has evolved dramatically in recent years. Early ML methods, like decision trees, linear regression, and K-Nearest Neighbour (KNN), provided foundational tools for data analysis but struggled with scalability as data volumes and model complexity grew. The advent of DL, utilizing artificial neural networks with multiple layers, marked a significant leap, enabling significant advances in areas such as image recognition, speech processing, and natural language understanding [6]. However, while these DL techniques have brought substantial breakthroughs, they come with challenges, including the need for massive computational resources, extensive training times, and large datasets, which continue to strain classical computing systems.

The growth of ML applications has driven demand for more powerful computational frameworks. Classical machines, despite improvements in hardware, are still inadequate for solving increasingly complex problems in areas like medical diagnostics, financial modeling, and climate prediction. As the complexity of data grows, classical systems are often overwhelmed by the sheer volume of computations required for training and inference in modern ML and DL models [1]. This bottleneck has motivated the search for new paradigms that could overcome these limitations and offer more efficient solutions to data-intensive tasks. Quantum Machine Learning (QML) combines the power of QC with the versatility of ML algorithms. The key advantage of QML lies in its ability to harness quantum computational advantages, such as superposition, entanglement, and quantum parallelism, to solve problems more efficiently than classical methods. By leveraging these quantum phenomena, QML has the potential to accelerate training times, improve classification accuracy, and enable new algorithms that classical computers would find infeasible [7].

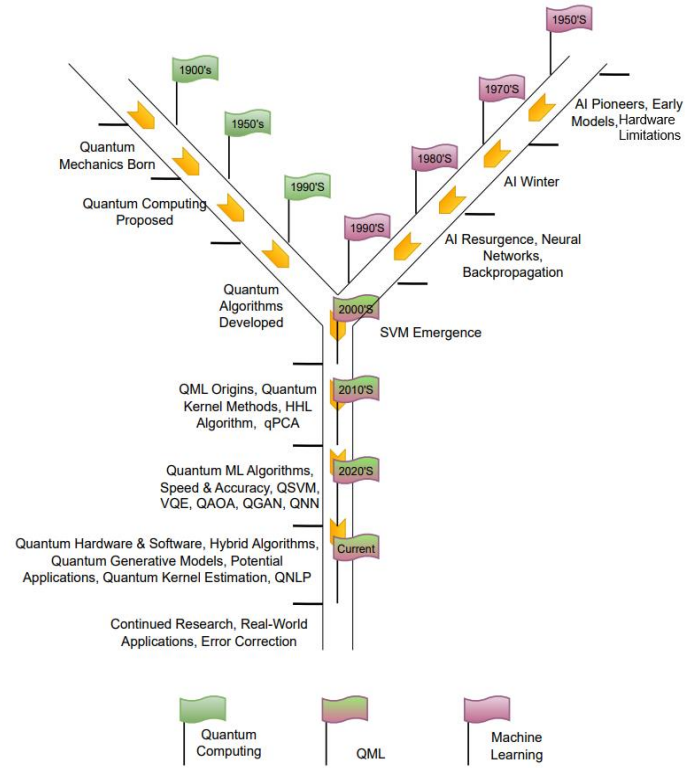

*Figure 1: Evolution and Convergence of Quantum Computing and Machine Learning Leading to Quantum Machine Learning.*

Figure.1 illustrates the evolution of quantum computing (QC), machine learning, and quantum machine learning (QML). Quantum computing originated in the 1900s with the birth of quantum mechanics and progressed through the proposal of quantum computing, the development of algorithms, and advancements in hardware, hybrid models, and practical applications. Meanwhile, machine learning began with early AI models, experienced setbacks during the "AI Winter," and later saw a resurgence with neural networks, the emergence of support vector machines (SVM), and the rise of modern AI techniques. QML emerged in the 2000s, integrating quantum kernel methods, quantum speedup algorithms like Quantum Support Vector Machine (QSVM), Variational Quantum Eigen (VQE), and Quantum Generative Adversarial Networks (QGAN), and adopting hybrid approaches. The timeline underscores continued research and real-world applications, driving the future of QML. One of the most promising areas of QML is the development of quantum versions of classical ML algorithms. For instance, SVMs have been shown to outperform their classical counterparts in certain tasks by reducing the number of steps required for classification. Similarly, quantum-enhanced nearest neighbor classifiers can accelerate data search and pattern recognition by exploiting quantum parallelism to perform simultaneous computations on multiple data points, offering a speed advantage over classical methods [8][9]. These quantum models could be particularly impactful in fields like big data analytics, where large datasets and complex pattern recognition tasks are the norm. In addition to speeding up computations, QML also presents the opportunity to solve problems that were previously intractable for classical ML methods. For example, quantum computers can be used to perform optimization tasks more efficiently, such as in the training of deep neural networks. Quantum-inspired neural networks have been proposed, where quantum algorithms are used to accelerate the optimization process, enabling faster convergence on complex tasks [7]. Furthermore, QML has the potential to improve data security and privacy through quantum encryption methods, which can enhance the confidentiality of sensitive data while still enabling complex computations [3]. Today, QML is at the forefront of research and development, driven by the need to overcome the limitations of classical systems. QC offers a new way to approach the problems faced by traditional computing, such as large-scale data processing, optimization, and model training. As quantum computers continue to improve, the integration of quantum algorithms into ML frameworks is expected to provide substantial performance enhancements, particularly in terms of computational speed and scalability [2].

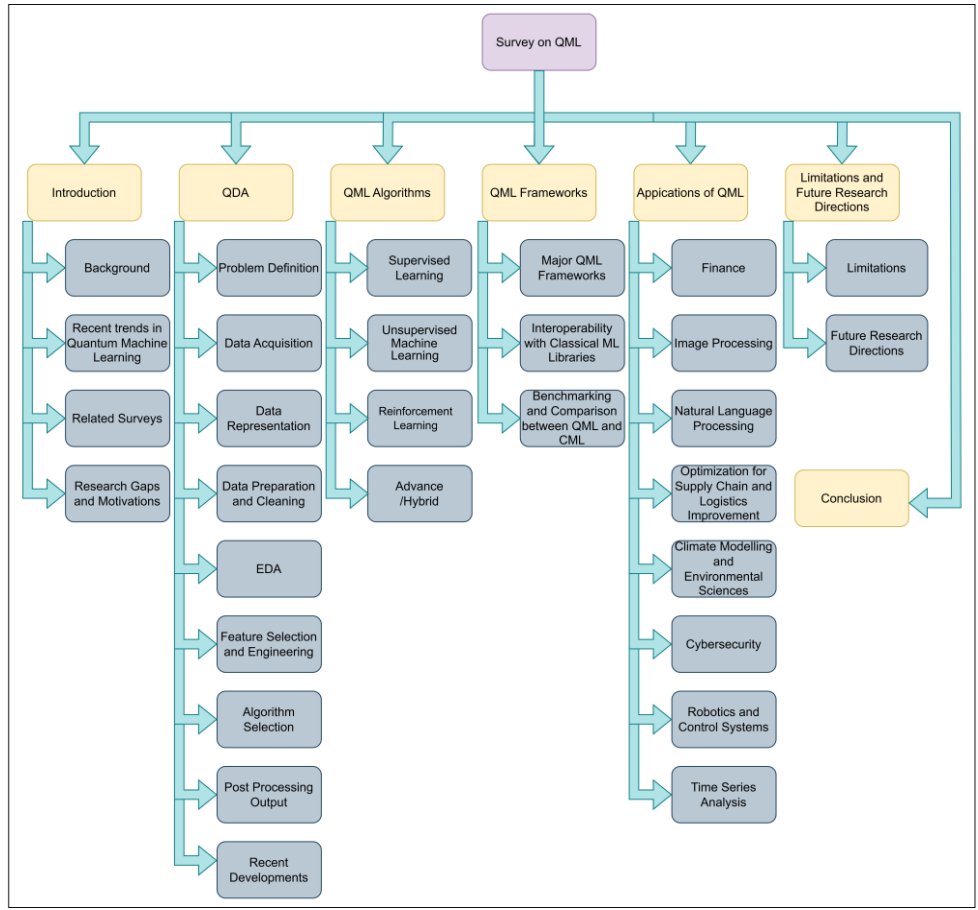

*Figure 2: Detailed Organization of the Paper.*

The exploration of QML is fueled by its potential to unlock new capabilities that are beyond the reach of classical systems. For example, quantum-enhanced algorithms could revolutionize industries such as finance, healthcare, and cybersecurity, where the ability to process vast amounts of data quickly and securely is paramount. The current focus on QML is not just an academic pursuit, but a critical step toward solving real-world problems that classical computing cannot address efficiently [10]. As quantum technologies mature, QML has the potential to redefine the landscape of ML, offering solutions that could transform various sectors and applications [7][8]. Figure 2 represents the detailed organization followed in the paper.

## 1.1. Background

This section provides the foundational understanding necessary for exploring QML, beginning with an introduction to QC and its key principles, contrasting it with classical computing to highlight its unique capabilities. It then transitions into a discussion of classical ML, outlining its methodologies, applications, and the challenges that motivate the integration of quantum techniques. The section further explores hybrid quantum-classical algorithms, emphasizing their potential to enhance computational efficiency by combining the strengths of both systems, before delving into hybrid QML, which aims to overcome the limitations of classical approaches. A review of related surveys offers insights into the current landscape of QML research, while the identification of research gaps and motivations provides a clear direction for future advancements in the field. This progression ensures a comprehensive understanding of both the theoretical foundations and practical implications of QML.

QC has emerged as a transformative paradigm capable of addressing computational challenges that are intractable for classical systems. By leveraging quantum phenomena such as superposition, entanglement, and quantum parallelism, QC offers potential advantages in domains requiring high-dimensional data processing, optimization, and pattern recognition. ML, a domain traditionally dominated by classical systems, has seen exponential growth, yet remains constrained by limitations in scalability, data representation, and computational intensity. These challenges create a compelling opportunity for QC to redefine ML workflows. However, while QC holds theoretical promise, current hardware in the NISQ era is limited by noise, decoherence, and qubit count. These constraints necessitate innovative approaches that bridge quantum potential with classical reliability, paving the way for hybrid quantum-classical algorithms. By combining the strengths of both paradigms, hybrid approaches represent a practical path forward, enabling early exploration of quantum advantages without being hindered by current hardware limitations.

1.1.1. Quantum Computing

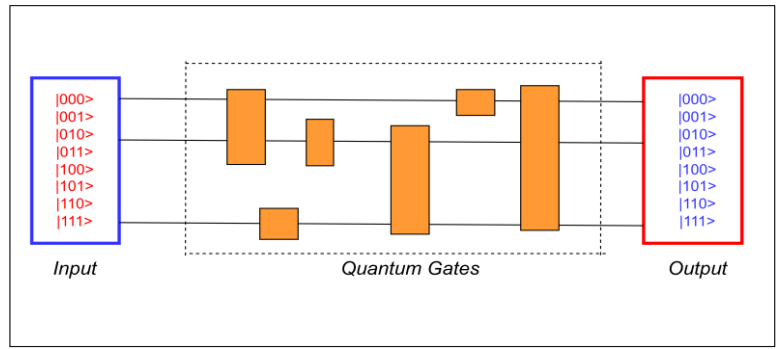

*Figure 3: Quantum circuit: transforming input qubits through gates to output states.*

QC has the potential to transform data processing by exploiting the unique principles of quantum mechanics. This section details key elements of QC that enable it to perform complex tasks beyond classical capabilities. Quantum circuit transforms input qubits through gates to output states is shown in Figure 3.

- Qubits: Qubits, the fundamental unit of quantum information, represent information in a quantum state. Unlike classical bits, qubits can exist in multiple states simultaneously as depicted in Figure 4, due to superposition, enabling exponential computational power [2]. In a classical system, adding a bit doubles the information capacity, but in a quantum system, adding a qubit scales capacity exponentially. This rapid scalability offers potential advantages in solving problems such as cryptographic functions, optimization, and large-scale simulations that classical computing struggles with [13].

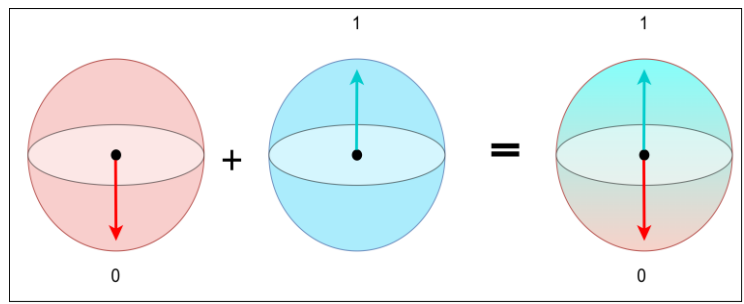

*Figure 4: Visualization of qubit states and their superposition on the Bloch sphere*

- Superposition: Superposition is the ability of a qubit to exist in multiple states (0 and 1) simultaneously as represented in Figure 5. This principle underlies the parallelism of QC, where a single qubit can hold more information than a classical bit. Superposition allows quantum computers to handle complex computations by exploring multiple possibilities at once, making them well-suited for optimization problems that require examining numerous configurations simultaneously [11]. Superposition is experimentally realized through quantum gates that manipulate qubits into mixed states, significantly enhancing computational performance in tasks that involve high-dimensional data [2].

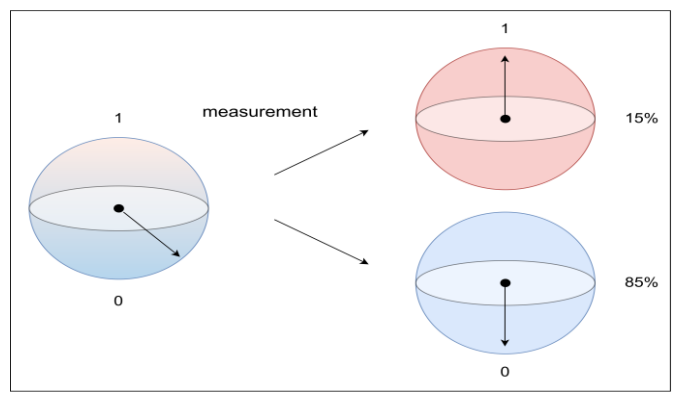

*Figure 5: Qubit measurement collapsing a superposition into definite states with probabilities.*

- Quantum Gates: Quantum gates are the fundamental operations applied to qubits, transforming them in ways that are fundamentally different from classical logic gates. Quantum gates are reversible, meaning they can change a qubit state without losing information, a property essential to maintain quantum coherence [12]. Some key gates shown in Figure 6 includes:
  a. Hadamard Gate: This gate creates superposition, allowing qubits to exist in a balanced state between 0 and 1, which is foundational for many quantum algorithms [2].
  b. CNOT Gate: The CNOT (Controlled NOT) gate is critical for entangling qubits, a requirement for many multi-qubit operations, as it links the state of one qubit to another, creating entangled states that hold interdependent information [14].
  c. Pauli Gates (X, Y, Z): These gates represent quantum rotations on different axes, applying transformations to qubit states. In quantum algorithms, these gates serve as core operations, creating necessary state transformations that support complex computations [2][12].

*Figure 6: Quantum gates and their matrix representation*

- Quantum Circuits: Quantum circuits are composed of quantum gates arranged in a sequence to perform specific computations. They form the architectural foundation of quantum algorithms by providing the framework within which quantum gates interact. Quantum circuits can be implemented on real quantum processors or simulated on classical computers, offering a testbed for designing and optimizing quantum algorithms represented in Figure 7. Practical algorithms like Grover's search and Shor's factoring use complex quantum circuits to demonstrate the power of quantum computation, specifically in domains that require high computational power [14][17].

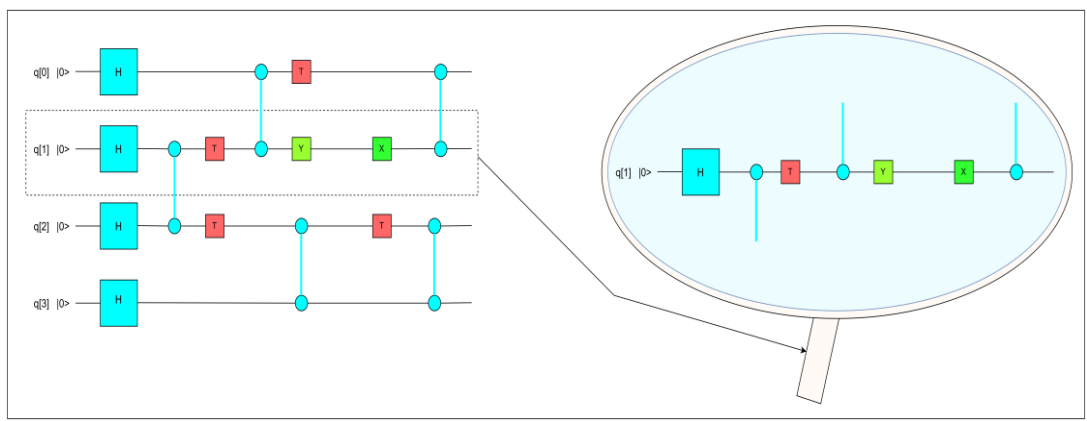

*Figure 7: 4x1 universal random quantum circuit & "bristle-brush" pattern formed by t gates applied to qubit q [1]*

- Quantum Correlations:
  a. Quantum Entanglement: Entanglement is a quantum phenomenon where qubits become linked, such that the state of one qubit directly influences the state of another, regardless of distance. This phenomenon enables secure communication protocols, like quantum teleportation, where information can be transferred without transmitting the actual particles, offering potential applications in quantum networking and cryptography [15][16]. Entanglement allows quantum systems to process and share information in fundamentally new ways, enabling exponential speedups in certain quantum algorithms [13].
  b. Quantum Decoherence: Quantum decoherence is the loss of quantum information as qubits interact with their environment. It disrupts the quantum states of superposition and entanglement, posing a significant challenge to maintaining stable computations. Decoherence becomes especially pronounced in physical quantum processors and limits the scalability of quantum algorithms in noisy environments, such as those in NISQ devices (NISQ computers) [18][19]. Controlling and mitigating decoherence is essential to the stability and reliability of QC, and ongoing research in error-correction techniques addresses these limitations [2].
- Quantum Noise and Error Mitigation: Quantum systems are inherently vulnerable to noise due to their sensitivity to environmental interference and the inherent uncertainty in quantum states:
  a. Uncertainty Principle: Heisenberg's uncertainty principle states that certain properties of a quantum particle, such as position and momentum, cannot be simultaneously measured with arbitrary precision, which inherently affects the stability of qubit states and limits computation fidelity [20].
  b. NISQ Era: In the current NISQ era, quantum processors operate with limited qubit coherence, requiring error-correction strategies to maintain computational accuracy in the presence of noise [17]. Researchers aim to leverage noise-resistant algorithms to achieve more practical applications within the NISQ framework by optimizing algorithms to tolerate certain error levels [18].
  c. Error Correction and Fault Tolerance: Quantum Error Correction (QEC) codes, like Shor's code and the surface code, offer structured approaches to detecting and correcting errors in quantum computations. These codes help in achieving fault-tolerant quantum computation by encoding logical qubits into multiple physical qubits, ensuring resilience against errors [21]. Fault-tolerant QC, which requires advanced error correction to enable stable, scalable computations over long periods, is a significant research area aimed at realizing practical and accurate quantum computers [19].

Together, these principles form the core of QC and underscore its potential to revolutionize computing by overcoming limitations faced by classical computers.

### 1.1.2. Classical Machine Learning

ML is a domain within AI that allows systems to learn from data and improve their performance without direct human intervention [22]. In essence, ML empowers machines to recognize patterns, predict future outcomes, and make decisions based on past experiences or data. The increasing availability of large datasets and advancements in computational power have accelerated the development and application of ML techniques, enabling them to solve complex problems across diverse sectors [24]. The significance of ML lies in its ability to automate tasks that traditionally required human intelligence, such as speech recognition, medical diagnostics, and decision-making in dynamic environments [23]. As ML continues to evolve, it promises to revolutionize industries ranging from healthcare to finance by enhancing accuracy, efficiency, and scalability [25]. This paper reviews key ML methodologies, applications, and challenges, providing insights into the future trajectory of the field.

ML can be broadly categorized into three primary types: supervised learning, unsupervised learning, and Reinforcement Learning (RL). Each of these categories addresses different kinds of problems and leverages different approaches to data analysis.

- Supervised Learning: Supervised learning involves training a model using labeled data, where each training sample is paired with a correct output label. The goal of supervised learning is to learn a mapping from inputs to outputs based on the provided examples. This approach is widely used for both classification and regression tasks. Classification tasks aim to categorize input data into predefined classes, such as image classification or spam detection. For example, a supervised model might classify images of animals into categories like "cat," "dog," or "bird" [24]. Regression tasks, on the other hand, focus on predicting a continuous value, such as predicting house prices or stock market trends based on historical data. Supervised learning is one of the most widely used ML techniques due to its simplicity and effectiveness. It has been applied in diverse applications, including healthcare diagnostics, where medical images are classified to detect diseases like cancer [22].
- Unsupervised Learning: Unsupervised learning differs from supervised learning in that it works with unlabeled data. The aim is to uncover hidden structures or patterns within the data without prior knowledge of what the output should be. Unsupervised learning is often used for clustering and dimensionality reduction tasks. Clustering involves grouping data points that are like each other. This can be applied in customer segmentation, where businesses group customers based on purchasing behavior or demographics [24]. Dimensionality reduction techniques, such as Principal Component Analysis (PCA), are used to reduce the number of variables under consideration, making complex datasets more manageable and interpretable [25]. Unsupervised learning plays a crucial role in data exploration, anomaly detection, and feature extraction, where it helps uncover patterns that were not initially obvious.
- Reinforcement Learning: RL involves an agent that interacts with its environment, learning to take actions to maximize some notion of cumulative reward. Unlike supervised and unsupervised learning, RL is based on a trial-and-error process, where the agent explores various actions and learns from the outcomes to optimize its strategy over time [24]. RL is particularly effective in dynamic environments requiring sequential decision-making, such as playing games, autonomous driving, and robotic control. A notable example is the AlphaGo system developed by DeepMind, which learned to play the board game Go at a superhuman level through RL [24]. The applications of RL are growing rapidly in areas like robotics, automated decision-making, and adaptive control systems.

Despite the many successes of ML, there are several challenges that need to be addressed for further advancement. One of the main challenges is the interpretability of ML models, especially DL models, which are often described as "black boxes." Efforts are being made to develop techniques that make these models more transparent and explainable [26]. Additionally, ML models are highly dependent on data quality. Issues such as data bias, incomplete datasets, and privacy concerns need to be carefully managed to ensure fair and ethical use of ML technologies. ML is a rapidly evolving field that has already had a profound impact across various sectors, from healthcare to finance. With continued advancements in algorithms, computational power, and data availability, the potential applications of ML are limitless. By addressing existing challenges and embracing emerging technologies, ML will continue to drive innovation, enhance decision-making, and improve efficiency across the globe.

1.2. Recent trends in Quantum Machine Learning

In this section, we explore the latest trends and advancements in the rapidly evolving field of Quantum Machine Learning (QML). As quantum computing continues to mature, its integration with machine learning has led to the development of innovative approaches that promise to revolutionize AI. We will examine recent breakthroughs, emerging techniques, and the growing synergy between quantum algorithms and classical machine learning methods. By analyzing current trends, this section provides insight into the ongoing transformations in QML, highlighting its potential to address complex problems more efficiently than classical systems.

1.2.1. Hybrid Quantum Classical Algorithms

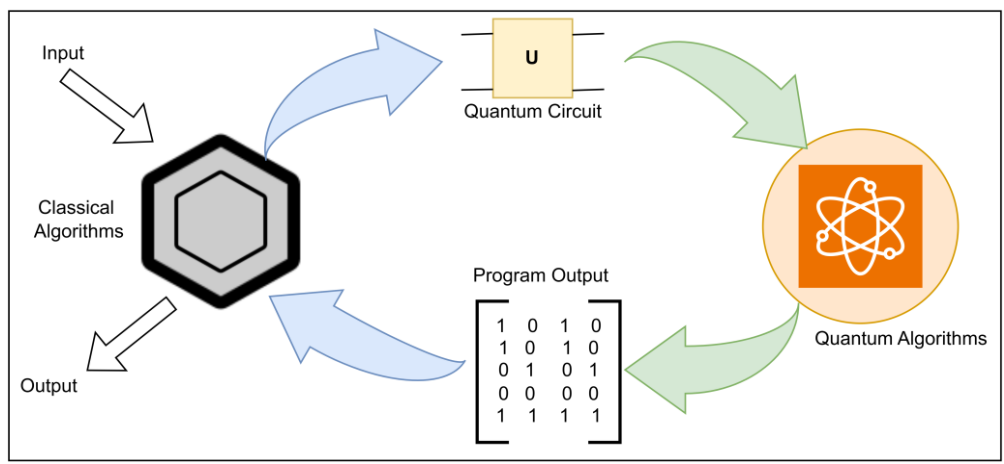

*Figure 8: Hybrid quantum algorithm solver via quantum circuit.*

In recent years, the integration of quantum and classical computational methods has garnered significant interest, particularly in the context of solving complex optimization problems and accelerating ML tasks. Hybrid quantum-classical algorithms are designed to leverage the strengths of both quantum and classical computing resources. This paper provides an overview of the theory, applications, and challenges associated with hybrid quantum-classical algorithms, drawing on various studies in the field. QC promises to revolutionize fields such as optimization, ML, and simulation by providing exponential speedup for certain types of problems. However, quantum hardware is still in its infancy, with limited qubits and high error rates. Classical computers, on the other hand, are well-established and capable of handling large-scale computations. Hybrid quantum-classical algorithms, which combine the strengths of both paradigms, have emerged as a potential solution to bridge the gap between current quantum hardware limitations and the need for practical quantum algorithms. The foundation of hybrid quantum-classical algorithms lies in the combination of quantum mechanics and classical optimization techniques shown in Figure 8. One notable example is the Variational Quantum Algorithm (VQA), where quantum computers are used to perform certain quantum operations, while classical computers optimize parameters of the quantum system. VQAs for solving problems like eigenvalue estimation and optimization were introduced, emphasizing their potential for near-term quantum devices [27]. These algorithms typically involve iterating between quantum operations and classical optimization steps, with the quantum computer evaluating the objective function and the classical optimizer refining the parameters.

A closely related concept is the Quantum Approximate Optimization Algorithm (QAOA), which was introduced for solving combinatorial optimization problems. QAOA uses a quantum circuit to generate a superposition of possible solutions, with a classical optimizer iteratively adjusting the parameters of the quantum circuit to maximize the probability of finding the optimal solution [28]. Hybrid quantum-classical algorithms have shown promising results in a variety of domains, including combinatorial optimization, ML, and data science. For instance, variational approaches for solving eigenvalue problems, which are essential in quantum chemistry and material science, have been demonstrated [29]. These methods combine quantum state preparation and measurement with classical optimization to improve convergence in challenging problems. In ML, quantum-enhanced algorithms such as Quantum Principal Component Analysis (QPCA) have been proposed to speed up data analysis tasks. QPCA leverages quantum mechanics to accelerate the computation of principal components, a fundamental operation in many ML algorithms [30]. Quantum-assisted learning has been further explored to enhance classical ML models by using quantum computers to learn graphical structures more efficiently [31]. Furthermore, hybrid quantum-classical algorithms are also being explored in data science applications, such as graph analytics and recommendation systems. A detailed review highlighted the growing role of these algorithms in addressing large-scale data analysis problems, which could potentially outperform classical algorithms in processing vast datasets [32]. Despite their potential, the implementation of hybrid quantum-classical algorithms faces several challenges. One key issue is the inherent noise in current quantum devices. As noted in the literature, noise and decoherence significantly limit the effectiveness of quantum operations, which can hinder the performance of hybrid algorithms [33]. This necessitates the development of error-correction techniques and noise-resilient quantum algorithms to improve the reliability of hybrid approaches. Another challenge lies in the optimization process itself. Classical optimization methods may struggle with high-dimensional quantum systems, leading to slow convergence rates. Difficulties in applying classical optimization techniques to quantum systems have been extensively reviewed, with some proposing quantum-inspired classical optimizers and adaptive learning strategies as alternatives [34]. The field of hybrid quantum-classical algorithms is still evolving, and several directions for future research are emerging. One promising avenue is the development of QML algorithms, which combine QC with classical ML models to enhance data processing capabilities. QML algorithms for both supervised and unsupervised learning tasks have been proposed with the goal of accelerating learning processes in large datasets [35]. Another exciting development is the exploration of hybrid quantum-classical search algorithms. Recent studies have explored quantum-classical search algorithms, which combine the quantum speedup from Grover's algorithm with classical optimization techniques to solve search problems more efficiently [36]. Furthermore, hybrid quantum-classical algorithms are being applied to various areas, including simulation, optimization, and ML, with several studies discussing the potential for future breakthroughs [37].

Hybrid quantum-classical algorithms are poised to play a central role in the future of QC, offering a practical means of exploiting quantum advantages while leveraging the robustness of classical computing. As quantum hardware continues to improve, these algorithms will likely become more efficient and widely applicable. Future research will need to address challenges such as noise, optimization, and scaling to fully realize the potential of hybrid quantum-classical systems.

1.2.2. Hybrid Quantum Machine Learning

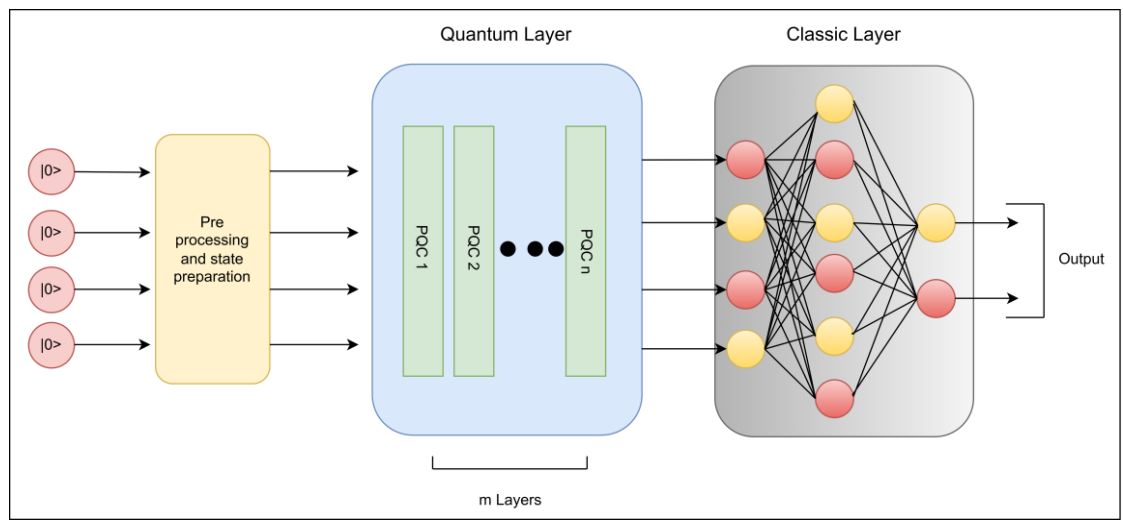

*Figure 9: Hybrid Quantum Machine Learning architecture passing through quantum layer*

Hybrid QML is an interdisciplinary field that merges QC with classical ML techniques to address problems that are difficult or even intractable for classical computers alone. The goal of hybrid QML is to combine the strengths of both quantum and classical systems in a complementary manner as represented in Figure 9. QC holds the potential for exponential speedups in specific areas, such as optimization and simulation, while classical ML techniques are more mature, robust, and well-suited for a wide range of practical tasks. This fusion of quantum and classical approaches has led to the development of several promising hybrid methods, which leverage quantum algorithms to improve classical ML models.

- Quantum-enhanced Feature Space: One of the central ideas in hybrid QML is the use of quantum algorithms to map data into higher-dimensional feature spaces. This is achieved through quantum kernel methods, which can provide more expressive models that classical algorithms may struggle to capture. By leveraging the power of quantum mechanics, such as quantum superposition and entanglement, these methods can potentially uncover complex relationships in data that would otherwise be difficult for classical systems to detect. Quantum kernel methods can be especially beneficial for classification tasks, where the ability to identify complex patterns in high-dimensional spaces is crucial [3][5].
- Quantum Circuit Learning: Quantum circuits can be used to optimize ML models, allowing for more efficient training and inference. These approaches typically involve encoding an optimization problem into a quantum circuit, which can then be evaluated using quantum parallelism. This allows quantum algorithms to explore multiple solution spaces simultaneously, potentially speeding up the search for optimal solutions compared to classical methods. For instance, quantum circuit learning can be applied to neural network architectures, where quantum gates are used to accelerate parts of the learning process, such as weight optimization or feature extraction [38][39].
- VQAs: VQAs are one of the most significant developments in hybrid QML, particularly in optimization and simulation tasks. Examples of such algorithms include the VQE solver and the QAOA. These algorithms combine quantum circuits with classical optimization routines. The quantum computer is used to evaluate and parameterize quantum circuits, while classical optimization techniques are employed to adjust the parameters to optimize the performance of the circuit. VQAs have shown promise in solving combinatorial optimization problems such as those encountered in scheduling, logistics, and financial portfolio management [27][29][40].
- Quantum-DL Hybrid: DL models, particularly neural networks, can benefit from the integration of quantum circuits. Quantum-enhanced DL approaches aim to speed up certain parts of the model, such as the extraction of features or the optimization of weights. Quantum circuits are employed in the initial layers of the model, while classical hardware is used for the deeper layers. This hybrid architecture seeks to combine the representational power of DL with the computational advantages of quantum systems. Recent work has explored applying quantum-enhanced neural networks to image processing, Natural Language Processing (NLP), and pattern recognition tasks, showing potential for improved accuracy and efficiency [41][42].

Hybrid QML approaches, particularly those using VQAs and QAOA, are highly effective for solving optimization problems. These algorithms are well-suited to tasks that involve finding optimal configurations in large solution spaces, such as resource scheduling, financial portfolio optimization, and logistics planning. Quantum systems can leverage quantum parallelism to evaluate multiple possibilities at once, potentially reducing the time required to find optimal or near-optimal solutions [31][32]. Hybrid models that use quantum-enhanced feature spaces, such as quantum kernel methods, offer significant advantages in classifying large datasets, especially in high-dimensional spaces. By mapping data into higher-dimensional quantum feature spaces, these models can capture more complex patterns than classical methods. QSVMs have shown the ability to classify big data more effectively than their classical counterparts in some cases [3][43]. One of the most promising applications of hybrid QML is in the field of quantum chemistry and materials science. By using quantum computers to simulate molecular systems, researchers can gain insights into chemical reactions, molecular structures, and material properties that would be computationally expensive or infeasible with classical methods. This has important implications for drug discovery, materials design, and chemical reaction modeling, where quantum-enhanced ML models can offer a more accurate and efficient way of simulating complex quantum interactions [5][44].

Hybrid QML is an emerging field that shows great potential but is still in its infancy. The primary challenge lies in building practical, scalable systems that can effectively harness the power of QC while integrating it seamlessly with classical ML techniques. Current quantum hardware is still in the NISQ era, where qubits are error-prone and limited in number. Despite these challenges, hybrid QML approaches are already providing insights into optimization, data classification, and simulation tasks that classical systems struggle with [45][33]. As quantum hardware improves and more sophisticated algorithms are developed, hybrid quantum-classical systems are expected to become increasingly important in the computational landscape. This hybrid approach can help bridge the gap between the theoretical potential of QC and the practical realities of using quantum systems for real-world ML tasks. Future breakthroughs in QEC, quantum hardware, and hybrid algorithm development could lead to significant advances in AI and ML, particularly in areas such as automated reasoning, drug discovery, and large-scale data analysis [33][46].

Hybrid QML represents a promising frontier in computational science, combining the strengths of QC and classical ML to tackle complex, computationally intensive problems. While achieving full-scale quantum supremacy in ML is still a long-term goal, hybrid QML approaches are already demonstrating potential advantages in optimization, classification, and simulation tasks [47]. As both quantum hardware and quantum algorithms continue to improve, hybrid QML is poised to make significant contributions to fields ranging from AI to quantum chemistry, providing a powerful tool for addressing some of the most challenging problems of the future.

### 1.3. Related Surveys

This section presents a comparative analysis of existing surveys on QML, focusing on their primary areas of research, methodologies, applications, limitations, and coverage of ML algorithms. By summarizing key surveys from various years in

Table 1, this analysis highlights the progression and trends in QML research while identifying existing gaps and opportunities for further exploration. The following table organizes these surveys systematically, providing insights into their contributions and limitations, and offering a clear understanding of how prior work has shaped the current landscape of QML.

*Table 1: Study of related surveys*

| Sr. No | Reference and Year | Primary Focus Area | Methodology | Application Covered | Coverage of ML Algorithms | Limitations |
|---|---|---|---|---|---|---|
| 1 | [48] & 2018 | QML from a classical perspective | Discusses hybrid classical-quantum approaches and theoretical frameworks for quantum-enhanced learning. | General classification and optimization tasks | Focuses on bridging classical ML with QC. Covers linear classifiers and kernel methods. | Lack of practical implementation on quantum hardware. |
| 2 | [49] & 2020 | QML algorithms | Provides a detailed taxonomy of QML algorithms and their computational advantages. | Algorithms for supervised and unsupervised learning | Discusses SVMs, clustering, and regression algorithms. | Limited exploration of hardware implementation challenges. |
| 3 | [50] & 2020 | QML for classification | Surveys classification algorithms using quantum principles and highlights use cases. | Applications in medical diagnosis and image recognition | Concentrates on classifiers, including quantum-enhanced decision trees and SVMs. | Limited focus on scalability and NISQ systems. |
| 4 | [51] & 2020 | Advances in QML | Reviews recent advancements and categorizes them by model and application. | Data classification, optimization, and RL | Covers RL, regression, and classification methods. | Limited discussion on experimental validation. |
| 5 | [52] & 2021 | Quantum DL | Explores the transition from basic QML techniques to quantum DL architectures. | Image recognition, optimization | Includes discussions on Quantum Neural Networks (QNN) and DL algorithms. | Theoretical focus; minimal practical results. |
| 6 | [53] & 2022 | Comprehensive survey on QML | Explores theoretical and experimental developments in QML alongside potential applications. | Financial modeling, medical imaging | Covers supervised, unsupervised, and RL. | Overlooks algorithmic scalability issues on large datasets. |
| 7 | [54] & 2023 | Overview of QML | Provides a broad overview of QML principles, focusing on both classical and quantum perspectives. | Pattern recognition, optimization | Discusses SVMs, quantum clustering, and regression methods. | Limited emphasis on real-world datasets. |
| 8 | [55] & 2024 | AI, ML, and DL in QC | Reviews trends and challenges at the intersection of AI, ML, and QC, with future research directions. | Industry 4.0 applications | Discusses hybrid algorithms, including ML models applied in quantum systems. | Lacks specificity in quantum-specific methodologies. |

| 9 | [56] & 2024 | Design and analysis of QML | Surveys the architecture and mathematical foundation of QML models. | High-energy physics, quantum chemistry | Covers logistic regression, kernel methods, and QNN. | Limited practical implementations discussed. |
| 10 | [57] & 2024 | Trends and challenges in QML | Provides a forward-looking review of challenges, trends, and research opportunities in QML. | Financial analysis, optimization | Discusses quantum algorithms for classification, regression, and clustering, emphasizing optimization approaches. | Overlooks algorithmic trade-offs in NISQ-era hardware. |

1.4. Research Gaps and Motivation

QML is rapidly emerging as a transformative paradigm that combines the computational potential of quantum mechanics with the adaptability of machine learning (ML) algorithms. While existing surveys provide valuable insights into QML, they often fail to address critical aspects of the pipeline that significantly affect algorithm performance and practical applicability. This survey aims to fill these gaps by offering a novel, comprehensive perspective that extends beyond algorithmic exploration to include quantum data analysis (QDA), preprocessing techniques, and application-specific implementations.

- Theoretical Coverage and Algorithmic Exploration: Most existing surveys (e.g., [48], [51], and [55]) focus primarily on widely known algorithms such as Quantum Support Vector Machines (QSVMs) and Quantum Neural Networks (QNNs). However, these surveys lack a comprehensive analysis of the broader QML landscape. They neglect underexplored areas such as quantum-enhanced feature engineering, advanced feature selection strategies, quantum random forests, and innovative ensemble learning techniques. Furthermore, essential topics like enhanced quantum data analysis Exploratory Quantum Data Analysis (EQDA), which emphasizes preprocessing strategies critical for downstream ML tasks, remain largely unaddressed.
- Challenges in Real-World Applicability: Previous works (e.g., [52], [53], and [57]) have primarily highlighted the theoretical potential of QML algorithms but have not sufficiently addressed their practical implementation in real-world scenarios, particularly with the limitations of NISQ devices. These studies provide a solid theoretical foundation but fall short in exploring how QML algorithms can be effectively deployed on current quantum hardware. Challenges such as noise, qubit coherence, and limited computational resources are inadequately addressed. Furthermore, existing surveys rarely delve into hybrid quantum-classical frameworks, which are critical for bridging the gap between theoretical promise and practical feasibility. These hybrid frameworks, which combine quantum processing with classical computing, offer solutions to current hardware constraints and pave the way for the practical adoption of QML. Additionally, the impact of quantum preprocessing methods like QDA on real-world applications remains underexplored, despite their potential to significantly enhance data quality and improve the efficiency of QML algorithms in domains such as healthcare, finance, and climate science.
- Data Representation and Encoding: Efficiently encoding classical data into quantum states remains one of the most significant challenges in QML, as noted by studies like [69]. Classical data, represented as bits, must be transformed into quantum states (qubits) to leverage quantum computing's unique properties, such as superposition and entanglement. However, the preprocessing steps needed to optimize these encodings for specific datasets or tasks are often overlooked in current research. Critical preprocessing techniques like dimensionality reduction, noise filtering, and feature selection are essential for ensuring that data is properly mapped into quantum states that maximize the advantages of quantum processing. Without careful preprocessing, the encoding process may fail to capture the relevant data features, hindering the scalability and performance of QML algorithms. Addressing these preprocessing challenges is critical to achieving meaningful quantum speedups in large-scale applications.
- Benchmarking and Performance Evaluation: Standardized benchmarking remains a significant gap in QML literature. While surveys like [57] acknowledge the importance of benchmarking for assessing QML algorithm performance, they rarely propose actionable frameworks that evaluate the entire QML pipeline. Existing benchmarks often focus narrowly on individual algorithms or computational speedups, neglecting the broader pipeline context, including data preprocessing. Effective benchmarking must consider quantum-enhanced preprocessing techniques like feature extraction and dimensionality reduction, which are essential for algorithm success. Without an integrated benchmarking approach, it is challenging to evaluate the practical performance of QML algorithms holistically.
- Neglect of Application-Specific Implementations: Current surveys often fail to explore how QML addresses domain-specific challenges in critical fields such as healthcare, finance, and climate science. For instance, applications like biomarker discovery in healthcare or long-term weather prediction in climate modelling are insufficiently covered, limiting QML's practical utility.

- Limited Focus on Circuit-Level Implementations: Many surveys omit detailed circuit diagrams for algorithms like QSVMs, QNNs, and Variational Quantum Algorithms (VQAs). This omission creates a barrier to understanding the feasibility of implementing these algorithms on quantum hardware.
- Inadequate Comparative Analysis: Current benchmarks often lack a thorough evaluation of QML algorithms against their classical counterparts. Metrics such as scalability, accuracy, and hardware constraints are rarely assessed comprehensively, leaving gaps in understanding the trade-offs between quantum and classical approaches.
- Challenges with Hybrid Quantum-Classical Frameworks: Hybrid systems are crucial for overcoming NISQ limitations, yet existing research often overlooks their potential. These frameworks play a vital role in bridging the gap between quantum capabilities and the practical constraints of current quantum hardware.

This survey bridges these gaps by presenting a cohesive framework that integrates quantum preprocessing as a foundational step in the QML pipeline. By emphasizing the interplay between preprocessing, data encoding, and algorithmic efficiency, it advances QML research beyond traditional algorithmic focus. The survey includes detailed circuit diagrams to facilitate practical understanding and implementation on quantum hardware. Additionally, it introduces a multi-dimensional benchmarking framework to evaluate computational performance, preprocessing efficiency, and pipeline scalability, ensuring a comprehensive assessment of QML algorithms. The motivation for this survey lies in advancing QML from a promising theoretical domain to a practical computational tool. The survey aims to unlock QML's potential by addressing gaps in data preprocessing, algorithm robustness, and interdisciplinary application. By exploring use cases in diverse fields like healthcare, finance, and climate science, it highlights how quantum preprocessing and hybrid frameworks can address complex challenges that classical ML struggles to solve. This focus on real-world applications makes the survey a critical resource for both academics and industry practitioners. The following are the contribution of the survey.

- Quantum Preprocessing Techniques: The survey highlights the novelty of quantum preprocessing techniques, such as quantum-enhanced clustering and dimensionality reduction. These methods significantly improve the quality and efficiency of downstream machine learning tasks by optimizing data handling.
- QDA as a Foundational Step: The survey positions Quantum Data Analysis (QDA) as a foundational step in the QML pipeline. Techniques like quantum clustering and Quantum Principal Component Analysis (QPCA) are emphasized for their ability to enhance data quality and optimize machine learning tasks.
- Hybrid Frameworks: The advantages of hybrid quantum-classical frameworks are showcased, leveraging current quantum hardware capabilities. Actionable insights are provided through visual aids and comparative analyses, bridging theoretical exploration with practical feasibility.
- Holistic QML Framework: A comprehensive algorithmic framework for QML is introduced, covering the entire pipeline, including data acquisition, encoding, algorithm selection, and post-processing. The framework ensures cohesion and efficiency throughout the workflow.
- Benchmarking: A multi-dimensional benchmarking is explored to evaluate QML performance. Key areas of focus include preprocessing, scalability, and overall efficiency, providing actionable metrics for assessing QML pipelines.
- Comparative Analysis: Quantum-enhanced methods are compared to classical preprocessing techniques, highlighting their superior scalability and accuracy.
- Domain-Specific Applications: Applications of QML in healthcare, finance, and climate modelling are explored in depth. These domain-specific studies showcase how QML addresses challenges and drives innovation in critical sectors.
- Applications and Their Practical Use Case Studies: The survey discusses real-world use cases of QML in solving complex problems in various domains. Detailed examples demonstrate the practical benefits and potential of QML applications.
- Circuit-Level Diagram of QML Algorithms: Circuit-level diagrams of QML algorithms are provided to illustrate their design and functioning. These diagrams serve as practical tools for understanding and implementing quantum machine learning solutions.

This survey integrates foundational concepts, practical implementations, and interdisciplinary applications, establishing itself as an essential resource for advancing QML research and fostering real-world adoption. Additionally, it offers actionable insights and detailed case studies across various industries—from healthcare to quantum chemistry—that highlight the significant advantages of incorporating QDA into QML workflows. These examples demonstrate how QDA addresses the limitations of classical data processing techniques effectively. By combining core principles, practical approaches, and cross-disciplinary perspectives, this survey becomes an invaluable reference for both researchers and practitioners seeking to unlock the full potential of QML in addressing some of the most complex computational challenges. In conclusion, this survey not only addresses critical gaps in existing literature but also emerges as a key resource for driving the evolution of QML. Its focus on practical implementation, comprehensive benchmarking, and domain-specific applications cements its role as a foundational guide for the future development of QML.

## 2. QUANTUM DATA ANALYSIS

This section explores QDA, a pivotal component of the QML pipeline that ensures high-quality, well-prepared data for quantum models. Beginning with problem definition and data acquisition, QDA emphasizes understanding the dataset and gathering data from classical or quantum sources. It then focuses on quantum data representation, where classical data is mapped into quantum states using encoding techniques, followed by data preparation and cleaning, which optimizes the data by noise filtering and dimensionality reduction. The role of EQDA is highlighted as it uncovers patterns and correlations using quantum techniques like QPCA and quantum clustering. This is followed by feature selection and engineering, which reduce data dimensionality and identify critical features using quantum-enhanced methods. The discussion then moves to algorithm selection, where suitable quantum algorithms such as QNN and QSVMs are applied, and concludes with post-processing, where quantum outputs are refined and interpreted for practical use. In essence, QDA provides the foundation for effective QML by addressing quantum hardware challenges and leveraging quantum computational advantages, enabling breakthroughs in domains like healthcare, finance, and scientific research.

QDA represents a groundbreaking approach to processing and interpreting large-scale, high-dimensional datasets by harnessing the power of QC. In traditional ML, data analysis is crucial for ensuring the effectiveness of the model. The performance of ML algorithms is heavily dependent on the quality, relevance, and preparation of the input data. This role is no less important when quantum algorithms are employed, as QC offers significant computational advantages, especially for tasks involving complex, high-dimensional datasets. However, despite these advantages, the importance of data analysis does not diminish in the quantum domain. Instead, QDA plays a similarly vital role within the QML pipeline, ensuring that quantum models receive high-quality, well-prepared data, which enables them to unlock their full potential. The integration of QDA within QML is essential to fully realize the benefits of quantum algorithms. Just as in traditional ML, where effective data preprocessing, feature extraction, and cleaning are necessary for high-performing models, QDA ensures that QML models are given the clean, relevant, and well-encoded data they need. The workflow of QDA is shown in form of flowchart in Figure 10. QDA not only amplifies the advantages of quantum algorithms but also addresses the challenges posed by quantum hardware limitations, such as noise, qubit coherence, and limited computational resources. With the help of quantum phenomena like superposition and entanglement, QDA can facilitate exponential speedups in data processing tasks, offering significant improvements over classical methods, especially in fields like healthcare, finance, and scientific research.

Although QML harnesses the power of QC for speedups, the role of QDA is indispensable in providing data that enhances the potential of quantum algorithms. The process of building a QML model, just like in traditional ML, begins with problem definition, which includes identifying the specific task, understanding the dataset, and selecting appropriate quantum algorithms. This is followed by data acquisition, where data from various sources—either classical or quantum-generated—is gathered. The quality and format of the acquired data significantly impact the downstream performance of QML algorithms, which makes this step pivotal in the overall pipeline. Next, quantum data representation becomes a critical step, as classical data must be mapped into quantum states, such as qubits, through methods like amplitude encoding or basis encoding. The success of quantum processing depends on how effectively this classical data is transformed into a quantum-compatible format, directly influencing the efficiency of quantum algorithms. After the data is represented, quantum data preparation and cleaning follow, which is where QDA techniques such as noise filtering, dimensionality reduction, and feature cleaning come into play. These steps are crucial to ensuring that the data is ready for quantum processing by eliminating noise and irrelevant data, which could otherwise impede the performance of quantum algorithms. EQDA further enriches the data preparation process by uncovering patterns, correlations, and outliers within quantum datasets. EQDA, which leverages quantum-enhanced techniques like quantum clustering and QPCA, helps in identifying key features that are most relevant for subsequent ML tasks. This is similar to classical EDA but enhanced by quantum computational techniques that allow for more complex and deeper analysis. The insights gained from EQDA guide decisions on feature selection, ensuring that only the most important data points are retained for further processing.

Feature selection and engineering are then applied to reduce the dimensionality of the dataset, focusing on the most informative features to improve the predictive power of the quantum model. Quantum-enhanced methods such as SVMs are particularly effective in identifying and selecting these critical features. This step not only boosts the model's accuracy but also makes it more efficient, especially when handling large datasets in real-world applications. After these steps, algorithm selection follows, where various QML algorithms can be chosen based on the use case and the characteristics of the dataset. Algorithms such as QNN, Quantum Random Forests, and SVMs are then applied to the preprocessed data, selected based on the problem at hand. Once the quantum model generates output, post-processing is the final step, where the results are interpreted and refined. This may involve translating quantum results into classical formats, correcting errors, or combining quantum outputs with classical data to produce actionable results. Effective post-processing ensures that quantum insights can be applied in practical scenarios, from predictive analytics to decision-making processes.

In essence, QDA is an essential and foundational component of the QML pipeline. From problem definition and data acquisition to preprocessing, exploratory analysis, feature engineering, and algorithm selection, QDA ensures that quantum algorithms are fed with high-quality data, enabling them to deliver optimal performance. As QC continues to evolve, the role of QDA in QML will only become more critical, allowing for enhanced scalability and more accurate results in real-world applications, such as personalized medicine, climate modeling, and financial forecasting. By integrating QDA into the QML workflow, we can bridge the gap between the theoretical advancements of quantum algorithms and their practical, real-world applications, ultimately realizing the full potential of QML.

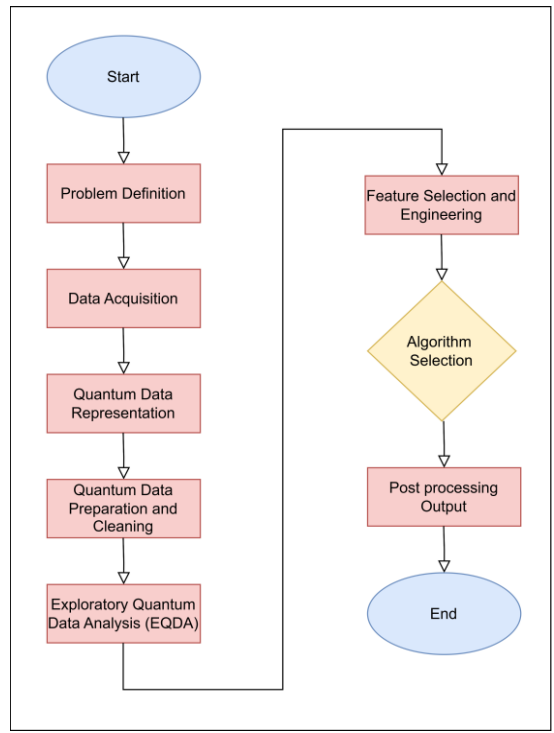

*Figure 10: Quantum Machine Learning Workflow: From Problem Definition to Post-Processing*

In climate modeling, classical data analysis struggles to process the immense volume of sensor and simulation data required to predict long-term weather patterns. Quantum algorithms, such as Grover's search and QAOA, can accelerate the identification of critical parameters, leading to faster and more accurate models. Similarly, in genome sequencing, QDA enables efficient identification of genetic variations across large datasets, paving the way for personalized medicine and drug discovery. By integrating QDA into data analysis pipelines, researchers can tackle computational bottlenecks and unlock new possibilities in data-intensive fields.

### 2.1.1. Problem Definition

Data analysis in domains like healthcare, finance, and scientific research often involves large-scale feature selection, optimization, and pattern recognition tasks. These tasks are computationally intensive, especially when datasets are high-dimensional or require complex, nonlinear modeling. Classical data analysis methods rely on sequential processing and have inherent limitations when faced with such computational demands. For example, feature selection in healthcare diagnostics, where identifying critical biomarkers for diseases like cancer involves sifting through vast genomic datasets, can become infeasible due to the time and resources required.

QDA addresses these challenges by leveraging the principles of QC. Quantum algorithms, such as the QAOA and Grover's search, use quantum superposition and entanglement to explore solution spaces exponentially faster than classical methods. QAOA, for instance, can efficiently solve combinatorial optimization problems by encoding the solution space in a quantum system and iteratively improving results through quantum circuits [28], [40]. Grover's search algorithm offers a quadratic speedup for unstructured search problems, significantly reducing the time required to identify optimal features or parameters in high-dimensional data [58]. In healthcare diagnostics, identifying the most relevant features, such as genetic biomarkers, is critical for developing personalized treatments. Classical methods for feature selection involve exhaustive search or heuristic approaches, which are computationally expensive as the dataset size grows. QDA, utilizing Grover's algorithm, can accelerate the search for these biomarkers by exploring multiple possibilities simultaneously. For example, Grover's algorithm enables faster identification of specific genetic variations associated with diseases, significantly reducing analysis time while maintaining high accuracy [58]. Additionally, QAOA has been applied to optimize feature subsets for predictive models, allowing researchers to focus on the most impactful variables. By leveraging QDA, medical researchers can overcome classical bottlenecks, improving the efficiency and precision of diagnostic tools and enabling quicker clinical decision-making [28], [40]. Quantum systems encode data in qubits, which allow representation in superposition states. Unlike classical bits, which can only exist in a state of 0 or 1, qubits can represent multiple states simultaneously, drastically reducing the storage and computational cost for high-dimensional datasets.

*Table 2: Comparison of classical system and quantum system*

| Aspect | Classical Systems | Quantum Systems |
|---|---|---|
| **Data Units** | Bits (0 or 1) | Qubits (superposition of 0 and 1) |
| **Representation** | Single value at a time | Exponentially large state space |
| **Storage Efficiency** | Linear growth with data size | Exponential encoding potential |
| **Example Capacity** | 3 bits can represent one of 8 possible states | 3 qubits can simultaneously represent all 8 states due to superposition |
| **Processing** | Sequential | Parallel |

Table 2 shows the comparison of classical system and quantum system based on various parameters. This integration of quantum systems into data representation and analysis pipelines offers transformative potential for addressing the computational challenges of high-dimensional datasets across diverse fields like healthcare and beyond.

2.1.2. Data Acquisition

Data acquisition involves collecting, transforming, and loading data from various sources into a usable format for further analysis. In classical systems, this process relies on Extract, Transform, Load (ETL) pipelines that standardize and clean data for applications such as ML or statistical modeling. While effective for smaller datasets, classical methods struggle to handle high-dimensional data, noise, and incomplete information efficiently, often requiring extensive computational resources. Quantum systems enhance data acquisition by leveraging quantum mechanical properties like superposition and entanglement to improve accuracy, speed, and scalability. Quantum sensors, for example, can achieve far greater precision and sensitivity than classical counterparts, making them ideal for applications requiring high-resolution measurements [15], [16]. Additionally, quantum algorithms like Grover's search [58] enable rapid data searches and efficient parameter selection, significantly reducing the computational complexity of identifying relevant features or parameters in large datasets. Quantum-enhanced error correction mechanisms further ensure data reliability, addressing issues like noise and missing values during the acquisition process [59]. By integrating QEC techniques, quantum systems can reduce the preprocessing required for noisy or incomplete datasets, improving the overall quality of acquired data. In medical diagnostics, selecting the most relevant biomarkers from vast genomic datasets is computationally expensive. Quantum algorithms such as the QAOA [28] can streamline this process by exploring the solution space efficiently. This capability significantly accelerates the identification of critical features for disease prediction models, enabling faster and more accurate diagnostic tools. Quantum sensors are being used for high-precision environmental data acquisition, such as monitoring greenhouse gas levels or detecting minute changes in atmospheric conditions. By utilizing quantum correlations, these sensors provide unparalleled sensitivity, allowing for real-time monitoring of climate change effects [15], [16].

2.1.3. Quantum Data Representation

In classical computing, data is typically represented as bits (binary values) and organized into structures like vectors, matrices, or tables. While effective for small-scale problems, these representations become inefficient when handling high-dimensional data. Classical methods often struggle with the curse of dimensionality, where computational resources grow exponentially with the size of the data. This limitation significantly affects tasks like feature selection, clustering, and optimization in fields like healthcare and finance. This demands for the encoding of the data as shown in Figure 11.

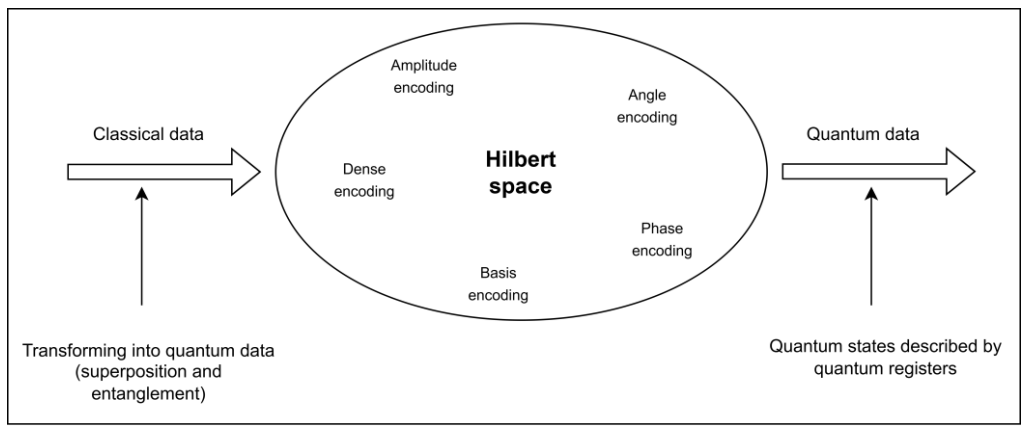

*Figure 11: Diagram illustrating various quantum data encoding methods for transforming classical data into quantum states within Hilbert space.*

Quantum systems address this challenge by representing data in quantum states using qubits. Unlike classical bits, qubits can exist in superposition, allowing quantum systems to encode and process exponentially larger datasets efficiently. For instance, an n-qubit system can encode $2^n$ states simultaneously, drastically reducing storage and computation requirements [13][21]. Quantum approaches also support unique encoding techniques such as amplitude encoding, basis encoding, and phase encoding, enabling efficient processing of high-dimensional data. Quantum data representation utilizes superposition and entanglement to address classical data processing challenges. Figure 12 shows some of the prominent encoding techniques.

- Amplitude Encoding: Amplitude encoding represents classical data as the amplitudes of a quantum state. By encoding $2n$ data points into n qubits, this method offers an exponential reduction in memory usage. It is particularly useful for processing high-dimensional datasets, as demonstrated in quantum algorithms like the QPCA. The data normalization requirement ensures efficient representation, making amplitude encoding suitable for quantum linear systems and ML tasks. For instance, in solving linear systems of equations, amplitude encoding enables exponential speedup over classical methods while maintaining accuracy and scalability. However, implementing this encoding can be challenging due to the need for precision in preparing quantum states. The method's efficiency and compactness are well-suited for quantum computers, especially in handling complex problems with large datasets efficiently [61].

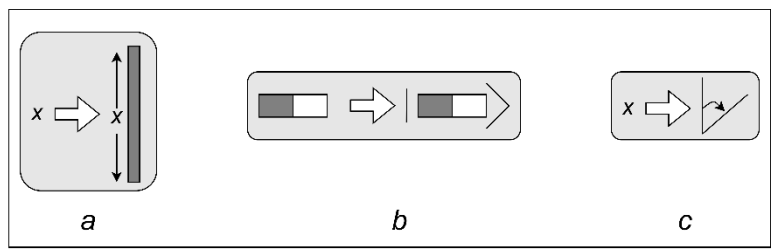

*Figure 12: Schematic representation of a). amplitude encoding method for converting classical data (x) into quantum states through mapping onto state amplitudes. b). of basis encoding technique for mapping classical binary data into quantum basis states through direct state correspondence. c). angle encoding technique for mapping classical binary data into quantum basis states through angular transformation.*

- Basis Encoding: Basis encoding maps classical data directly onto the computational basis states of qubits. Each bit of classical information corresponds to a specific qubit, making it ideal for binary data representation. For instance, a three-bit binary string 101 is encoded as the quantum state |101>. This straightforward encoding scheme enables efficient implementation of Boolean functions and binary operations on quantum systems. Basis encoding is advantageous due to its simplicity and compatibility with current quantum hardware. However, it is constrained by the number of qubits available, limiting its scalability for larger datasets. Despite these limitations, basis encoding has been effectively utilized in quantum-enhanced feature spaces, enabling faster computations and novel approaches in ML applications [62].
- Phase Encoding: Phase encoding leverages the phase rotation of quantum states to represent classical data. By encoding data values into the phase of a quantum state, it offers an efficient method for representing continuous variables. This approach is particularly effective in hybrid quantum-classical systems, where phase rotation is used to encode features for VQAs. For instance, quantum linear systems exploit phase encoding to perform complex computations, achieving exponential speedup for certain problems. While this method ensures efficient utilization of qubits, it requires precise control over phase rotation, which can be technically challenging. Despite this, its integration into quantum systems demonstrates its potential for advancing tasks like optimization and data analysis [61].
- Angle Encoding: Angle encoding maps classical data to rotation angles of quantum gates applied to qubits. Each data point is represented by the angular rotation around one or more axes, enabling an intuitive and compact representation. This method has proven effective in QML, particularly in hybrid models like quantum-classical Convolutional Neural Networks (CNN). By applying parameterized rotation gates, angle encoding facilitates feature mapping and dimensionality reduction. For example, learnable rotation angles allow dynamic adjustments to the encoded data, improving the adaptability of QNN to complex datasets. However, the method may require numerous qubits and gates for encoding large datasets, posing a challenge for implementation. Nevertheless, its compatibility with current variational algorithms makes it a promising technique for quantum data representation [63].
- Dense Encoding: Dense encoding exploits quantum entanglement to encode multiple classical bits into fewer qubits, offering an efficient data representation strategy. For example, a single entangled qubit pair can encode two classical bits of information, enabling more compact data transmission and storage. This method is widely used in quantum communication, where dense coding enhances data throughput and reliability. In experimental setups, dense encoding has demonstrated its potential for optimizing resource use, particularly in distributed quantum systems. While it requires pre-established entanglement, the benefits of reduced quantum resource consumption and increased data capacity outweigh its limitations. Dense encoding's utility extends to scenarios requiring high data efficiency, including secure quantum networks and fault-tolerant computation [64].

These quantum encoding methods enable data-intensive tasks, such as feature selection, clustering, and pattern recognition, to be performed more efficiently. Quantum systems also allow for the integration of native quantum data, bypassing classical-to-quantum conversion processes and directly leveraging quantum-native datasets from sensors or simulators [30][59]. Feature selection in medical datasets often involves identifying critical biomarkers from vast amounts of genomic data. Classical systems struggle with the computational cost of evaluating all possible feature subsets. Quantum algorithms like Grover's search and QAOA enable rapid exploration of solution spaces, significantly reducing the complexity and time required for feature selection [28][58]. For example, QDA has been used to prioritize genetic markers associated with specific diseases, enabling more accurate and personalized treatments. In fields like finance, where portfolio optimization requires analyzing vast datasets of market trends and asset correlations, QPCA can identify dominant factors faster than classical PCA. By encoding data into quantum amplitudes, QPCA leverages quantum parallelism to process large-scale financial data efficiently [30][61]. Quantum data representation has been applied to model the interactions of numerous climate variables, such as temperature, humidity, and $CO_2$ levels. By encoding these variables into quantum states, researchers can simulate complex systems more efficiently, enabling faster predictions of long-term climate patterns [28][31].

### 2.1.4. Quantum Data Preparation and Cleaning

Data preparation and cleaning are foundational steps in any data analysis pipeline, ensuring that datasets are consistent, accurate, and suitable for further processing. In classical systems, this involves tasks such as normalization, standardization, handling missing values, and outlier detection. These steps mitigate errors and biases that could compromise the quality of results. In quantum systems, data preparation faces additional challenges due to the sensitivity of qubits to environmental noise and decoherence. These factors can distort quantum states, making data cleaning not only a necessity but also a critical determinant of the system's overall performance. Addressing these issues requires specialized techniques, such as QEC, hybrid frameworks, and quantum feature mapping. QC enhances data preparation and cleaning in the following ways:

- Techniques like Shor's Code protect quantum data by encoding it redundantly, ensuring resilience against errors caused by decoherence and noise [21][59]. These methods maintain the integrity of quantum states during computations.
- Classical data is embedded into quantum-enhanced feature spaces through quantum feature mapping, allowing algorithms to filter noise while preserving key data characteristics. This approach improves the robustness of models by reducing the influence of corrupted inputs [62].
- Hybrid systems combine QEC with classical statistical methods to enhance data cleaning. Classical preprocessing can normalize raw data, while quantum algorithms fine-tune and process it, ensuring a seamless and resilient workflow [31].
- Quantum data cleaning can accelerate the identification of relevant biomarkers from noisy genomic datasets. For instance, hybrid frameworks can preprocess large datasets to remove redundancies and then apply quantum algorithms like the QAOA for efficient feature selection [28][40]. This not only reduces computational complexity but also enhances diagnostic accuracy by focusing on the most critical features.
- Quantum-enhanced sensors collect high-dimensional data with exceptional precision but are prone to noise. QEC ensures that environmental datasets, such as those used for climate modeling, remain reliable despite these challenges [15][16]. Combining this with classical imputation techniques creates robust and actionable insights for environmental studies.

Despite these advancements, quantum data cleaning faces limitations due to the noise sensitivity of quantum systems and the resource-intensive nature of encoding high-dimensional classical data into quantum states. Ongoing research into fault-tolerant QC and efficient hybrid frameworks is crucial to overcoming these barriers [21][59]. As quantum technologies evolve, enhanced data preparation and cleaning methods will play a pivotal role in unlocking the full potential of QDA.

### 2.1.5. Exploratory Quantum Data Analysis

EQDA is a quantum-enhanced approach to traditional exploratory data analysis (EDA) that leverages the principles of quantum mechanics to process and analyze data. EQDA is used to uncover patterns, identify outliers, and reduce dimensionality in complex datasets, all while exploiting quantum advantages like parallelism and quantum superposition. Unlike classical methods, which analyze data one configuration at a time, quantum algorithms can examine multiple possibilities simultaneously, thus accelerating the identification of underlying structures in large-scale datasets.

Traditional EDA techniques such as PCA and clustering often face limitations when dealing with high-dimensional datasets due to the computational overhead involved in processing large amounts of data. Quantum approaches can significantly enhance these techniques by leveraging quantum parallelism, allowing simultaneous exploration of multiple features or data points. For instance, QPCA [30] accelerates the extraction of principal components from large datasets by using quantum systems to perform calculations exponentially faster than classical methods. Quantum-enhanced clustering algorithms, such as the quantum version of k-means (q-means) [35], exploit quantum interference to find cluster centers with reduced complexity, offering a quadratic speedup compared to classical approaches. These quantum enhancements can be particularly impactful in tasks like image recognition, genomic data clustering, and market analysis, where datasets are often too large and complex for classical methods to handle efficiently.

A prime example of the potential of EQDA lies in the clustering of genomic data, where classical k-means clustering often struggles with the complexity of high-dimensional data. The quantum k-means algorithm [35] offers a quantum-enhanced approach to unsupervised ML, which can reduce the computational cost of clustering by leveraging quantum superposition and entanglement. This allows for the processing of larger datasets with greater speed and efficiency, making it a valuable tool in fields like genomics and bioinformatics, where rapid analysis of vast genetic datasets is essential. In the realm of finance, EQDA can help in pattern recognition for market predictions. Classical methods rely on analyzing past data trends to forecast future behaviors, but these methods often fail to account for complex, non-linear relationships in data. Quantum-enhanced data analysis can reveal hidden correlations between multiple variables simultaneously, allowing for more accurate predictions and risk assessments. Researchers are actively working on several solutions to improve the practicality of EQDA. One major area of focus is the development of QEC techniques, which aim to minimize the impact of noise in quantum computations. These techniques, such as Shor's code [59], could make quantum algorithms more stable and reliable in real-world applications.

Moreover, advancements in quantum hardware are expected to improve the scalability of QDA. As quantum processors with more qubits and higher coherence times become available, EQDA will be able to handle larger datasets more effectively. Furthermore, hybrid quantum-classical algorithms [40] are being explored to integrate the power of quantum systems with the reliability of classical computing, thereby improving the overall efficiency and applicability of EQDA in various domains.

### 2.1.6. Feature Selection and Engineering

Feature selection is a crucial step in data analysis, where the goal is to identify the most relevant variables or features from a dataset to improve model performance, reduce overfitting, and decrease computational cost. In classical ML, algorithms like SVMs rely on preselected features, and the process can become computationally expensive when dealing with high-dimensional datasets. In such scenarios, selecting important features becomes a bottleneck, especially for healthcare diagnostics, where large medical datasets must be analyzed for patterns related to disease prediction. QC introduces significant advantages to feature selection through quantum-enhanced algorithms that can efficiently process high-dimensional data spaces. Quantum approaches, such as the SVM, utilize quantum states to encode data in higher-dimensional spaces where patterns may be more easily separated. This allows for the exploration of more complex relationships within the data that classical methods may miss, offering potential exponential speedups. Quantum-enhanced kernel methods, such as those used in QSVM, enable the computation of inner products in high-dimensional spaces faster than classical counterparts, making them particularly effective in selecting relevant features for ML tasks. The use of quantum systems allows for the parallel processing of multiple potential feature combinations. Quantum algorithms like Grover's search (for searching unstructured databases) and QAOA have been shown to significantly speed up the search for optimal feature subsets by leveraging quantum parallelism and superposition [15][28]. Quantum circuits can handle exponentially larger datasets and model higher-dimensional spaces more efficiently than classical algorithms [31].

In healthcare diagnostics, QSVM has shown potential in improving the accuracy of classifying medical images or genomic data. For example, in the context of cancer diagnosis, QSVM can identify critical biomarkers in genomic datasets more efficiently than classical SVMs. Classical SVMs may require the use of a kernel trick to map data into a higher-dimensional feature space, but SVMs can accomplish this task more efficiently due to their ability to operate in exponentially larger feature spaces, thereby improving classification accuracy and reducing training time. Research has demonstrated that quantum feature spaces can offer advantages in situations where classical methods face scalability issues [35][62]. Despite the promising capabilities of quantum-enhanced feature selection, several challenges remain. Quantum systems are still vulnerable to noise, and qubits in current quantum hardware are prone to decoherence, which can degrade the performance of quantum algorithms. NISQ devices, which are the current state of quantum hardware, are limited by qubit coherence times and gate fidelities [59]. These limitations affect the reliability of quantum circuits used for feature selection and optimization tasks.

Additionally, the scalability of data representation in quantum systems remains a challenge. Encoding large datasets into quantum states requires complex quantum circuits, and as the dimensionality of the data grows, so do the demands on quantum resources [61]. Therefore, efficient encoding techniques and error-correction protocols are essential to ensure the scalability of quantum feature selection methods. Finally, while quantum algorithms offer speedup, they often still depend on classical preprocessing steps. For example, data normalization, outlier detection, and feature extraction still typically rely on classical methods before being fed into quantum systems. This hybrid approach is essential to make quantum feature selection techniques practical with current quantum hardware, but it also adds complexity to the process. In conclusion, while quantum-enhanced feature selection offers substantial promise, it is still an emerging field. The development of more robust quantum hardware and the refinement of hybrid frameworks will play a critical role in realizing the full potential of QDA in practical applications such as healthcare, finance, and beyond.

### 2.1.7. Algorithm Selection

Algorithm selection in QC involves choosing the most suitable quantum algorithm for a specific task or problem. In the context of ML, this process ensures the chosen algorithm can solve the problem more efficiently compared to classical methods. For example, in healthcare diagnostics, selecting an optimal algorithm for feature selection could significantly reduce the computational complexity and enhance performance. Quantum algorithms improve traditional methods by utilizing quantum properties such as superposition, entanglement, and interference. These properties allow quantum systems to perform computations more efficiently. The QAOA has been shown to offer significant improvements in solving combinatorial optimization problems compared to classical algorithms [28]. Quantum algorithms like the SVM and Quantum Discriminant Analysis leverage quantum states to process large feature spaces more efficiently, leading to faster and more accurate solutions in various ML tasks, such as feature selection in healthcare diagnostics [62].

Quantum-Enhanced Feature Selection: In healthcare diagnostics, quantum algorithms like QDA can reduce the complexity of feature selection from a large set of variables. This is especially beneficial in cases where classical feature selection methods are computationally expensive. By using quantum algorithms, the system can process larger datasets and identify the most relevant features more efficiently [62].

SVM in Drug Discovery: QSVMs have been used to classify chemical compounds based on their molecular structure in drug discovery. Quantum-enhanced versions of SVM outperform classical SVMs by efficiently handling high-dimensional data, a common challenge in molecular biology [62].

### 2.1.8. PostProcessing Ouput

The Post-Processing Output phase in QML is vital for translating quantum algorithm outputs into actionable insights. Since quantum computations often yield complex probabilistic data, this step ensures that raw quantum results are processed, refined, and presented in a format comprehensible to researchers, practitioners, or stakeholders. Post-processing in quantum computations is a multi-faceted process that ensures raw quantum outputs are transformed into meaningful and actionable results. The first step results decoding, where the probabilities derived from quantum state measurements are interpreted into usable forms such as class labels, cluster indices, or regression predictions. For instance, in classification tasks, the category corresponding to the highest probability measurement is selected, while in optimization problems, the quantum state that represents the most optimal solution is extracted and validated. An example of this is the use of quantum variational algorithms for regression, where the measured quantum amplitudes are translated into numerical predictions. Quantum systems, however, face challenges from hardware imperfections, environmental noise, and limited qubit coherence times, making error mitigation a critical step. Techniques like error extrapolation adjust measured outputs by extrapolating to an idealized, noiseless scenario. Noise-aware modelling incorporates hardware noise profiles into calculations, while data smoothing filters out anomalies from quantum measurements. These methods significantly enhance the accuracy and reliability of the quantum results.

Once processed, the quantum outputs undergo evaluation and validation to ensure their accuracy, scalability, and practical applicability. This includes comparing quantum models against classical algorithms to identify computational advantages, applying cross-validation techniques to test results across multiple datasets, and analysing performance using metrics such as accuracy, precision, recall, or mean squared error. Validation ensures that the quantum solution is aligned with real-world requirements and free from bias or overfitting. Data visualization plays a key role in interpreting quantum results and communicating findings effectively. Tools generate graphs, charts, and 3D models to present insights. For example, decision boundaries for classification tasks can be visualized, error trends can be depicted using bar graphs or line charts, and optimization landscapes can be illustrated through heatmaps or 3D plots. These visualizations make it easier for stakeholders and researchers to comprehend the practical implications of quantum results.

Finally, integration with classical systems enables a hybrid approach, leveraging the strengths of both paradigms. Quantum outputs can be fed back into classical models through feedback loops for iterative improvement, and they are formatted for compatibility with standard software systems to enable seamless downstream processing. Together, these steps ensure that quantum computations are effectively translated into practical applications across various industries.

## 2.2. Recent Developments

QDA is rapidly emerging as a transformative field, leveraging the principles of quantum mechanics to overcome the limitations of classical methods in handling complex and large-scale datasets. Table 3 provides an insight of the surveys done related to QDA.Traditional data analysis techniques often struggle with the growing computational demands associated with big data and high-dimensional datasets. QC presents new opportunities by utilizing quantum phenomena like superposition and entanglement, which enable faster and more efficient processing [65]. The work also emphasizes the persistent obstacles in quantum systems, such as noise, limited qubit resources, and specialized hardware requirements. These challenges highlight the need for continued advancements in both hardware and algorithmic design to fully realize the potential of QC in data analytics.

The comparison between classical and quantum methods [66] contrasts traditional techniques with emerging quantum algorithms. Classical algorithms, particularly in real-time data processing and high-dimensional spaces, often struggle due to their sequential nature. In contrast, quantum algorithms exploit parallelism and state-space complexity, offering exponential speed-ups in tasks such as clustering, PCA, and real-time analytics. Despite these advantages, quantum systems face significant practical deployment challenges, including decoherence and high error rates, which can hinder their effectiveness. Hybrid quantum-classical solutions [66], offer a promising pathway to mitigate these limitations while maximizing computational efficiency by integrating the strengths of both paradigms. In advanced data analysis, quantum algorithms hold promise for geometric and topological tasks. Quantum algorithms designed for tasks like persistent homology and the computation of Betti numbers are critical for identifying structures and patterns in high-dimensional datasets and have shown significant speed-ups over classical techniques [67]. Similarly, QCNN [68] leverage the entanglement properties of quantum systems to address the curse of dimensionality in high-energy physics datasets. QCNNs demonstrate enhanced pattern recognition capabilities, offering the potential to revolutionize data analysis in fields such as genomics and particle physics.

A key challenge in QML is the effective encoding of classical data into quantum systems [69]. Various encoding techniques, such as amplitude encoding, basis encoding, and dense angle encoding, form the foundation for QML. Their comparative analysis highlights the sensitivity of amplitude encoding to noise, which necessitates robust error correction mechanisms. The choice of encoding has a direct impact on the scalability and performance of quantum algorithms, making this an active area of research [70]. Further exploration of quantum algorithms like the Quantum Fourier Transform and Grover's algorithm enhances data representation and signal processing, streamlining preprocessing in quantum workflows and demonstrating their potential to transform data analysis pipelines. QML models, such as SVMs and Variational Quantum Classifiers (VQC), extend QC's impact beyond data representation to pattern recognition and optimization tasks. Quantum kernel methods in QSVMs and VQCs enable handling complex datasets with overlapping distributions, improving classification accuracy and reducing training times compared to classical models [71]. Furthermore, quantum optimization techniques, including quantum annealing and hybrid quantum-classical algorithms, excel in navigating vast solution spaces, making them ideal for tasks like feature selection and hyperparameter tuning.

Despite these advancements, the practical realization of QDA is constrained by several challenges. Scalability issues of current quantum hardware, including the trade-offs between qubit quality and quantity [65]. To overcome these limitations, further research into hardware development and error correction techniques is essential. Hybrid quantum-classical systems [66] provide a viable solution by integrating the computational power of quantum systems with classical systems to enhance data analysis capabilities. Finally, benchmarking quantum algorithms is crucial for advancing the field. Establishing reliable metrics to compare the performance of quantum and classical methods will help identify where QC offers a tangible advantage [67]. Such benchmarks will drive innovation and guide researchers in optimizing quantum techniques for practical, real-world applications. The collective research from [65]– [71] highlights both the remarkable progress and the remaining challenges in the field of QDA. As quantum technologies continue to evolve, their integration into data analytics is poised to redefine how we process and interpret complex datasets, offering unprecedented computational capabilities.

*Table 3: Related Surveys in Quantum Data Analysis*

| Sr. No | Reference and Year | Contribution | Research Gap | Quantum Advantage |
|---|---|---|---|---|
| 1 | [65] & 2016 | Surveyed applications of QC in big data analytics, focusing on superposition and entanglement. | Highlighted limitations in current quantum hardware, such as noise and limited qubits. | Potential exponential speed-ups in big data tasks, including clustering and pattern recognition. |
| 2 | [67] & 2016 | Developed quantum algorithms for topological and geometric analysis, focusing on high-dimensional data structures. | Lack of practical implementation and benchmarks for comparing quantum and classical algorithms. | Significant speed-ups in analyzing high-dimensional datasets for pattern and structure detection. |
| 3 | [70] & 2020 | Explored quantum algorithms like QFT and Grover's algorithm for efficient data representation and signal processing. | Limited applicability of algorithms due to hardware constraints and error rates. | Enhanced preprocessing and efficient data searching capabilities. |
| 4 | [68] & 2022 | Proposed QCNN for analyzing high-energy physics datasets. | QCNN performance on NISQ devices remains uncertain. | Addressed the curse of dimensionality and improved pattern recognition for specialized datasets. |
| 5 | [66] & 2024 | Explored differences between classical and quantum approaches for real-time and high-dimensional data streams. | Limited exploration of hybrid quantum-classical solutions for overcoming quantum hardware limitations. | Enhanced parallelism and state-space complexity for real-time analytics and dimensionality reduction. |

| 6 | [69] & 2024 | Compared classical-to-quantum encoding techniques and their impact on ML accuracy. | Encoding strategies are highly sensitive to noise, limiting scalability for larger datasets. | Improved data representation, ensuring compactness while retaining significant data features for quantum tasks. |
| 7 | [71] & 2024 | Explores the integration of quantum computing with machine learning techniques to address complex AI challenges in data analysis. | Limited applicability of algorithms due to hardware constraints and error rates. | Potential speedup in processing and optimizing AI models |

3. QUANTUM MACHINE LEARNING ALGORITHMS

This section provides an in-depth exploration of various QML algorithms that harness the power of QC to enhance traditional ML methods. As QC evolves, it offers new possibilities for developing algorithms that can tackle complex, high-dimensional problems more efficiently than classical counterparts. Each algorithm discussed in this section represents a quantum-enhanced approach to solving problems in ML, ranging from regression and classification to clustering and RL. The section covers both well-established algorithms, such as Quantum Linear Regression and SVMs, and more advanced techniques like QGAN and Quantum Reinforcement Learning (QRL). Additionally, hybrid quantum-classical models are explored, emphasizing their potential to combine the strengths of both quantum and classical systems. By examining the principles, applications, and potential advantages of each algorithm, this section offers a comprehensive overview of how QC is revolutionizing the field of ML. Notably, this survey paper explores many QML algorithms as shown in Figure.13 at the circuit level, primarily using Qiskit, offering a centralized resource for understanding QC's transformative impact on ML.

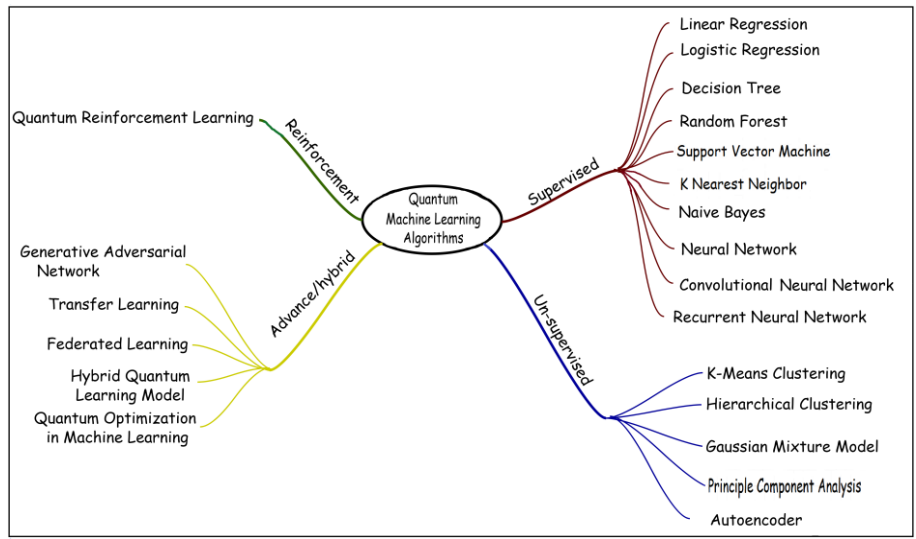

*Figure 13: Comprehensive taxonomy of Quantum Machine Learning algorithms categorizing supervised, unsupervised, reinforcement learning, and advanced/hybrid approaches for quantum computing applications.*

3.1. Supervised Machine Learning

Supervised learning, a core component of machine learning, has seen transformative potential in the realm of Quantum Machine Learning (QML). In this paradigm, quantum algorithms are employed to enhance the performance of classical supervised learning methods, especially for tasks that involve large datasets and complex decision boundaries. This subsection explores the integration of quantum computing with supervised learning, focusing on key quantum algorithms, their advantages over classical approaches, and their applications. Quantum algorithms for supervised learning can accelerate the training and prediction processes by leveraging the power of quantum mechanics, particularly through quantum parallelism and entanglement. Quantum-enhanced versions of classical algorithms such as SVM, KNN, and decision trees are a few examples where quantum computing has shown promise in improving speed and accuracy.

One notable quantum algorithm is the QSVM, which utilizes quantum properties like superposition and entanglement to perform faster data classification tasks. Unlike classical SVM, which relies on solving optimization problems to find the maximum margin hyperplane, QSVM uses quantum feature maps to project data into higher-dimensional spaces. This projection allows the quantum computer to exploit quantum parallelism to process and classify data more efficiently, particularly in cases where the data is non-linearly separable. Another quantum approach, Quantum K-Nearest Neighbour

(QKNN), improves the classical k-NN algorithm by employing quantum algorithms for faster distance computation. The speedup in distance calculation, which is central to k-NN, comes from quantum algorithms like the Quantum Phase Estimation (QPE) and Amplitude Amplification, which can provide an exponential reduction in the number of required steps for solving nearest neighbor searches in high-dimensional spaces. Quantum-enhanced linear regression and logistic regression algorithms also showcase the potential of quantum machine learning for supervised tasks. In these cases, quantum computing accelerates matrix inversion, which is crucial for solving the regression models, and can significantly reduce the time complexity of solving these problems in high-dimensional datasets. Quantum variants of regression are particularly useful in fields like finance, where predictive models require the handling of massive amounts of data for risk assessment and pricing.

However, despite the promise of quantum supervised learning, the practical implementation on current NISQ devices remains challenging. These devices, although powerful, are still prone to noise and errors that can affect the reliability of quantum algorithms. Researchers are actively developing hybrid quantum-classical models that combine the strengths of both paradigms. For example, quantum computers can handle the feature mapping and kernel computation in algorithms like QSVM, while classical computers can perform the rest of the computation, creating a complementary synergy that leverages the advantages of both quantum and classical systems. Overall, quantum supervised learning presents a transformative shift in how machine learning tasks, especially those involving large, complex datasets, are approached. By harnessing the quantum capabilities of superposition, entanglement, and quantum parallelism, these algorithms promise to offer speedups and improvements in accuracy that classical methods may not achieve, especially as quantum hardware continues to improve.

### 3.1.1. Linear Regression

Linear regression is a fundamental statistical method widely used for predictive modeling, where the goal is to establish a relationship between a dependent variable and one or more independent variables. Classically, linear regression is solved by minimizing the sum of squared residuals, typically through closed-form solutions or optimization techniques like gradient descent. However, as the scale and complexity of data grow, classical methods face limitations, especially when dealing with high-dimensional data or large datasets that require significant computational resources. Classical algorithms rely on matrix operations, which are computationally expensive when working with large-scale problems, especially in real-time applications. QC, with its potential to leverage quantum mechanical phenomena such as superposition and entanglement, presents an opportunity to speed up these computational processes. Quantum linear regression aims to address the challenges faced by classical algorithms by harnessing quantum parallelism. Quantum computers can process multiple data points simultaneously due to the nature of quantum bits (qubits), which can exist in multiple states at once. This parallelism offers the potential for exponential speedup, particularly for regression tasks that involve large matrices or complex optimizations. Quantum linear regression, therefore, promises to accelerate the prediction process and handle more complex models efficiently, which is crucial for modern ML and data analytics tasks.

The classical and quantum approaches to linear regression are not mutually exclusive. In fact, a hybrid model that combines classical and quantum techniques often yields the best results. Quantum methods can be used to perform computationally expensive tasks such as matrix inversion or solving linear systems, while classical methods can still be applied for other steps, such as data preprocessing or simple optimization routines. For example, a classical computer could handle feature selection and data processing, while QC could tackle the matrix inversion or optimization phases, providing a synergistic solution to the regression problem. This combination of classical and quantum techniques is referred to as quantum-classical hybrid approaches, and it has been demonstrated to offer improvements in terms of both speed and accuracy in many ML tasks [72][73]. Quantum methods for linear regression implement various algorithms, such as quantum algorithms for matrix inversion or quantum phase estimation, which reduce the computational complexity of solving linear regression models. For example, quantum algorithms for linear regression use quantum operations that can solve systems of linear equations more efficiently than classical counterparts. Quantum phase estimation is one such technique that can be used to find eigenvalues of a matrix, which is central to solving regression equations efficiently. Another approach involves adiabatic QC, where the quantum system evolves slowly towards the minimum of a cost function that represents the best fit for the regression model, achieving optimal solutions without the need for iterative optimization steps. Additionally, techniques such as quantum state tomography can be integrated with quantum linear regression to improve the accuracy of predictions by reconstructing quantum states more effectively [74][75][76].

The applications of quantum linear regression extend across various domains where large datasets need to be processed quickly and accurately. One of the key areas is in QML, where quantum linear regression can be applied to classification and regression tasks, improving the scalability of ML models. In fields like finance, quantum linear regression could be used for real-time risk analysis, where the ability to process large datasets rapidly can significantly reduce the time required to make predictions. Another promising application is in quantum chemistry, where quantum linear regression can help predict molecular properties, which is computationally intensive for classical computers due to the complexity of quantum mechanical systems. Moreover, quantum linear regression could be used in fields like image processing, optimization problems, and even medical diagnostics, where high-dimensional data needs to be analyzed in a short time [77]. Despite the promising potential of quantum linear regression, there are still several challenges that need to be addressed. The first challenge lies in the current limitations of quantum hardware, including qubit coherence times and gate fidelities. Quantum computers are still in the nascent stages of development, and the scalability of quantum linear regression algorithms is yet to be fully realized in practice. Moreover, quantum algorithms often require specialized knowledge to implement, which can limit their accessibility for a wider range of practitioners. Another challenge is noise, which can interfere with quantum calculations and reduce the accuracy of results. Addressing these hardware and algorithmic challenges will be crucial for the widespread adoption of quantum linear regression techniques.

Research Horizons in quantum linear regression include the development of more robust quantum algorithms that can better handle noise and errors, as well as advancements in quantum hardware that can support larger and more complex models. Additionally, research is ongoing to create quantum-classical hybrid models that combine the strengths of both paradigms to achieve faster, more accurate regression models. There is also a growing interest in exploring new quantum algorithms that can be more easily implemented on current quantum devices, expanding the practical applications of quantum linear regression for real-world problems [78][79].

### 3.1.2. Logistic Regression

Logistic regression is a foundational algorithm in classical ML, used primarily for binary classification tasks. It works by fitting a linear model to data, applying the sigmoid function to produce outputs between 0 and 1, which can be interpreted as probabilities. In its classical form, logistic regression involves optimization techniques such as gradient descent to find the best-fitting model parameters. While logistic regression is effective in many scenarios, it struggles with very high-dimensional data or complex relationships between features. As the number of features increases, the optimization process becomes computationally expensive and prone to overfitting, particularly when the data size grows. Figure 14 shows the quantum circuit implementation of logistic regression.

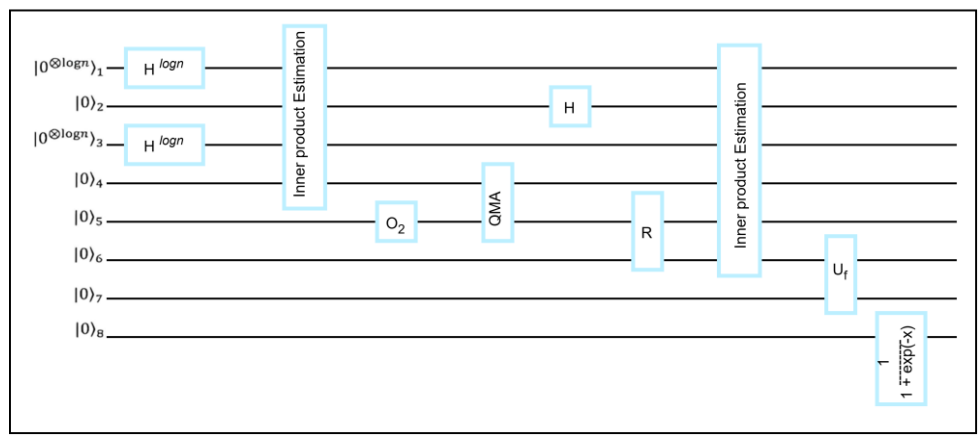

*Figure 14: Quantum circuit implementation of logistic regression featuring Hadamard initialization, inner product estimation modules, and quantum phase estimation for parameter optimization and binary classification [80].*

QC, with its ability to exploit quantum superposition, entanglement, and parallelism, offers a potential solution to the limitations of classical logistic regression. Quantum algorithms have the potential to process information much faster than classical algorithms, especially when dealing with large, high-dimensional datasets. Quantum logistic regression leverages quantum mechanics to speed up the training process, potentially offering an exponential advantage over classical methods. For instance, quantum algorithms can perform matrix inversion or optimization tasks more efficiently, which is crucial in the logistic regression training process. These advantages arise because quantum computers can handle and process exponentially large datasets through superposition, enabling faster convergence in finding the model parameters compared to classical optimization techniques [80][81]. While quantum logistic regression holds promise, it is often implemented in conjunction with classical methods, creating a hybrid approach that leverages the strengths of both paradigms. In quantum-classical hybrid models, quantum algorithms can be used for specific tasks, such as evaluating the quantum kernel in kernel-based logistic regression, while classical computers handle other tasks, like parameter optimization. This hybridization makes quantum logistic regression feasible on current quantum hardware, as the quantum component handles the computationally expensive parts, and the classical component manages other aspects of the ML pipeline. Combining both classical and quantum techniques allows practitioners to overcome the current limitations of quantum computers while still benefiting from their potential speedup in specific operations [80][81]. In practical implementations of quantum logistic regression, quantum circuits are designed to process the logistic regression function efficiently. For example, quantum versions of optimization techniques like Newton's method or gradient descent can be employed to estimate the model parameters. Quantum kernel methods are also integrated into logistic regression to handle non-linear classification tasks. These methods use QC to calculate kernel matrices that map data points into a higher-dimensional space, where linear separation of data becomes possible. Quantum algorithms, such as the quantum kernel logistic regression (QKLR) method, are increasingly used in ML for their ability to perform complex computations more efficiently than classical counterparts. Additionally, quantum state tomography and quantum data encoding techniques can be used to improve the training accuracy and efficiency of logistic regression models [80][82].

The potential applications of quantum logistic regression are vast. In finance, quantum logistic regression could be used for tasks like fraud detection or credit scoring, where large and complex datasets need to be processed quickly. In healthcare, it could be used to predict disease outcomes, such as cancer prognosis, based on high-dimensional genomic data. Quantum logistic regression could also play a role in NLP tasks, such as sentiment analysis or text classification, where the feature space is large and non-linear. Additionally, in fields like image recognition, quantum logistic regression could improve classification accuracy, particularly when dealing with large, high-dimensional image data. The ability to handle such large-scale data efficiently opens new possibilities for QML in real-world applications [81]. However, despite the promising potential of quantum logistic regression, several challenges remain. One of the main obstacles is the current state of quantum hardware. The quantum computers available today are NISQ devices, which are still limited in terms of qubit coherence, gate fidelity, and the number of qubits available for computation. This makes large-scale quantum logistic regression difficult to implement on current hardware. Additionally, quantum algorithms for logistic regression often require large quantum circuits, which can be difficult to scale on today's quantum devices. Another challenge is the need for error correction and mitigation techniques, which are essential to ensure the reliability of quantum computations. Lastly, quantum-classical hybrid models require careful coordination between the quantum and classical components, which can introduce additional complexity to the overall system [80].

Looking ahead, the future of quantum logistic regression will depend on advances in quantum hardware, algorithm development, and hybrid quantum-classical approaches. As quantum hardware improves, with better qubit coherence times and error correction methods, the ability to implement quantum logistic regression on real-world datasets will become more feasible. Furthermore, research into new quantum algorithms that reduce the complexity of implementing logistic regression and optimize quantum-classical hybrid workflows will drive the field forward. As these advancements unfold, quantum logistic regression is poised to play an increasingly important role in various fields, including finance, healthcare, and ML, offering the potential for more efficient and accurate predictive models [82].

### 3.1.3. Decision Trees

In classical computing, decision trees are powerful tools for classification and regression tasks. They function by recursively splitting data based on attribute values, forming a tree structure where each internal node represents a decision rule, and leaf nodes provide final predictions. Despite their interpretability and efficiency, classical decision trees face limitations, such as computational inefficiencies with large datasets and the inability to fully exploit complex data patterns in high-dimensional spaces. These challenges have sparked interest in augmenting classical methods with quantum mechanics, where the inherent properties of superposition and entanglement promise significant computational advantages [83], [84]. QC introduces a fundamentally different paradigm by leveraging qubits to process and represent information. Decision trees have been adapted into this framework to form quantum decision trees, where the data and computation reside in quantum states. Quantum algorithms provide a speedup for constructing and traversing decision trees, particularly for problems where classical trees would face exponential growth in complexity. Early studies demonstrated quantum computation's applicability in decision tree tasks, showing that quantum methods can significantly reduce the query complexity required to solve problems, such as those modeled in decision tree algorithms [84], [85]. The merging of classical and quantum paradigms has further enhanced decision tree capabilities. Hybrid models combine classical preprocessing steps with quantum speedup for tree construction or inference. For instance, classical methods might be used for feature selection, while quantum algorithms handle tree traversal or optimization. These hybrid approaches bridge the gap between classical and QC, making quantum decision tree methods more practical for real-world applications [86], [87]. Quantum methods in decision trees have been implemented using various techniques. Grover's search algorithm, for instance, is used for optimizing tree traversal by reducing the number of required steps to identify specific outcomes. Quantum logic-based frameworks like QLDT employ principles of quantum mechanics for both data representation and tree construction, which have been shown to improve efficiency in terms of computational resources [88]. Such implementations also use entangled states to evaluate multiple branches of the tree simultaneously, thereby offering a speedup over classical counterparts. Figure 15 shows the quantum circuit representation of decision tree.

Applications of quantum decision trees span various domains, including data classification, regression, and cryptography. They have been utilized in QML tasks for faster training and inference, enabling the analysis of massive datasets that are intractable for classical methods. For instance, hybrid quantum-classical approaches have improved data mining and security protocols by offering enhanced speed and accuracy in decision-making processes [86]. Additionally, quantum decision trees have shown potential in agnostic learning scenarios, where they achieve higher efficiency in improper learning tasks compared to classical models [89]. Despite these advancements, several challenges persist. Practical quantum decision tree implementations are constrained by current quantum hardware limitations, such as qubit noise, decoherence, and limited qubit counts. Furthermore, developing quantum algorithms that can be efficiently executed on NISQ devices remains a pressing concern. Scalability and integration with classical systems are other barriers, requiring substantial advancements in hybrid architectures and error correction techniques.

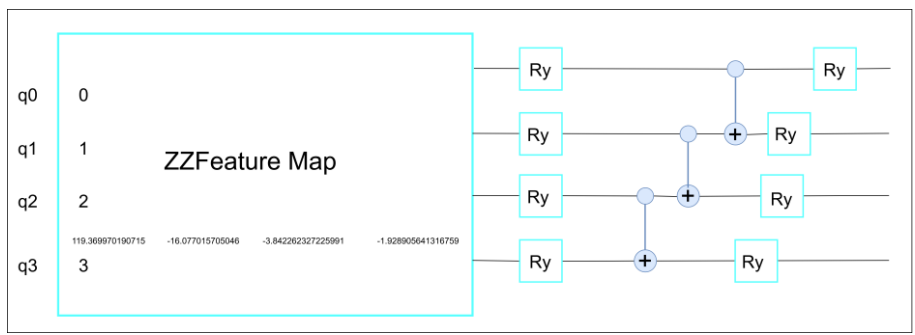

*Figure 15: Quantum circuit representation of a decision tree node using ZZFeatureMap for data encoding, which employs parameterized $R_y$ rotations and entanglement to capture feature relationships, in a quantum random forest implementation for feature-based classification [91].*

Research Horizons involve refining quantum algorithms for decision trees to achieve better performance and robustness on NISQ devices. Research into fault-tolerant QC could alleviate hardware limitations, allowing more complex quantum decision trees to be realized. Additionally, exploring novel quantum-inspired algorithms that can operate effectively in classical systems may yield new insights into optimizing decision-making frameworks. With continued advancements, quantum decision trees hold the potential to revolutionize computational approaches across various scientific and industrial domains.

### 3.1.4. Random Forests

Classical random forests are ensemble learning models that combine multiple decision trees to improve classification and regression tasks. Each tree is trained on a subset of the dataset, and predictions are aggregated to produce a more accurate result. The model's robustness to overfitting and ability to handle complex datasets make it a popular choice in classical ML. However, as datasets grow and complexity, the computational cost of training and inference in random forests increases significantly. These limitations have led researchers to explore QC to enhance the efficiency and scalability of random forest models [90]. QC offers unique advantages due to its principles of superposition, entanglement, and quantum parallelism. These properties allow quantum systems to process and represent data in ways that are fundamentally different from classical approaches. Quantum random forests leverage these capabilities to reduce the computational burden associated with large-scale data analysis. For instance, quantum algorithms can accelerate the construction of decision trees within a forest and optimize the aggregation of predictions. This speedup is especially beneficial for high-dimensional data and problems requiring intensive computation, such as those in cryptography and medical imaging [92], [93]. The integration of classical and quantum paradigms has resulted in hybrid models like quantum random forests. These models often use classical methods for preprocessing or feature extraction and quantum methods for tree construction and ensemble aggregation. For example, kernels based on quantum mechanics have been incorporated into random forest models to improve classification accuracy by exploiting quantum-enhanced feature spaces. Such hybrid approaches allow quantum random forests to leverage the strengths of both computing paradigms, making them feasible even on current NISQ devices [90], [94]. Implementations of quantum random forests have utilized a variety of quantum techniques. One approach is the use of Grover's search algorithm to enhance tree construction by efficiently identifying optimal splits.

Another is the application of quantum kernels, which transform data into higher-dimensional spaces for improved separability. Some algorithms, such as QC-Forest, combine classical and quantum resources to accelerate the retraining of random forests while maintaining high accuracy. These implementations demonstrate the practical potential of quantum random forests in handling computationally demanding tasks [94]. Quantum random forests have found applications across diverse fields. In QML, they have been employed for classification and regression tasks, offering speedup and enhanced accuracy in areas such as NLP and financial modeling. In healthcare, hybrid quantum-classical approaches using random forests have shown promise in predicting disease progression and analyzing medical imaging data. Furthermore, they have been used in quantum cryptography for tasks like optimizing parameter selection in Quantum Key Distribution (QKD) systems, showcasing their relevance in both theoretical and applied domains [92]. Despite their potential, quantum random forests face several challenges. Current quantum hardware limitations, including qubit noise, decoherence, and scalability constraints, hinder the full realization of their capabilities. Developing quantum algorithms that perform effectively on NISQ devices while being compatible with classical preprocessing steps remains an open problem. Additionally, integrating quantum random forests into existing workflows requires overcoming barriers related to interoperability and error correction. These challenges highlight the need for continued advancements in QC hardware and algorithm design [94].

Research Horizons for quantum random forests include refining hybrid models to maximize the utility of quantum resources while addressing hardware limitations. Research into error correction techniques and fault-tolerant QC could unlock the full potential of these models. Additionally, exploring novel quantum-inspired algorithms that mimic quantum behavior on classical systems may provide practical insights into optimizing random forest performance. As QC technology evolves, quantum random forests are poised to play a transformative role in addressing computational challenges across scientific and industrial domains.

### 3.1.5. Support Vector Machine

Classical SVMs are widely used supervised learning models for classification and regression tasks. They operate by finding an optimal hyperplane that maximizes the margin between different classes in a dataset. This is achieved by transforming data into a higher-dimensional space using kernel functions, enabling the separation of linearly inseparable data. While classical SVMs are robust and effective, their performance deteriorates with large-scale or high-dimensional datasets due to computational overhead in constructing the kernel matrix and solving optimization problems. These challenges motivate the exploration of QSVMs, which leverage quantum computational principles to address these limitations [7], [95]. QC introduces a new paradigm where information is processed using qubits that can exist in superposition states. QSVMs take advantage of this quantum parallelism and the ability to encode and manipulate data in high-dimensional Hilbert spaces. Quantum speedups arise in constructing and evaluating kernel matrices, as well as in solving optimization problems using quantum algorithms. These capabilities allow QSVMs to handle tasks involving large datasets and complex features more efficiently than their classical counterparts. For instance, quantum-enhanced kernels can explore feature spaces that are intractable for classical machines, providing a significant edge in classification tasks [7], [96]. The integration of classical and quantum approaches has led to hybrid QSVM models. These models typically preprocess data using classical techniques before feeding it into quantum circuits for kernel computation and optimization. Such hybrid systems make QSVMs compatible with existing quantum hardware, particularly NISQ devices. Recent studies have demonstrated experimental realizations of QSVMs, proving their viability for real-world applications. These implementations often use variational quantum circuits to optimize kernel functions, striking a balance between quantum efficiency and classical practicality [95], [97]. QSVM implementations rely on several quantum algorithms. One notable approach is the quantum kernel method, where data is mapped into a quantum feature space using a quantum circuit. The inner product of quantum states represents the kernel values, enabling efficient computation in high dimensions. Another approach avoids iterative optimization by solving the SVM dual problem directly on a quantum computer, reducing computational overhead. Variational kernel training, an advanced technique, further enhances QSVM performance by adaptively optimizing the quantum kernel parameters for specific datasets [97].

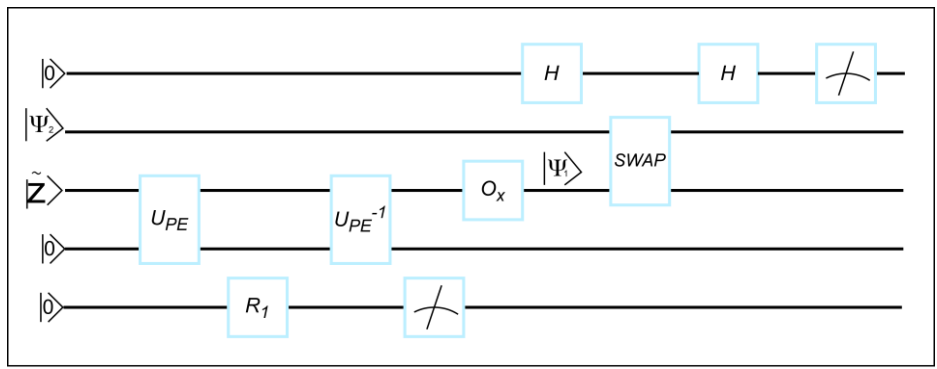

*Figure 16: Quantum circuit for Support Vector Machine classification showing phase estimation (UPE), controlled operations, and SWAP test for computing kernel functions and decision boundaries [98].*

Figure 16 shows the Quantum circuit for Support Vector Machine classification. Applications of QSVMs span various domains, offering advantages in tasks requiring rapid and accurate classification. In healthcare, QSVMs have been used for medical imaging, such as classifying brain tumors with improved precision. In finance, they assist in fraud detection and credit risk analysis by efficiently processing large datasets. QSVMs also play a critical role in quantum-enhanced ML systems, enabling breakthroughs in NLP and image recognition. Their ability to classify noisy and high-dimensional data makes them valuable for problems in quantum chemistry and material science as well [99], [100]. Despite their promise, QSVMs face challenges related to quantum hardware constraints and algorithmic complexity. Current quantum computers suffer from qubit noise, limited coherence times, and scalability issues, which restrict the size of problems that QSVMs can address. Additionally, the quantum advantage of QSVMs is closely tied to the efficiency of quantum kernel computations and the ability to generalize well in real-world datasets. Developing error mitigation techniques and scalable quantum circuits is essential for realizing the full potential of QSVMs. Hybrid quantum-classical frameworks can mitigate some of these issues by leveraging classical preprocessing and postprocessing steps [101], [102].

Future research on QSVMs aims to overcome hardware and algorithmic limitations. Advances in QEC and fault-tolerant QC are expected to significantly expand the applicability of QSVMs. Exploring novel quantum kernels tailored for specific applications and datasets will further enhance their performance. Additionally, improving the interoperability between classical and quantum components in hybrid models will make QSVMs more accessible and effective for industrial use. As quantum technology matures, QSVMs are poised to become indispensable tools for solving complex classification and regression problems across diverse fields.

### 3.1.6. K-Nearest Neighbor

Classical KNN are a simple and effective algorithm used for classification and regression tasks. It operates by identifying the k closest data points (neighbors) to a given query point, based on a chosen distance metric, and then assigns the majority class or computes the mean value from the neighbors. While classical KNN is easy to implement and works well with small datasets, its efficiency declines with large datasets due to high computational costs in distance calculations and sorting operations. These limitations motivate the exploration of QC for KNN, leading to the development of QKNN algorithms [103], [104]. Figure 17 represents the quantum circuit for KNN classification. QC offers significant speedup for problems involving large-scale data by leveraging superposition, entanglement, and quantum parallelism. In QKNN, quantum algorithms accelerate distance calculations and nearest-neighbor searches, enabling efficient processing of high-dimensional and large datasets. For instance, Grover's search algorithm is often used to optimize the neighbor selection process, drastically reducing the computational complexity compared to classical approaches. These quantum capabilities make QKNN a compelling alternative for applications requiring rapid classification and decision-making [103], [105]. The integration of classical and quantum approaches has resulted in hybrid QKNN models. In these systems, data preprocessing and feature extraction are often carried out using classical methods, while quantum resources handle the computationally intensive neighbor identification and classification tasks. This hybrid design makes QKNN implementations compatible with current NISQ devices. Furthermore, hybrid frameworks can incorporate divide-and-conquer strategies to decompose complex classification problems into smaller, more manageable quantum computations, enhancing scalability and efficiency [105], [106]. QKNN implementations rely on a variety of quantum techniques. Quantum distance metrics, such as Hamming distance and polar distance, are computed using quantum circuits to measure similarity between data points efficiently. Additionally, quantum state preparation techniques allow encoding of high-dimensional data into quantum states for parallel processing. Advanced methods, like variational approaches, are employed for dynamic k-value selection and neighbor weighting, optimizing QKNN performance for specific datasets. These techniques enable QKNN to outperform classical KNN in scenarios involving large and complex datasets [104].

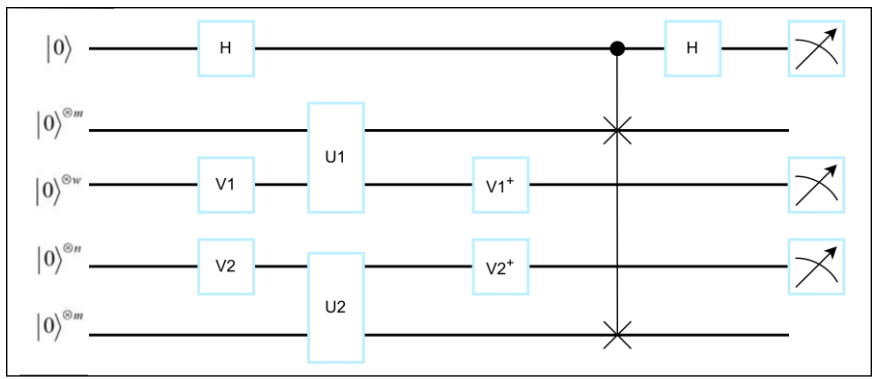

*Figure 17: Quantum circuit diagram for k-nearest neighbors classification using controlled-swap operations and unitary transformations to compute quantum state similarities [107].*

Applications of QKNN span numerous domains. In image processing, QKNN has been used for object recognition and image classification, demonstrating higher accuracy and efficiency compared to classical methods. It has also been applied in medical diagnosis, where rapid and accurate classification of data is crucial for early detection of diseases. Furthermore, QKNN algorithms have shown promise in fields such as NLP, finance, and quantum chemistry, where they provide speedup and improved performance for large-scale classification tasks [106], [108]. Despite its advantages, QKNN faces several challenges. Current quantum hardware is limited by qubit noise, decoherence, and insufficient scalability, restricting the size of datasets that QKNN can handle effectively. Preparing quantum states for high-dimensional data remains a non-trivial task, often requiring extensive classical preprocessing. Additionally, achieving robust generalization and minimizing classification errors in real-world applications require further optimization of quantum distance metrics and neighbor selection strategies. These challenges highlight the need for advancements in both quantum hardware and algorithm design [105].

Research Horizons for QKNN include improving quantum state preparation techniques to handle high-dimensional and noisy data more effectively. Advances in QEC and fault-tolerant QC are expected to expand the practical applicability of QKNN. Developing novel quantum algorithms for adaptive k-value selection and integrating QKNN with other QML models can further enhance its capabilities. As QC technology matures, QKNN has the potential to revolutionize classification and regression tasks across diverse industries and research fields [104], [109].

### 3.1.7. Naive Bayes

Classical Naive Bayes is a simple and effective probabilistic algorithm for classification tasks. It operates under the assumption that features are conditionally independent given the class label. Using Bayes' theorem, it calculates the posterior probabilities of classes for a given data point and assigns the class with the highest probability. While Naive Bayes is computationally efficient and works well with small datasets, its performance suffers when dealing with high-dimensional data or when the independence assumption does not hold. To address these limitations, researchers have explored quantum approaches, leading to the development of Quantum Naive Bayes (QNB) [110], [111]. QC introduces the potential for significant speedup in probabilistic computations by leveraging superposition and entanglement. QNB exploits quantum algorithms to efficiently calculate probabilities and likelihoods, even for high-dimensional datasets. The use of quantum parallelism enables the simultaneous computation of probabilities for multiple classes, reducing the time complexity compared to classical methods. Additionally, quantum techniques can overcome some independence assumptions by representing data in quantum states, allowing richer feature interactions to be considered [111], [112]. The merger of classical and quantum methodologies in QNB involves a hybrid approach. Classical preprocessing steps are often used to prepare data, such as normalizing features or selecting relevant variables. Once the data is ready, quantum circuits are employed to encode the probability distributions and compute the posterior probabilities efficiently. This integration makes QNB adaptable to NISQ devices while leveraging quantum speedups for critical computations. Such hybrid designs enable practical implementations of QNB on current quantum hardware [110]. Implementation of QNB primarily involves quantum state preparation, probability computation, and classification decision-making. Quantum circuits are designed to represent Bayesian networks, where qubits encode the conditional probabilities and relationships between features and class labels. Algorithms such as amplitude amplification and Grover's search enhance the probability computation process, ensuring efficient retrieval of the maximum posterior probability. Furthermore, advanced quantum techniques can incorporate cost-sensitive or adaptive mechanisms, tailoring the algorithm for specific applications, such as imbalanced datasets or real-time decision-making [111]. Figure 18 represents Quantum circuit representation of naive Bayes classifier.

Applications of QNB extend across various fields. In image classification, QNB has demonstrated improved accuracy and efficiency compared to classical methods by effectively managing high-dimensional data. It has also been applied in NLP for text classification tasks, such as spam detection and sentiment analysis. In medical diagnosis, QNB offers potential advantages in probabilistic inference from complex, high-dimensional medical data, aiding in early disease detection and personalized medicine. These applications underscore the versatility and potential of QNB in handling diverse classification problems [112]. Despite its promise, QNB faces several challenges. Current quantum hardware is limited by qubit coherence times, gate fidelity, and the scalability of quantum circuits. Preparing quantum states to encode complex probability distributions is computationally intensive and often requires sophisticated techniques. Additionally, the practical implementation of QNB algorithms depends on the availability of efficient QEC methods to mitigate the impact of noise. These challenges necessitate advancements in both quantum hardware and algorithm design to realize the full potential of QNB [110], [112].

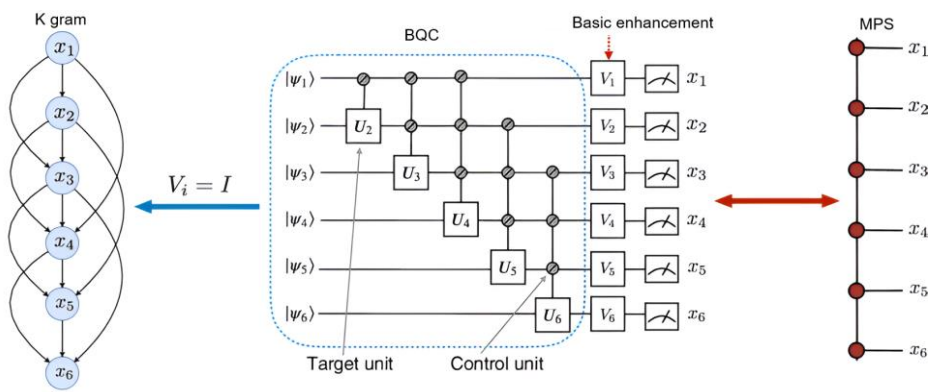

*Figure 18: Quantum circuit representation of naive Bayes classifier showing transformation from K-gram structure to Matrix Product State (MPS) through Basic Quantum Classifier with target and control units [113].*

Research Horizons for QNB focus on improving quantum circuit designs for better scalability and robustness. Research into novel quantum algorithms for state preparation and probability computation is critical for enhancing QNB performance. Hybrid frameworks that seamlessly integrate classical and quantum components can make QNB more accessible and effective for real-world applications. As quantum technology advances, QNB is expected to play a pivotal role in probabilistic reasoning and classification tasks across industries, providing significant computational advantages [111].

### 3.1.8. Neural networks

Classical neural networks are powerful computational models designed to approximate complex functions and relationships within data. These networks consist of layers of interconnected nodes (neurons) that process and transform input data through weighted connections and activation functions. Over decades, classical NNs have been successfully applied to tasks like image recognition, NLP, and generative modeling. However, their performance is limited by the exponential growth in resource requirements for high-dimensional data and complex architectures. These challenges have motivated the exploration of QC to extend the capabilities of neural networks, resulting in the development of QNN [114], [115]. Quantum Neural Network architecture is shown in Figure 19. QC introduces principles such as superposition, entanglement, and quantum parallelism, enabling a new computational paradigm for neural networks. Unlike classical NNs, QNNs encode data into quantum states, allowing multiple computations to occur simultaneously. This parallelism provides exponential speedup for specific tasks, such as high-dimensional data processing and optimization. Additionally, quantum systems can naturally represent and manipulate complex entangled features, offering a fundamentally different approach to learning and decision-making that is infeasible with classical systems [116], [117]. The integration of classical and quantum methodologies forms the foundation of QNNs. In hybrid QNN architectures, classical systems handle data preprocessing and parameter updates, while quantum processors perform core computations like matrix multiplications, non-linear transformations, and feature extraction. This collaboration maximizes the advantages of both paradigms, ensuring practical implementations on current NISQ devices. Hybrid frameworks allow scalable and adaptive learning, making QNNs suitable for diverse applications while accommodating the constraints of quantum hardware [115], [118]. Implementing QNNs involves designing quantum circuits to mimic neural network layers. Each layer represents a quantum operation, such as unitary transformations, that processes quantum-encoded inputs. Training QNNs requires gradient-based optimization techniques adapted for quantum systems, often leveraging parameterized quantum circuits. Advanced strategies, such as VQAs and equivariant quantum networks, improve learning efficiency by tailoring the architecture to the underlying symmetry of data. Additionally, overparameterization theories in QNNs suggest that increasing the number of parameters can enhance learning capacity without overfitting, which is a significant departure from classical NNs [119].

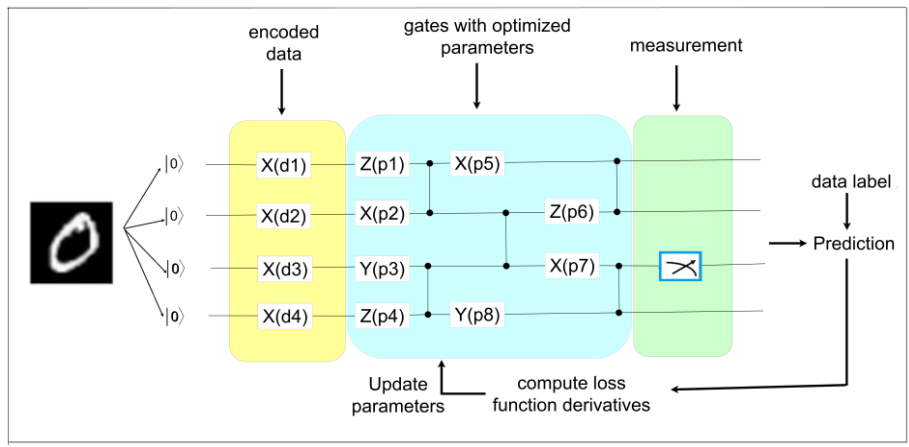

*Figure 19: Quantum Neural Network architecture showing data encoding, parameterized quantum gates, and measurement operations for image classification, with parameter updating through loss function optimization [120].*

Applications of QNNs are expanding rapidly. In generative learning, QNNs have demonstrated the ability to create high-quality synthetic data, which is crucial for tasks like data augmentation and anomaly detection. In healthcare, QNN-based multimodal systems have been employed for intelligent diagnosis, integrating data from diverse modalities such as imaging and genomic data to provide accurate predictions. QNNs have also shown promise in quantum chemistry, enabling efficient simulation of molecular systems and solving optimization problems in drug discovery and materials design. Their versatility positions QNNs as transformative tools across industries [118], [121]. Despite their promise, QNNs face significant challenges. Current quantum hardware is constrained by qubit noise, limited coherence times, and gate fidelity issues, restricting the size and complexity of QNN architectures. Quantum state preparation and encoding high-dimensional classical data into quantum systems are non-trivial and resource-intensive. Moreover, training QNNs is computationally expensive due to the lack of efficient quantum gradient computation methods. These challenges highlight the need for innovations in quantum hardware, error correction techniques, and quantum optimization algorithms [117], [122].

Research Horizons for QNN research focus on addressing scalability and reliability issues. Development of robust QEC mechanisms and advancements in quantum hardware are critical to enabling larger and deeper QNNs. Research into hybrid quantum-classical training algorithms and efficient encoding strategies will enhance their practical applicability. Additionally, exploring novel architectures like equivariant and graph-based QNNs can expand their usability in domains such as network science and cryptography. As quantum technologies evolve, QNNs are poised to revolutionize AI by combining the strengths of quantum mechanics and neural computation [119], [123].

### 3.1.9. Convolutional Neural Network

Classical CNN are widely used for image recognition and analysis due to their ability to extract hierarchical spatial features through convolutional layers. These networks excel in tasks such as object detection, medical imaging, and satellite data analysis. However, as the complexity of image data increases, the computational resources required by classical CNNs grow exponentially. Moreover, traditional CNNs often struggle with certain types of feature correlations that require non-linear transformations. These limitations have driven the exploration of QC to enhance CNN architectures, leading to the development of QNN [124], [125]. QC leverages the principles of superposition and entanglement to process information in fundamentally new ways. In QCNN, quantum circuits replace or augment classical convolutional layers to encode and process data as quantum states. This integration allows for richer feature extraction and improved handling of high-dimensional data while potentially reducing computational complexity. The ability of quantum systems to explore a vast feature space in parallel provides a significant advantage over classical CNNs, particularly for datasets with intricate patterns and correlations [124], [126]. QCNN represent a hybrid architecture combining classical and quantum components. Classical preprocessing prepares input data for quantum encoding, where quantum circuits, such as parameterized unitary operations, perform feature extraction. The processed quantum features are then fed back into a classical neural network for further processing and decision-making. This collaboration ensures that QNNs can operate on current NISQ devices while benefiting from quantum-enhanced computation. Such hybrid systems also allow for incremental adoption of quantum technologies in classical ML pipelines [125], [127]. Implementation of QNNs typically involves designing quantum circuits tailored for specific tasks. For example, variational quantum convolutional layers utilize parameterized quantum gates to optimize feature extraction dynamically. Enhanced image encoding strategies, such as amplitude encoding or angle encoding, ensure efficient utilization of quantum resources. Advanced designs also explore federated quantum learning, allowing distributed quantum computations across devices while maintaining data privacy. These innovations address practical constraints and expand the applicability of QNNs across diverse domains [126].

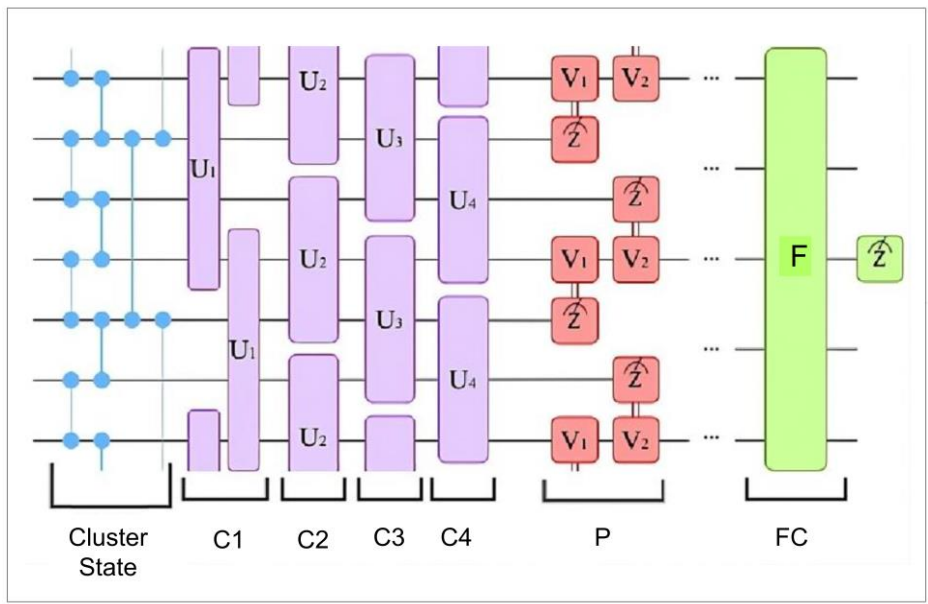

*Figure 20: Quantum Convolutional Neural Network architecture illustrating the sequence of quantum convolution layers (C1-C4), pooling operations (P), and final classification layer (FC) with intermediate unitary transformations [128].*

Figure 20 represents the QCNN architecture. Applications of QCNN are diverse and rapidly growing. In healthcare, QNNs have been used for automated detection of diseases like rheumatoid arthritis using thermal imaging. Earth observation tasks, such as environmental monitoring and disaster management, benefit from QNNs' ability to process satellite imagery efficiently. In federated learning scenarios, QNNs enable collaborative training across multiple quantum systems, preserving data privacy while achieving superior performance. Furthermore, QNNs have demonstrated efficacy in classical image recognition tasks, such as MNIST digit classification, showcasing their versatility and robustness [129], [130]. Despite their potential, QCNN face significant challenges. Quantum hardware limitations, including qubit noise and limited coherence times, restrict the size and depth of quantum circuits. Efficient encoding of high-dimensional image data into quantum systems remains a computational bottleneck. Additionally, training QNNs is resource-intensive due to the need for specialized quantum simulators or access to physical quantum devices. Addressing these challenges requires advancements in quantum hardware, error correction, and optimization algorithms tailored for QML [127].

The future of QCNN lies in overcoming scalability and reliability hurdles. Progress in QEC and the development of fault-tolerant quantum systems will enable deeper and more complex QNN architectures. Enhanced encoding techniques and hybrid training algorithms promise to improve the efficiency and scalability of QNNs. Exploration of domain-specific quantum architectures, such as those tailored for medical imaging or Earth observation, will further expand their impact. As quantum technologies mature, QCNN are poised to transform fields reliant on complex pattern recognition and large-scale data processing [130], [131].

### 3.1.10. Recurrent Neural Networks (RNN)

Classical RNN are a foundational tool in sequential data processing, leveraging their ability to model temporal dependencies and relationships. Applications such as NLP, time series forecasting, and video analysis benefit from the architecture's capacity to retain information across time steps. However, classical RNNs often struggle with issues like vanishing gradients, high computational demands, and limited scalability when handling large or complex temporal datasets. These limitations motivate exploring QC as an alternative, particularly to enhance computational efficiency and representational power [132], [133]. QC introduces the principles of superposition and entanglement, which enable exponential parallelism in processing information. When applied to RNN, quantum systems can potentially represent and process sequential data in higher-dimensional feature spaces wi0074h fewer resources. Quantum Recurrent Neural Networks (QRNN) extend classical RNNs by incorporating quantum circuits into their structure, enabling the modeling of intricate temporal dependencies and patterns beyond the scope of classical systems. This transition leverages quantum advantages in speed and representational richness [133], [134]. The merger of classical and quantum paradigms in QRNNs follows a hybrid architecture. In this framework, temporal data is encoded into quantum states using quantum circuits designed to model sequential dependencies. These circuits, often parameterized for adaptability, serve as the core of the recurrent structure. Classical layers handle input preprocessing and output decoding, allowing QRNNs to operate effectively on NISQ devices while bridging the gap between quantum hardware and real-world applications. This integration ensures QRNNs remain feasible and adaptable during the transitional phase of QC development [135], [136]. Implementing QRNNs involves several quantum methods tailored to sequential learning. Variational quantum circuits are commonly used to encode temporal dynamics, optimizing the representation of sequential data through iterative parameter adjustments. Reservoir computing architectures leverage quantum systems to store and process temporal information efficiently, reducing the reliance on gradient-based optimization. Tensor-network-based approaches also provide scalable methods for representing temporal data in QRNNs. Such implementations address the complexity of training quantum systems while maintaining their temporal modeling capabilities [134], [137].

The applications of QRNNs span various domains requiring advanced sequential learning. In physics, QRNNs reconstruct quantum dynamics by modeling time-dependent systems with high precision. In signal processing, they enhance tasks like speech recognition and sequential RL by exploiting quantum-enhanced feature spaces. Additionally, QRNNs show promise in temporal anomaly detection, enabling applications in finance, healthcare, and cybersecurity. These advantages highlight the potential of QRNNs in tasks where classical RNNs face scalability and efficiency challenges [132], [138]. Despite their potential, QRNNs face significant challenges that must be addressed to unlock their full potential. Current quantum hardware limitations, including qubit coherence times, gate fidelity, and noise, restrict the size and complexity of QRNN architectures. Efficient encoding of sequential data into quantum systems remains computationally intensive, and the lack of robust optimization methods tailored for quantum sequential learning hinders training scalability. Additionally, the interpretability of quantum models, particularly in the temporal domain, poses a challenge for broader adoption. Addressing these obstacles will require advancements in quantum hardware, algorithm design, and hybrid optimization techniques [133], [136].

Research Horizons for QRNNs focus on improving scalability and practicality. Development of fault-tolerant quantum hardware will enable deeper and more expressive QRNN architectures. Integration of domain-specific quantum circuits tailored for applications such as language modeling or quantum state prediction will enhance their versatility. Exploring hybrid quantum-classical training methods, such as RL and transfer learning, will improve the efficiency and adaptability of QRNNs. As these advancements progress, QRNNs are poised to become transformative tools in sequential data processing and modeling [137], [138].

### 3.2. Unsupervised Machine Learning

Unsupervised learning, where the algorithm is tasked with finding hidden patterns or structures in unlabelled data, stands as one of the most promising applications of QML. The ability to detect patterns, clusters, or features in high-dimensional datasets without prior labelling can benefit from the computational power of quantum algorithms. This subsection delves into the integration of quantum algorithms with unsupervised learning tasks, focusing on how quantum computing can enhance clustering, dimensionality reduction, and anomaly detection. In classical machine learning, unsupervised learning tasks such as clustering and dimensionality reduction can become computationally intensive, especially when dealing with large datasets.

Quantum computing offers the potential to significantly accelerate these tasks by utilizing quantum properties like superposition, entanglement, and interference, which allow for faster processing and more efficient exploration of complex datasets.

One of the most notable applications of QML in unsupervised learning is quantum clustering. Quantum clustering algorithms, such as the Quantum K-Means algorithm, exploit quantum parallelism to speed up the identification of clusters in large datasets. In classical K-Means clustering, the algorithm iteratively assigns data points to clusters based on proximity and updates centroids, which can be computationally expensive. The quantum version of K-Means, using quantum superposition to explore potential cluster configurations simultaneously, offers an exponential speedup in finding optimal cluster assignments in high-dimensional spaces. This becomes particularly advantageous in areas like genomics, where datasets are both large and complex. Another important quantum approach in unsupervised learning QPCA, which is an extension of classical PCA. PCA is widely used for dimensionality reduction, helping to reduce the number of variables in large datasets while preserving essential information. The quantum counterpart, QPCA, leverages quantum mechanics to find the principal components of a dataset exponentially faster than its classical counterpart. This speedup is particularly valuable in fields like image recognition and data compression, where handling and processing vast amounts of high-dimensional data is a key challenge. Anomaly detection is another critical unsupervised task that can benefit from quantum computing. Quantum anomaly detection algorithms, such as those based on the QSVM framework which utilizes quantum-enhanced kernels to identify outliers in datasets more efficiently. These quantum techniques can detect anomalies in high-dimensional data with higher accuracy, making them suitable for applications like fraud detection, network security, and industrial monitoring, where spotting rare, unusual events is crucial.

Despite the promising potential of quantum unsupervised learning, challenges remain, especially in the practical implementation of these algorithms on current NISQ devices. These devices are susceptible to errors and noise, which can hinder the stability and accuracy of quantum algorithms. However, hybrid quantum-classical approaches are being developed to address these challenges. In these approaches, quantum algorithms can handle data encoding, feature mapping, or kernel computation, while classical computers manage the optimization and post-processing tasks, providing a robust solution to the noise issues inherent in NISQ devices. In conclusion, quantum-enhanced unsupervised learning represents an exciting frontier in machine learning. By leveraging quantum algorithms, unsupervised learning tasks like clustering, dimensionality reduction, and anomaly detection can achieve significant speedups and improvements in accuracy. As quantum hardware continues to evolve, the application of QML in unsupervised learning will become increasingly powerful, offering new opportunities to extract insights from complex, high-dimensional data across various domains, including healthcare, finance, and cybersecurity.

### 3.2.1. K-Means Clustering

Classical K-means clustering is a widely used unsupervised ML algorithm that groups data points into clusters based on their similarities, minimizing intra-cluster variance. Its efficiency in clustering tasks like customer segmentation, image compression, and anomaly detection is well-known. However, classical K-means struggles with high-dimensional data, convergence on local minima, and computational scalability as data sizes grow. These challenges have driven interest in quantum-enhanced methods that leverage QC's unique capabilities to address these limitations [139], [140]. QC offers a paradigm shift through principles like superposition and entanglement, enabling the exploration of multiple cluster configurations simultaneously. In Quantum K-means clustering, quantum circuits replace classical distance calculations and centroid updates, offering significant computational speedups, especially for large datasets. This shift is motivated by the potential for logarithmic scaling of distance computations and cluster assignments, as opposed to the polynomial scaling in classical methods [141], [142]. The transition from classical to quantum clustering is realized through hybrid approaches. These methods combine classical preprocessing with quantum-enhanced clustering steps. For instance, data encoding into quantum states is achieved using methods such as amplitude encoding or variational circuits, ensuring compatibility with quantum hardware. Quantum routines are then employed for distance computation or centroid adjustment, leveraging quantum parallelism. This hybrid structure addresses the limitations of current NISQ devices while enabling practical implementations [142], [143]. Implementing Quantum K-means requires efficient quantum methods to perform core clustering tasks. Algorithms often use quantum distance measures like the Manhattan or Euclidean distances, calculated via quantum circuits. Variational quantum circuits are employed for centroid updates, optimizing cluster representations iteratively. Hybrid algorithms, where quantum and classical computations alternate, have proven effective in reducing noise sensitivity while maintaining computational advantages. These implementations ensure the algorithm's adaptability to both quantum hardware limitations and complex clustering scenarios [140], [143].

Applications of Quantum K-means span diverse domains where large-scale or high-dimensional data clustering is critical. In healthcare, it facilitates the identification of disease patterns by clustering patient data efficiently. In energy systems, quantum clustering optimizes energy grid classifications, enhancing reliability and efficiency. Quantum K-means also finds applications in finance, where it aids in portfolio clustering and fraud detection. These applications demonstrate its ability to surpass classical clustering methods in both speed and precision [139], [140]. Despite its promise, Quantum K-means faces challenges that must be addressed for widespread adoption. Encoding high-dimensional data into quantum states remains resource-intensive, and existing quantum hardware suffers from noise and limited qubit availability. Moreover, ensuring the stability and scalability of quantum clustering algorithms on NISQ devices is an ongoing concern. Another challenge lies in designing interpretable quantum algorithms, as understanding quantum-processed clusters can be non-trivial. Addressing these challenges will require advancements in hardware, optimization techniques, and hybrid algorithm designs [142], [143].

Research Horizons in Quantum K-means research emphasize hardware improvements and algorithmic innovations. Development of error-corrected quantum systems will enable more complex clustering tasks. Exploration of advanced encoding methods, such as tensor-network representations, can improve algorithm efficiency. Additionally, integrating quantum K-means with other QML methods, like QNN, holds the potential for creating comprehensive quantum-driven data analysis pipelines. As these advancements unfold, Quantum K-means clustering is poised to become a cornerstone in the QML landscape [142], [143].

### 3.2.2. Hierarchal Clustering

Classical hierarchical clustering is a widely used method for organizing data into a hierarchy of nested clusters. It works by either agglomeratively merging smaller clusters or divisively splitting larger clusters based on a predefined distance metric, such as Euclidean or Manhattan distance. While effective in many applications, classical hierarchical clustering is computationally expensive for large datasets due to the iterative calculation of pairwise distances and cluster assignments. This limitation becomes critical as the size of the dataset increases, making the classical approach infeasible for modern, high-dimensional data analysis. QC introduces a paradigm shift by leveraging quantum principles, such as superposition and entanglement, to solve problems more efficiently than classical methods. In the context of hierarchical clustering, QC offers the potential for exponential speedups in calculating distances and processing large datasets. For instance, quantum states can encode high-dimensional data, and quantum algorithms can perform distance calculations and similarity checks in parallel. These capabilities address the computational bottlenecks of classical hierarchical clustering, making it more scalable and efficient. The merging of classical and quantum approaches has led to hybrid quantum hierarchical clustering methods. Classical algorithms provide a robust foundation for defining clustering frameworks, while quantum algorithms optimize specific sub-tasks, such as distance calculations or cluster centroid updates. For example, quantum annealing has been employed for combinatorial clustering problems, where the quantum system finds the global minimum of a cost function that represents optimal cluster assignments. Similarly, quantum state preparation techniques enable the encoding of data relationships in quantum states, facilitating faster similarity assessments [144], [145]. Implementation of quantum hierarchical clustering relies on quantum hardware and algorithms designed to exploit quantum properties. Techniques such as quantum annealing or variational quantum circuits are commonly used. In these methods, data points are encoded into quantum states, and clustering operations are performed using quantum gates or annealing processes. Advanced algorithms like the quantum nearest cluster centroid method utilize quantum registers to store centroids, and quantum circuits to iteratively optimize cluster assignments. These implementations are further enhanced by hybrid models, where classical optimization techniques refine quantum-derived clustering outputs [145], [146].

Applications of quantum hierarchical clustering span various fields, including biology, finance, and image processing. In bioinformatics, quantum clustering aids in identifying gene expression patterns by efficiently analyzing large genomic datasets. In finance, it helps detect hierarchical structures in stock correlations, enabling better risk management. Quantum hierarchical clustering is also applied in image segmentation tasks, where quantum-enhanced computations identify clusters of pixels for feature extraction. Its scalability and efficiency make it suitable for analyzing complex, high-dimensional datasets across diverse domains [146], [147]. Challenges remain in implementing quantum hierarchical clustering. Current quantum hardware suffers from noise and limited qubit availability, restricting the size and complexity of datasets that can be processed. Furthermore, encoding large datasets into quantum states and maintaining coherence during computations are non-trivial tasks. Developing error-corrected quantum systems and efficient quantum algorithms is essential to overcome these limitations. Interpretability of quantum clustering results is another challenge, as understanding the quantum-processed clusters often requires specialized expertise [147].

Research Horizons in quantum hierarchical clustering include improving quantum hardware capabilities and designing more efficient hybrid algorithms. Advanced QEC techniques will enable more stable computations, while innovations in quantum state preparation and circuit design will enhance scalability. Integrating quantum clustering with other QML methods, such as QNN, holds the potential for developing comprehensive data analysis frameworks. As QC matures, quantum hierarchical clustering is expected to become a cornerstone for tackling complex clustering problems in big data.

### 3.2.3. Gaussian Mixture Model (GMM)

GMM are widely used in classical ML for probabilistic clustering and density estimation. They represent the data as a weighted sum of multiple Gaussian distributions, enabling flexible modeling of complex datasets. Classical GMMs employ the Expectation-Maximization algorithm to iteratively optimize the parameters of these distributions, maximizing the likelihood of the observed data. However, as the data's dimensionality and size increase, the computational cost of performing EM becomes prohibitive, especially for large-scale datasets. This limitation prompts the exploration of QC for accelerating GMM computations. QC offers inherent advantages in handling high-dimensional datasets due to its exponential scaling properties and the ability to leverage quantum superposition and entanglement for parallel computation. Quantum GMMs exploit these capabilities, particularly for the EM algorithm. By encoding data into quantum states, quantum algorithms can compute distances, update parameters, and evaluate probabilities exponentially faster in certain scenarios. For instance, the quantum EM algorithm achieves speed-ups in computing probabilities and parameter updates by utilizing quantum amplitude amplification and estimation techniques, significantly reducing the computational complexity compared to its classical counterpart [148], [149]. The hybridization of classical and quantum methodologies involves leveraging quantum algorithms to perform the computationally intensive parts of the EM algorithm while retaining classical methods for other tasks. For example, quantum subroutines can optimize the log-likelihood calculation and parameter estimation, while classical processing handles data preprocessing and post-analysis. This integration ensures practical applicability, as current quantum devices are still limited by noise and scalability [148], [150]. Quantum GMM implementation typically uses quantum circuits to encode data and perform operations like amplitude estimation and sampling. Variational approaches, wherein parameters are optimized on classical computers while quantum devices execute specific quantum operations, are also employed. Techniques such as using Gaussian mixture priors in VQAs have shown potential in addressing issues like barren plateaus, which are common in deep QNN [150].

Applications of quantum GMMs span various domains, including image and speech recognition, financial modeling, and medical diagnostics. These models excel in clustering high-dimensional data, where classical GMMs often struggle with computational inefficiencies. Additionally, they are valuable for quantum noise modeling, helping improve quantum communication channels and error correction [151].

Despite their promise, quantum GMMs face challenges, including the noise and decoherence in current quantum hardware, limitations in encoding large datasets efficiently, and ensuring convergence of the quantum EM algorithm. Research Horizons include developing fault-tolerant quantum devices, optimizing hybrid quantum-classical frameworks, and designing novel quantum algorithms to enhance the scalability and reliability of GMM implementations [148], [151].

### 3.2.4. Principal Component Analysis

Classical PCA is a foundational dimensionality reduction technique used to identify patterns in data by projecting it onto orthogonal components that capture maximum variance. By using eigenvalue decomposition or singular value decomposition, PCA reduces data dimensionality while preserving essential information. However, as datasets grow and complexity, classical PCA faces computational bottlenecks due to the polynomial scaling of time complexity with the number of dimensions and data points. This limitation becomes especially challenging for large-scale datasets in fields like genomics, finance, and image analysis. QPCA addresses these computational challenges by utilizing quantum mechanics to achieve potential exponential speedups in specific scenarios. QPCA encodes the covariance matrix of a dataset into a quantum state, leveraging quantum algorithms such as quantum phase estimation to extract eigenvalues and eigenvectors efficiently. The work by Lloyd et al. demonstrated the foundational algorithm for QPCA, showing that quantum systems can analyze the covariance matrix in logarithmic time concerning its size, given an efficient quantum state preparation process [152]. This efficiency stems from the ability of quantum computers to process and represent exponentially large data spaces using qubits. The integration of classical and quantum methodologies combines the strengths of both approaches. Classical methods handle pre-processing tasks like data cleaning and normalization, while quantum algorithms execute the computationally intensive PCA operations. Hybrid quantum-classical frameworks leverage parameterized quantum circuits to approximate principal components, providing flexibility for NISQ devices. Experiments conducted on existing quantum hardware have validated the theoretical speedups, such as the use of parameterized quantum circuits for QPCA implementations [152]. These experiments highlight the feasibility of QPCA even with the limitations of current quantum systems. Figure 21 represents the quantum circuit for PCA.

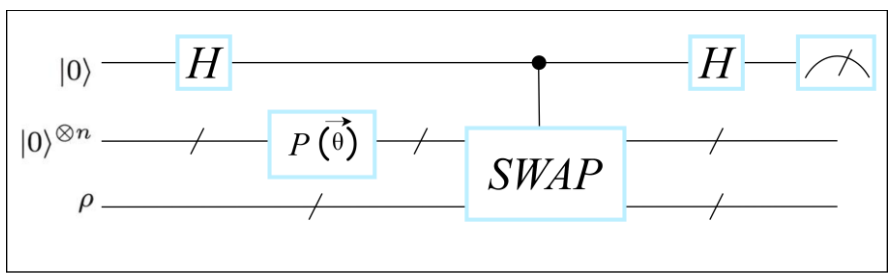

*Figure 21: Quantum circuit for Principal Component Analysis implementing state preparation, controlled-SWAP operations, and measurements to extract principal components through quantum phase estimation [153].*

The applications of QPCA span a wide array of fields. It is particularly valuable in quantum data compression, where high-dimensional quantum states are projected onto lower-dimensional subspaces without significant loss of information. QPCA also aids in anomaly detection, where it identifies patterns that deviate from the norm, making it useful in cybersecurity and fault detection. Moreover, the technique enhances classical ML algorithms by reducing the complexity of input data, improving model efficiency and interpretability [154]. In physics and material sciences, QPCA has been utilized for analyzing quantum systems and detecting emergent properties. Despite its potential, QPCA faces significant challenges. Efficient quantum state preparation remains a critical bottleneck, as the exponential speedups depend on the ability to encode data efficiently into quantum states [155]. The presence of noise and decoherence in current quantum hardware limits the practical implementation of QPCA algorithms. Additionally, the assumptions of sparsity and specific matrix structures restrict the general applicability of QPCA to real-world datasets [156]. Addressing these issues requires advancements in quantum hardware, error correction methods, and algorithmic design.

Future research directions for QPCA focus on developing robust hybrid frameworks that maximize the use of classical preprocessing while leveraging quantum capabilities for computationally intensive tasks. Efforts are also directed towards optimizing quantum state preparation techniques and extending QPCA applications to broader domains, such as quantum chemistry and advanced image processing. As QC hardware evolves, QPCA is expected to transition from a theoretical framework to a widely adopted tool for high-dimensional data analysis [155], [156].

### 3.2.5. Autoencoders

Quantum autoencoders are emerging as a pivotal tool in QML for compressing quantum data, denoising quantum states, and mitigating noise in quantum systems. Inspired by classical autoencoders, which are neural networks designed to learn efficient encodings for input data by compressing it into a latent space and reconstructing it with minimal loss, quantum autoencoders operate directly on quantum states. They aim to identify reduced representations of quantum information while preserving the essential characteristics of the data. Classical methods face limitations when handling high-dimensional quantum systems due to the exponential growth of classical data representations for quantum states. This limitation has fueled the need for quantum-native solutions. Quantum autoencoders address the inefficiencies of classical approaches by leveraging the principles of quantum mechanics. They exploit the inherent high dimensionality of quantum Hilbert spaces to compress quantum states more effectively than their classical counterparts. Early implementations demonstrated the feasibility of quantum autoencoders for reducing the qubit requirements of quantum data storage, significantly optimizing memory usage [157]. Additionally, quantum autoencoders are capable of directly handling entangled states, making them highly suitable for quantum communication and quantum cloud computing scenarios [158]. This efficiency is particularly advantageous for cloud QC, where communication overhead between local and cloud environments can be minimized through compressed quantum representations [157]. Hybrid quantum-classical implementations of quantum autoencoders have also been explored to capitalize on the strengths of both paradigms. Classical pre- and post-processing steps, such as feature extraction or evaluation of cost functions, can complement the quantum optimization process. In practical implementations, variational quantum circuits are often used as the core of quantum autoencoders. These circuits iteratively optimize parameters to learn efficient encodings. Experimental validations on real quantum hardware, including superconducting qubits and trapped ions, have shown that quantum autoencoders can effectively compress quantum data while being robust to noise [159]. Advanced techniques such as quantum adders and genetic algorithms have been employed to enhance the training process, ensuring scalability and precision [160]. Figure 22 illustrates the architecture of a quantum autoencoder, comprising an encoder ($\varepsilon$) and a decoder ($\mathcal{D}$). The figure highlights the intermediate quantum states and measurement operations used for dimensionality reduction and data reconstruction.

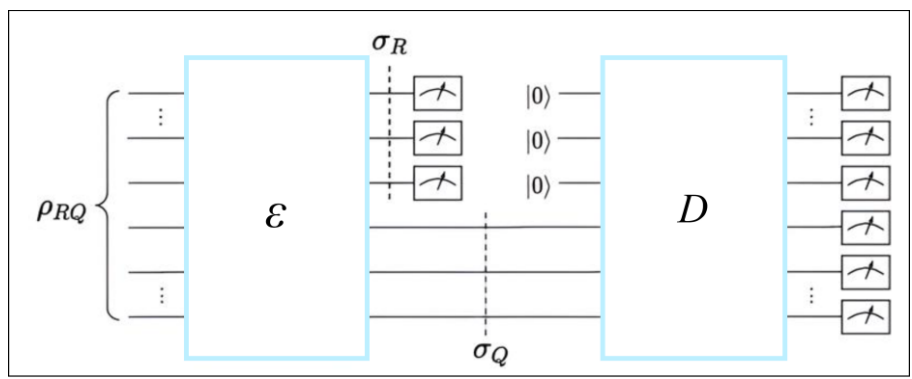

*Figure 22: Quantum autoencoder architecture showing the encoder ($\varepsilon$) and decoder (D) operations with intermediate quantum states and measurement operations for dimensionality reduction and data reconstruction [161].*

The applications of quantum autoencoders are diverse and impactful. They are particularly suited for noise reduction in quantum states, a critical requirement for fault-tolerant QC. By isolating and discarding noise components, quantum autoencoders improve the fidelity of quantum operations. In quantum communication, these tools enable more efficient data transmission by compressing quantum information before sending it across noisy channels. Additionally, quantum autoencoders have been utilized for anomaly detection in time-series data, demonstrating their potential in applications requiring high precision [158]. Beyond these, their role in error correction has been highlighted, where they help identify and mitigate error patterns in quantum circuits [159]. Despite their potential, quantum autoencoders face significant challenges. Training these systems requires sophisticated optimization techniques, as the parameter space grows exponentially with the number of qubits. Moreover, NISQ devices introduce constraints such as decoherence and gate fidelity, which can limit the effectiveness of quantum autoencoders [160]. Efficient state preparation and measurement pose additional hurdles, as they are often resource-intensive processes. Furthermore, extending quantum autoencoders to higher-dimensional quantum systems or real-time applications demands scalable solutions.

Future research in quantum autoencoders focuses on enhancing their robustness to noise and scalability. Techniques such as hybrid quantum-classical frameworks, noise-resilient variational circuits, and error-mitigation strategies are being explored. Moreover, advancements in quantum hardware, including improvements in qubit coherence times and gate fidelities, will play a crucial role in expanding the applicability of quantum autoencoders. Research is also directed toward integrating quantum autoencoders with other quantum algorithms to create end-to-end quantum data processing pipelines for applications such as QML and quantum sensing. These developments promise to further establish quantum autoencoders as indispensable tools in the evolving quantum technology landscape [158][160].

## 3.3. Reinforcement Learning

QRL integrates quantum mechanics into RL, a classical approach where agents learn optimal policies through interaction with an environment. Classical RL methods, such as Q-learning and policy gradient algorithms, rely on iterative updates to achieve convergence, but these techniques face significant challenges in high-dimensional state and action spaces due to computational and memory constraints. QC offers potential solutions to these limitations by leveraging quantum superposition, entanglement, and parallelism to accelerate learning and optimize resource usage. Classical RL uses value functions, policy optimization, and exploration-exploitation trade-offs to achieve decision-making goals. However, the computational overhead increases exponentially with the complexity of the problem. For instance, in large Markov decision processes, the iterative evaluation of state-action pairs becomes inefficient. QRL addresses this inefficiency by enabling faster computation of value functions and policies. Quantum states, through superposition, allow simultaneous evaluation of multiple states, and quantum entanglement aids in representing correlations between actions and their outcomes more compactly [162][163]. QRL represents a hybrid paradigm where classical methods guide quantum optimizations. For example, quantum-enhanced Q-learning leverages quantum memory and computation to store and process large state-action spaces more efficiently. Classical control strategies can be paired with quantum hardware to achieve better performance in tasks such as model-free RL. In these setups, quantum annealers or gate-based quantum computers help solve optimization sub-problems like Bellman updates or reward maximization [164][165]. QRL methods are implemented through quantum variational algorithms and quantum-enhanced neural networks. VQAs, such as the VQE solver, are adapted for RL tasks by encoding policies or value functions into parameterized quantum circuits. Similarly, hybrid quantum-classical methods integrate classical neural networks for policy evaluation with quantum subroutines to enhance search spaces or optimize cost functions. Tools like PennyLane and IBM Qiskit enable researchers to design and execute QRL models on quantum simulators or hardware. These implementations demonstrate improved convergence rates and better scalability for high-dimensional environments [166][167].

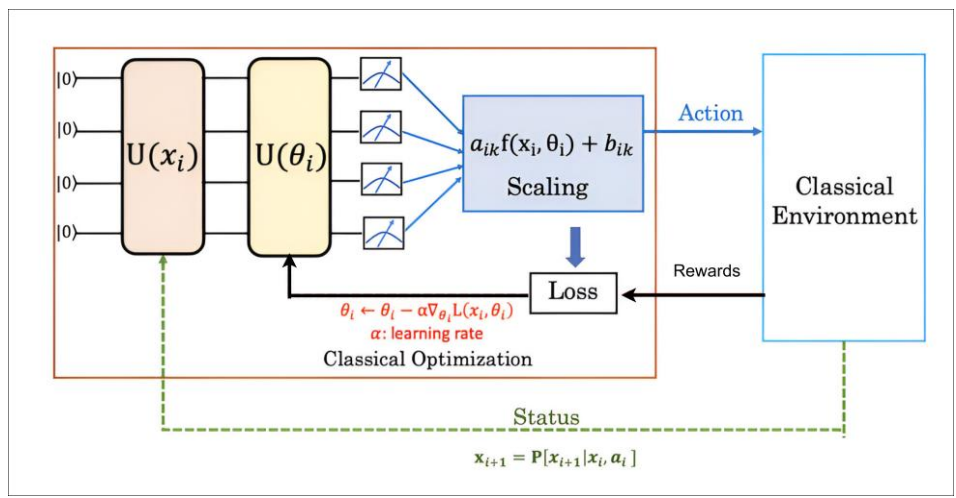

*Figure 23: Quantum-classical hybrid architecture for reinforcement learning, showing quantum circuit processing for state-action mapping, classical optimization, and environment interaction with feedback loop for policy updates [168].*

Quantum-classical hybrid architecture for reinforcement learning is shown in Figure 23. Applications of QRL span robotics, autonomous systems, and quantum control. In robotics, QRL has been employed to optimize navigation and control strategies in environments where classical methods are computationally expensive. In quantum control, QRL aids in designing protocols for maintaining coherence in quantum systems or optimizing the execution of quantum gates. Moreover, QRL has been applied to portfolio optimization and other decision-making problems in finance, demonstrating its versatility in solving real-world tasks with complex state-action representations [162][164]. Despite its promise, QRL faces challenges. Training quantum models requires robust quantum hardware, which is currently limited by noise, decoherence, and gate fidelity. The quantum advantage depends on efficient state preparation and error correction, both of which demand significant resources. Scalability is another concern, as current quantum processors are limited in qubit count, constraining the complexity of QRL tasks that can be solved practically. Furthermore, QRL algorithms often assume idealized quantum states and operations, which may not reflect the imperfections in real-world hardware [164]. Figure 24 depicts the quantum-classical hybrid architecture for reinforcement learning, illustrating the quantum circuit's role in state-action mapping, the classical optimization process, and the interaction with the classical environment, including a feedback loop for policy updates.

Research Horizons for QRL include improving noise resilience through error-mitigation techniques and developing more efficient quantum-classical hybrid architectures. As quantum hardware evolves, expanding QRL to larger and more complex environments will become feasible. Another promising area is exploring the interplay between QRL and other QML methods, such as quantum-enhanced generative models, to address broader categories of problems. These advancements aim to solidify QRL as a foundational approach in QML [163][169].

3.4. Hybrid Machine Learning Algorithms

Hybrid Machine Learning (ML) algorithms, which combine the strengths of both classical and quantum computing, represent one of the most promising avenues for advancing Quantum Machine Learning (QML). These algorithms leverage the power of quantum computers to handle specific computationally intensive tasks while utilizing classical computing for tasks that are better suited for existing classical hardware. This section explores the integration of quantum and classical components in machine learning workflows, focusing on how hybrid algorithms can address the limitations of both paradigms and enhance performance across various applications. Quantum computers, with their ability to process and manipulate quantum states, excel in areas such as optimization, feature mapping, and kernel methods. However, the current state of quantum hardware, particularly NISQ devices, limits their practical application to large-scale, complex ML tasks. In contrast, classical computers are well-established and efficient in handling data processing, optimization, and post-processing tasks. By combining the two, hybrid machine learning algorithms can capitalize on quantum speedups for specific tasks while maintaining the stability and scalability of classical systems.

One prominent hybrid approach is Quantum-Classical SVM. In this framework, a quantum computer is used to perform the feature mapping step, which involves encoding the classical data into a higher-dimensional quantum space. Quantum feature maps can leverage quantum entanglement and superposition to process data in ways that classical algorithms cannot, enabling the system to identify complex patterns in the data. The classical computer then handles the rest of the SVM algorithm, including optimization of the decision boundary. This combination allows for enhanced performance in classification tasks, especially when dealing with large or non-linearly separable datasets. Another hybrid algorithm gaining attention is Quantum-enhanced Reinforcement Learning. Reinforcement learning, which involves an agent learning to make decisions by interacting with an environment, often requires extensive computational resources due to the large number of states and actions involved. Quantum-enhanced reinforcement learning uses quantum techniques for value function approximation, state encoding, and policy optimization. Quantum computers can offer exponential speedups in solving certain types of Markov Decision Processes (MDPs), which form the basis of reinforcement learning tasks. Classical computers can still handle other aspects of the algorithm, such as exploring the environment and updating policies. This hybrid approach can significantly accelerate learning in complex environments, such as robotics and game playing.

Hybrid models are also being explored in QNNs, where quantum computers are used for certain aspects of training neural networks, such as weight optimization and feature encoding. Quantum neural networks combine quantum gates and classical neurons to create models that can potentially outperform classical neural networks, especially when the dataset is vast and the relationships between the data points are highly complex. By using quantum computing to perform parallel processing of features and classical methods for optimizing the training process, hybrid QNNs aim to improve accuracy and speed compared to purely classical neural networks. Additionally, hybrid quantum-classical optimization algorithms are making significant strides in fields like finance and logistics, where optimization problems are often complex and involve many variables. Quantum optimization algorithms, such as the QAOA, can handle large combinatorial optimization problems more efficiently than classical approaches. When combined with classical methods like simulated annealing or gradient descent, these hybrid approaches can provide better solutions in less time, especially for problems such as portfolio optimization, route planning, and supply chain management.

The key benefit of hybrid machine learning algorithms is their ability to address the limitations of current quantum hardware. NISQ devices, while powerful, are still prone to noise and errors, limiting their ability to scale for large, complex machine learning tasks. By integrating classical components that can handle large-scale data processing and error correction, hybrid systems can mitigate the challenges of noisy quantum hardware. Moreover, as quantum devices continue to improve, the classical components of these hybrid systems may be gradually replaced with more advanced quantum techniques, leading to fully quantum-enhanced machine learning algorithms in the future. In conclusion, hybrid machine learning algorithms represent a critical development in Quantum Machine Learning, offering a practical and scalable solution to the current limitations of quantum computing. By combining the computational power of quantum systems with the reliability and scalability of classical systems, these hybrid models provide enhanced performance for a wide range of machine learning tasks, from classification and reinforcement learning to optimization and neural networks. As both quantum and classical technologies continue to evolve, the potential for hybrid quantum-classical approaches to revolutionize machine learning remains substantial, driving innovation across industries such as finance, healthcare, and logistics.

### 3.4.1. Generative Adversarial Networks(GAN)

QGAN extend the classical framework of GAN into the quantum domain, offering a novel approach to generative modeling. Classical GANs, comprising a generator and a discriminator, are widely used for tasks like image generation, data synthesis, and anomaly detection. However, their performance on high-dimensional data or complex distributions is constrained by computational limitations. QC promises exponential speedups in optimization and sampling, making QGANs a compelling advancement for such tasks. The classical GAN framework relies on the generator to create data samples that mimic a target distribution and the discriminator to distinguish between real and generated data. While powerful, these models often require extensive resources to train due to high-dimensional parameter spaces and non-convex optimization landscapes. QGANs exploit quantum properties such as superposition and entanglement to address these issues. For example, quantum generators can encode complex probability distributions more efficiently than classical ones, while quantum discriminators leverage quantum states to perform classification tasks with potentially fewer resources [170][171][172]. The integration of classical and quantum components in QGANs creates a hybrid paradigm that balances the strengths of both realms. Classical GAN frameworks often guide the design of QGANs, with quantum components integrated for enhanced performance. For instance, the generator may be implemented as a parameterized quantum circuit, while the discriminator remains classical or is augmented with quantum capabilities. Such hybrid architectures are particularly effective in NISQ environments, where quantum resources are limited [173][174]. Implementing QGANs involves designing quantum circuits to represent the generator and discriminator. Parameterized quantum circuits are optimized using variational techniques, where the parameters are updated based on feedback from the discriminator's performance. Hybrid quantum-classical optimizers, such as gradient descent augmented with quantum evaluations, are commonly used. Platforms like Qiskit, TensorFlow Quantum (TFQ), and Pennylane provide tools for implementing QGANs, enabling experiments on quantum simulators and hardware. Studies have demonstrated the potential of QGANs in tasks such as generating high-resolution images, simulating quantum states, and learning complex data distributions [175][176]. Figure 24 the architecture of a Quantum Generative Adversarial Network (QGAN) is illustrated, showcasing the adversarial training loop between the quantum generator and the classical discriminator, along with feedback optimization for real/fake data classification.

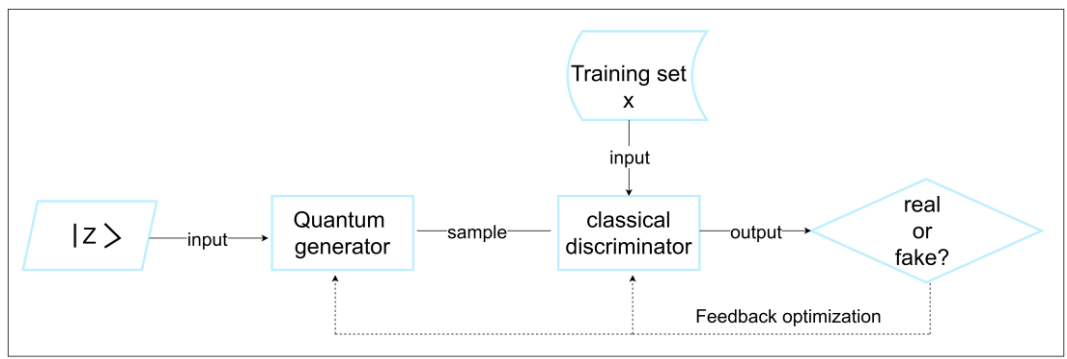

*Figure 24: Quantum Generative Adversarial Networks architecture illustrating the adversarial training loop between quantum generator and classical discriminator with feedback optimization for real/fake data classification [177].*

Applications of QGANs are diverse and impactful. In image generation, QGANs have shown promise in creating high-resolution and photorealistic images, outperforming classical GANs in certain scenarios. In finance, they are used for generating synthetic financial data and improving risk modeling. QGANs are also applied in quantum chemistry for simulating molecular structures and in healthcare for generating synthetic medical data, addressing data privacy concerns. Moreover, QGANs have been employed in optimizing quantum experiments by learning complex distributions efficiently [171][178]. Despite their potential, QGANs face significant challenges. Training QGANs requires quantum devices with high fidelity and low noise, which are not yet widely available. The optimization of quantum circuits is computationally intensive and prone to barren plateaus, where gradients vanish, hindering convergence. Furthermore, hybrid architectures demand seamless integration between quantum and classical systems, which introduces communication overhead. Scalability remains a critical concern, as current quantum hardware supports only a limited number of qubits and operations [179][180].

Research Horizons for QGANs include improving noise resilience and scalability. Developing error-mitigation techniques and optimizing quantum circuit designs can enhance performance on NISQ devices. Exploring fully QGAN architectures, where both generator and discriminator are implemented quantum-mechanically, is another promising avenue. Additionally, extending QGANs to handle broader classes of data distributions and improving hybrid optimization strategies will expand their applicability. As quantum hardware and algorithms advance, QGANs are poised to become a transformative tool in generative modeling and QML [171][175].

### 3.4.2. Transfer Learning

Quantum Transfer Learning (QTL) merges classical transfer learning techniques with QC to leverage quantum advantages in resource optimization and model performance. Classical transfer learning involves reusing a pre-trained model on a similar task to save computational resources and improve training efficiency. While effective, classical approaches face limitations in high-dimensional data and tasks requiring significant computational resources, such as quantum simulations. QC offers capabilities like enhanced parallelism and faster optimization, making it a natural candidate to extend transfer learning's scope. The move to QTL is driven by the promise of exponential speedups and the ability to handle quantum data natively. Classical neural networks trained on conventional hardware struggle with the complexity and structure of quantum datasets. In contrast, quantum models, using properties like superposition and entanglement, can process such datasets more efficiently. Additionally, quantum-enhanced models can integrate with classical frameworks, allowing hybrid approaches where classical networks perform initial tasks, and quantum networks refine or extend the results [181][182]. The integration of classical and quantum paradigms in transfer learning typically involves hybrid models. For example, a classical convolutional neural network might extract features from image data, which are then passed to a quantum network for further processing or decision-making. This architecture allows efficient handling of classical data while capitalizing on quantum benefits for complex computations. Such models have demonstrated success in applications ranging from medical diagnostics to NLP [183][184]. Implementing QTL involves several key steps. Initially, a classical model is pre-trained on a related task or dataset. This model's learned parameters are then transferred to a quantum model, which fine-tunes them for a specific quantum task. The quantum layer, often implemented using parameterized quantum circuits, operates on data encoded as quantum states. Hybrid optimization techniques are employed to train the combined model, with the classical and quantum components iteratively updating their parameters. Platforms like TFQ and Pennylane enable researchers to build and train such hybrid models [185][186]. Figure 25 presents various configurations of quantum transfer learning, demonstrating the integration of classical (yellow) and quantum (green) components in the network architecture.

The Figure 25 highlights the transition from a generic dataset ($D_A$) and task ($T_A$) to a specific dataset ($D_B$) and task ($T_B$). Pre-trained layers ($A'$) are reused and combined with trainable layers ($B$), enabling the adaptation of knowledge to new tasks. Different configurations are explored, including classical-to-quantum, quantum-to-classical, and fully quantum approaches, illustrating the flexibility of quantum transfer learning in optimizing network performance for diverse applications.

Applications of QTL are rapidly expanding across domains. In healthcare, QTL has been employed for tasks like breast cancer detection, leveraging quantum-enhanced models to improve classification accuracy. In NLP, hybrid models have demonstrated improved text classification by integrating pre-trained classical embeddings with quantum decision layers. In quantum chemistry, QTL accelerates molecular simulations by combining classical pre-training with quantum computation, enhancing both accuracy and efficiency. Similarly, QTL has shown promise in image classification and anomaly detection, where classical pre-trained networks are fine-tuned using quantum components for better generalization [182][187]. Despite its promise, QTL faces significant challenges. The primary barrier is the limited availability of quantum hardware with sufficient qubits and low error rates. Current NISQ devices restrict the scalability of quantum layers in hybrid models. Additionally, integrating classical and quantum components requires careful management of communication overhead, as transferring data between systems can be computationally expensive. Optimization techniques for hybrid models are still evolving, with issues like barren plateaus and vanishing gradients posing significant hurdles [188][189].

Research Horizons in QTL research include developing more efficient quantum encoding strategies and exploring fully QTL frameworks. Error-mitigation techniques and quantum-specific pre-training approaches are also critical for improving the robustness of QTL models. As quantum hardware matures, the scalability of QTL models is expected to increase, enabling their application to broader and more complex tasks. Advances in quantum-classical hybrid frameworks will further enhance the feasibility of deploying QTL solutions in real-world scenarios [186][189].

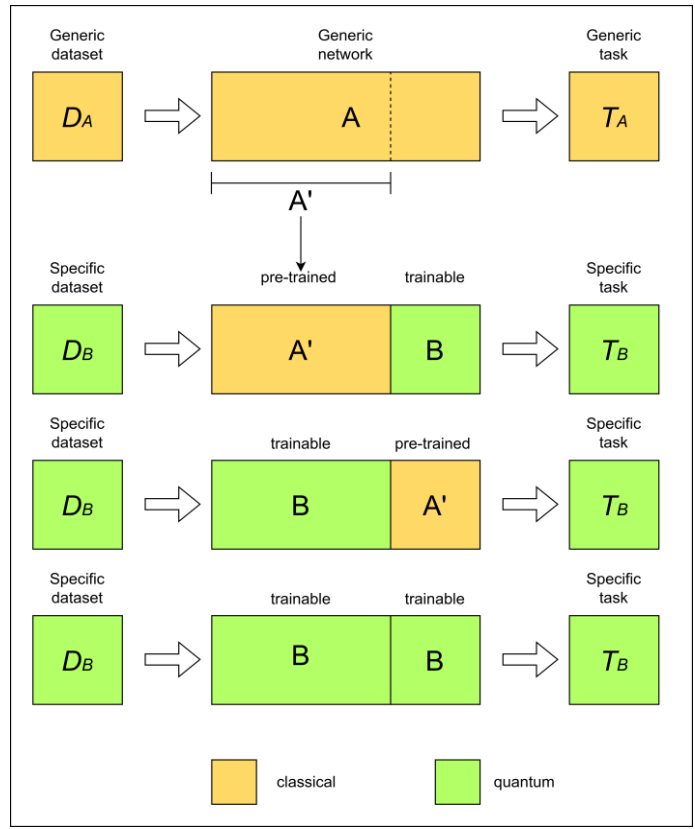

*Figure 25: Illustration of quantum transfer learning variations showing different combinations of classical (yellow) and quantum (green) components for network architecture, depicting transitions from generic to specific tasks.*

### 3.4.3. Quantum Federated Learning

Classical federated learning is a distributed approach to ML that enables collaborative model training across decentralized data sources without sharing raw data. It enhances data privacy and reduces communication overhead by exchanging only model updates. This framework is widely used in privacy-sensitive applications like healthcare and finance. However, classical federated learning faces challenges when dealing with high-dimensional data, heterogeneous datasets, and scalability. Additionally, optimizing and aggregating model updates across distributed systems can strain computational resources and introduce latency in communication networks. QC introduces significant potential improvements to federated learning, leveraging principles like superposition and entanglement for enhanced computational efficiency and data privacy. Quantum Federated Learning (QFL) integrates quantum algorithms to process data more efficiently, particularly in high-dimensional feature spaces, and to secure communication between clients through quantum cryptographic methods. Unlike classical models, QFL can exploit QNN to learn complex patterns faster while maintaining robust data privacy. The transition to QFL addresses scalability and optimization challenges that classical methods encounter, particularly in decentralized networks with resource constraints [190][191]. The integration of classical and quantum frameworks is achieved through hybrid models where classical federated architectures interact with quantum components. In such systems, classical clients perform preliminary data preprocessing and communicate with a central quantum server. The quantum server, equipped with QNNs or quantum optimization algorithms, processes the aggregated updates, optimizing the global model more effectively. This hybrid approach capitalizes on classical systems' robustness while integrating quantum capabilities for speed and accuracy. Frameworks like FedQNN, which utilize QNN in federated systems, illustrate the potential for this integration, showcasing improvements in tasks like image recognition and secure communications [192][193]. Implementing QFL involves adapting classical federated learning protocols to quantum architectures. Clients encode their data into quantum states using quantum encoding methods, such as amplitude encoding, and train QNNs locally. These local updates are transmitted to a central quantum server for aggregation using quantum-secure communication protocols, minimizing risks associated with data interception. Advanced methods like blind QC allow clients to retain data privacy by obfuscating their updates during transmission. The aggregated model is then shared back with clients for further refinements in subsequent iterations. Platforms like TFQ facilitate the integration of quantum components into federated learning systems [194]. Figure 27 illustrates the framework for quantum federated learning. The process begins with initializing parameters ($\theta$), which are sent to local quantum computers for training. Each local quantum device updates the parameters ($\theta_t \rightarrow \theta_{t+1}$) based on its local dataset while ensuring data privacy. The updated parameters are then transmitted to a central server for global parameter aggregation ($\phi_{t+1}$), resulting in a refined global model.

This updated global model is subsequently redistributed to the local quantum computers for further iterations of training, maintaining a federated and privacy-preserving learning process.

Applications of QFL span a range of domains, particularly where data privacy and scalability are critical. In healthcare, QFL enables collaborative training of diagnostic models across hospitals without sharing sensitive patient data. In IoT networks, QFL facilitates efficient communication and model training among decentralized edge devices, enhancing security in 6G-enabled networks. Quantum approaches are also employed in financial systems to detect fraud collaboratively while preserving user confidentiality. Other applications include image classification, NLP, and optimization tasks in supply chain management and logistics [196][197]. Despite its potential, QFL faces several challenges. The foremost limitation is the reliance on NISQ devices, which restrict the scalability and reliability of quantum components. The integration of quantum and classical systems introduces significant communication overhead and synchronization challenges, particularly in large-scale networks. Optimization methods for hybrid systems are still in their infancy, with issues like barren plateaus and convergence instability impacting performance. Moreover, ensuring compatibility between classical and quantum cryptographic methods for secure communication remains an area of active research [198][199].

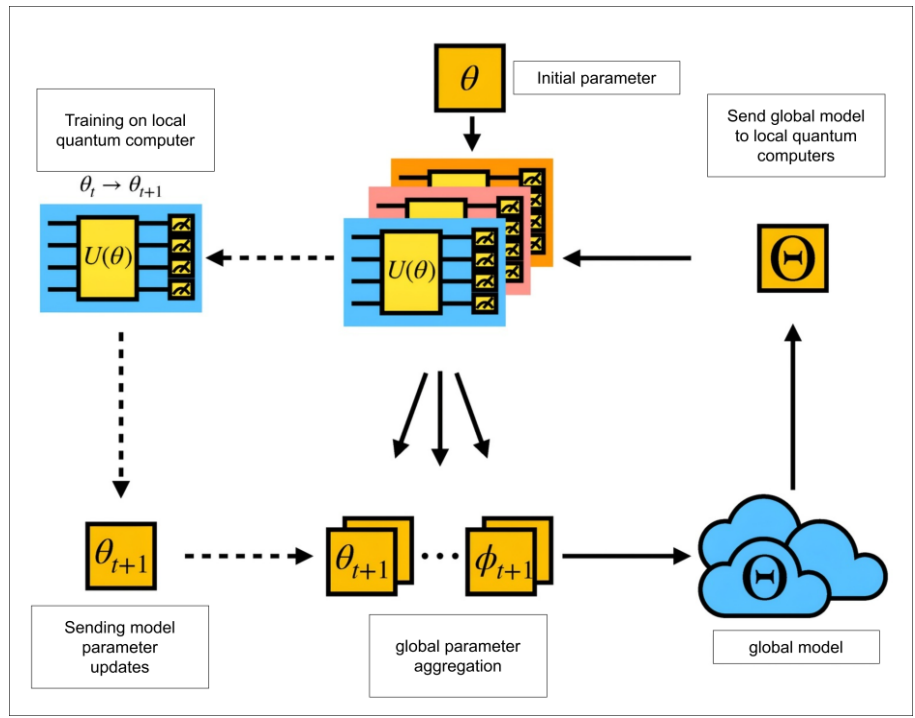

*Figure 26: Quantum federated learning framework showing distributed training across local quantum computers, parameter aggregation, and global model updates while maintaining data privacy [195].*

Figure 26 represents the QFL framework. Future research directions in QFL include the development of error-mitigation techniques to improve quantum device reliability and the exploration of fully quantum federated systems that eliminate classical intermediaries. Enhanced quantum communication protocols, such as quantum repeaters, will be crucial for scaling QFL systems across geographically distributed networks. Additionally, advances in QNN architectures and hybrid optimization methods will further enhance the performance and applicability of QFL in real-world scenarios. As quantum hardware evolves, the scope of QFL is expected to expand significantly, unlocking new opportunities in decentralized learning systems [193] [199].

### 3.4.4. Hybrid Quantum-Classical Learning Models

Classical ML models have achieved remarkable success in various domains, leveraging advanced neural network architectures to process large datasets. However, these methods often encounter limitations in terms of scalability, computational efficiency, and the ability to model complex quantum phenomena. As datasets grow and computational problems become more intricate, classical approaches may struggle to meet the increasing demands for speed and accuracy. QC offers an alternative by leveraging quantum mechanics to solve problems more efficiently. Quantum systems can process information using quantum bits (qubits), exploiting properties like superposition and entanglement. These features allow quantum computers to perform parallel computations, potentially surpassing classical systems in solving high-dimensional problems and optimization tasks. However, due to the constraints of NISQ devices, standalone quantum algorithms may not yet be practical for large-scale applications. This limitation necessitates the integration of classical and quantum approaches to capitalize on the strengths of both paradigms [200][201]. Hybrid quantum-classical learning models merge classical algorithms with quantum components, creating frameworks that utilize quantum advantages while relying on classical computational stability. These models typically consist of classical pre- and post-processing steps combined with quantum circuits for critical computational tasks. For instance, hybrid quantum-classical neural networks integrate quantum layers into classical architectures, using quantum circuits for feature transformation or optimization. Such integration not only enhances computational efficiency but also addresses challenges like barren plateaus and quantum noise by leveraging classical methods for error mitigation [47][203]. The implementation of hybrid quantum-classical models involves designing workflows where quantum components handle high-dimensional or computationally intensive subtasks. For example, quantum layers can process specific features extracted from classical datasets, transforming them into quantum states for advanced computations. Hybrid frameworks like Quantum-Train further improve efficiency by focusing on model compression and optimization during quantum computations. Tools like TFQ and PennyLane provide development environments for implementing such systems, enabling researchers to simulate and optimize hybrid models even with limited quantum hardware [204]. Figure 28 illustrates the architecture of a hybrid quantum-classical learning model. The figure demonstrates the process beginning with data encoding and state preparation in the quantum domain, followed by the preparation of a cluster state circuit. Quantum convolutional (QConv), quantum pooling (QPool), and parameterized quantum circuits Post-Quantum Cryptography (PQC) are employed for feature extraction. These extracted features are subsequently fed into a classical neural network, which performs further processing and decision-making. This integration highlights the synergy between quantum and classical components in achieving enhanced learning performance. It is shown in Figure 27 that how a Hybrid Quantum-Classical Learning Model combines quantum and classical components.

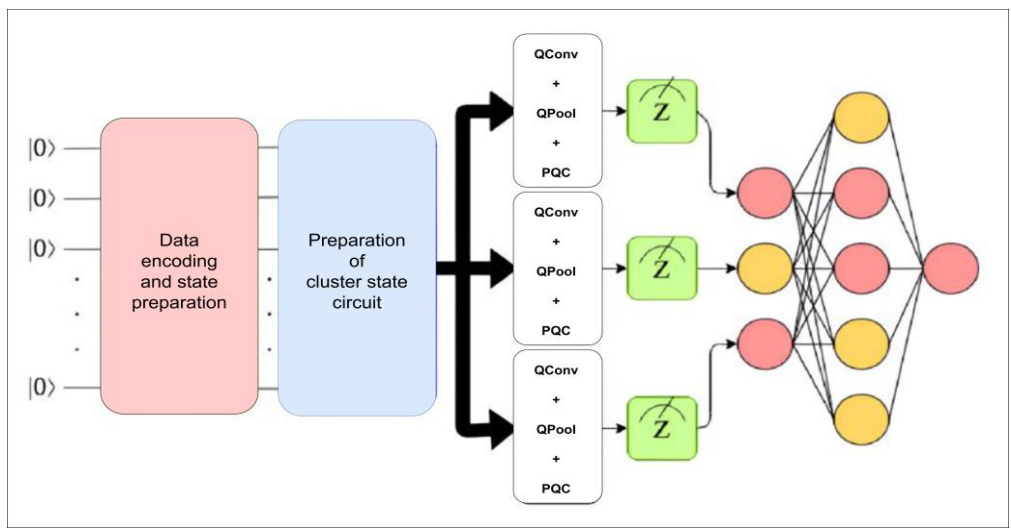

*Figure 27: A Hybrid Quantum-Classical Learning Model combines quantum and classical components. Quantum circuits perform data encoding, state preparation, and feature extraction. The extracted features are then fed into a classical neural network for further processing [205].*

Applications of hybrid quantum-classical models are extensive and span domains where classical methods have limitations. In healthcare, hybrid models improve image classification and anomaly detection, facilitating tasks like disease diagnosis. In finance, they enhance predictive modeling for market analysis and fraud detection. These models are also pivotal in NLP, where quantum circuits can capture complex semantic relationships within text data. Additionally, hybrid approaches have shown promise in materials science for simulating molecular structures and in supply chain optimization for solving logistics problems [200][207]. Despite their promise, hybrid models face significant challenges. The reliance on NISQ devices limits scalability and introduces noise, affecting the reliability of quantum computations. Furthermore, integrating quantum components into classical architectures requires specialized expertise and tools, increasing development complexity. Communication between classical and quantum systems can also lead to bottlenecks, particularly in large-scale applications. Moreover, there is a need for standardized frameworks to evaluate the performance of hybrid models and optimize their implementation [203][47].

Research Horizons in hybrid quantum-classical learning involve enhancing quantum hardware to reduce noise and improve scalability. Advances in QEC techniques and the development of more efficient quantum algorithms are critical to overcoming current limitations. Research into hybrid optimization methods, such as quantum-inspired metaheuristics, could further enhance model performance. Additionally, expanding the availability of QC resources through cloud-based platforms will democratize access, fostering innovation in hybrid learning models. As these advancements materialize, hybrid approaches are expected to play a pivotal role in bridging the gap between classical and QC [201][47].

### 3.4.5. Optimization in Machine Learning

Classical optimization techniques play a critical role in ML, enabling efficient solutions to complex problems in diverse domains like predictive modeling, pattern recognition, and decision-making. Algorithms such as gradient descent, genetic algorithms, and simulated annealing have been pivotal in minimizing error functions and tuning model parameters. However, as datasets grow and complexity, classical methods encounter limitations related to scalability, convergence speed, and computational resource constraints, particularly for non-convex or high-dimensional optimization problems [209][210]. QC introduces unique capabilities that address these challenges, leveraging principles like superposition, entanglement, and quantum parallelism. These properties allow quantum systems to explore multiple solutions simultaneously, enabling faster convergence for certain optimization problems. Quantum algorithms such as the QAOA and Grover's search algorithm exemplify this advantage by outperforming their classical counterparts in specific cases, such as combinatorial optimization and search tasks. The integration of quantum techniques into ML frameworks is driven by their potential to enhance the efficiency and accuracy of optimization processes [211][206]. The convergence of classical and quantum optimization methods in ML is a strategic approach to overcome the current limitations of quantum hardware. Hybrid frameworks utilize classical components for preprocessing and postprocessing tasks, delegating the computationally intensive optimization steps to quantum systems. For instance, classical optimization techniques may generate initial parameters, which are subsequently refined using quantum algorithms to achieve superior performance. This synergy not only mitigates quantum noise and resource limitations but also makes quantum advantages accessible on today's NISQ devices [214]. Quantum optimization methods in ML have been implemented across various domains, demonstrating their versatility. For example, quantum-enhanced gradient descent leverages quantum states to estimate gradients more efficiently, accelerating training for neural networks. Similarly, quantum annealing techniques optimize objective functions in RL and feature selection tasks. Tools like TFQ and Qiskit provide platforms for implementing quantum optimization methods, facilitating experimentation and integration into existing workflows. These advancements bridge the gap between theoretical potential and practical applicability, enabling broader adoption of quantum approaches in ML [215][216].

Applications of quantum optimization in ML span numerous fields. In finance, these methods improve portfolio optimization and risk analysis by solving high-dimensional problems with multiple constraints. In healthcare, quantum optimization aids in drug discovery and personalized medicine by efficiently analyzing complex molecular interactions. Additionally, quantum-enhanced feature selection and model optimization streamline predictive analytics in sectors like retail, logistics, and climate modeling. These applications underscore the transformative impact of quantum optimization on real-world problem-solving [209][211]. Challenges remain in the adoption and scaling of quantum optimization techniques. Quantum hardware limitations, including qubit coherence times and error rates, restrict the size and complexity of problems that can be addressed. Developing quantum algorithms that generalize well across diverse problem domains is another obstacle. Furthermore, ensuring seamless integration with classical ML frameworks requires advancements in hybrid algorithms and interoperability tools. Addressing these challenges will be crucial for realizing the full potential of quantum optimization in ML [206][214].

Research Horizons include improving quantum hardware to support larger-scale computations and reducing noise through advanced error correction techniques. Research into hybrid quantum-classical algorithms is likely to yield more robust and scalable solutions, combining the strengths of both paradigms. Moreover, exploring quantum-inspired optimization techniques on classical systems can provide interim benefits while quantum hardware matures. As these developments progress, quantum optimization is poised to redefine the landscape of ML, offering unparalleled capabilities for tackling complex optimization problems [216].

## 4. QUANTUM MACHINE LEARNING FRAMEWORKS

This section explores the major QML frameworks, their interoperability with classical ML tools, and strategies for scaling and deploying quantum models. It also discusses hybrid workflows that combine quantum and classical computations, as well as benchmarking methods to evaluate QML algorithms against their classical counterparts. By addressing these aspects, this section highlights the role of robust QML frameworks in advancing scalable and practical quantum-enhanced ML solutions. Figure 28 is the depiction of quantum-classical hybrid workflow.

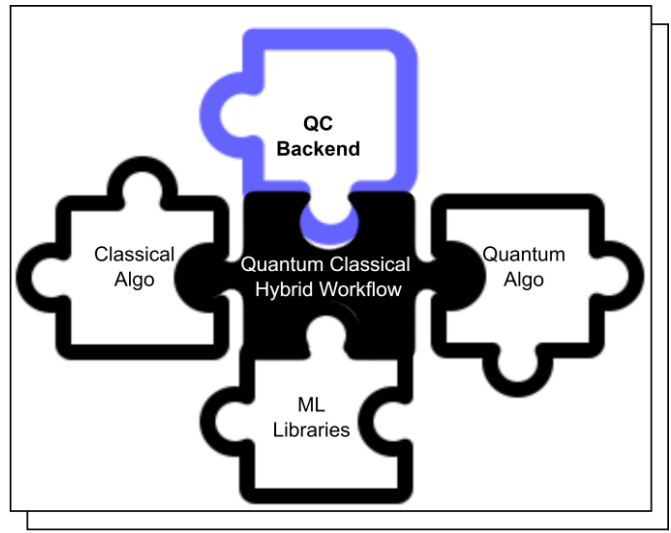

Figure 28: *Puzzle-piece representation of quantum-classical hybrid workflow integrating classical algorithms, quantum circuits (Quantum Computing backend), Machine Learning libraries, and quantum algorithms for comprehensive quantum computing implementation. A robust Quantum Machine Learning framework incorporates several key components that enable seamless integration between quantum and classical systems.*

QML frameworks serve as the foundational infrastructure for integrating QC with classical ML, enabling hybrid solutions that harness the strengths of both paradigms. These frameworks provide essential components such as QC backends, classical ML libraries, and hybrid workflows to facilitate seamless development and execution of QML algorithms. Quantum backends, including quantum hardware (e.g., IBM Q, Rigetti) and simulators (e.g., Qiskit Aer, Cirq), enable the implementation of quantum operations, while classical ML libraries like TensorFlow and PyTorch support data preprocessing, model training, and evaluation.

- QC Backend: The QC backend serves as the foundation of a QML framework, enabling the execution of quantum operations, the simulation of quantum algorithms, and the integration of quantum hardware resources. Quantum hardware, such as IBM Q, Rigetti, and IonQ, facilitates the execution of quantum circuits on real quantum devices, leveraging quantum bits (qubits) to solve problems beyond the reach of classical systems. To ensure flexibility and scalability, a robust framework should support multiple quantum hardware providers. Additionally, quantum simulators, like Qiskit Aer, Cirq, and QuTiP, play a crucial role in emulating quantum hardware, allowing developers to test and debug algorithms in a controlled environment without requiring physical quantum devices, which are still maturing. Complementing these are specialized quantum programming languages and frameworks, such as Qiskit, Cirq, and PennyLane, which provide the necessary tools for defining quantum algorithms, formulating circuits, and managing quantum operations effectively.
- Classical ML Libraries Integration: To unlock the full potential of QML, a framework must seamlessly integrate classical ML techniques, enabling a hybrid workflow that combines the strengths of both quantum and classical domains. Classical ML libraries such as TensorFlow, PyTorch, Scikit-learn, and XGBoost play a critical role in tasks like model training, evaluation, and hyperparameter tuning. Effective communication between quantum and classical components is vital to leveraging the capabilities of both technologies. Data preprocessing and feature engineering are essential steps in preparing data for quantum models, employing classical techniques such as normalization, feature scaling, and dimensionality reduction (e.g., PCA) to ensure only the most relevant features are encoded into quantum states. Classical ML also supports model training and evaluation, handling tasks such as hyperparameter optimization and performance assessment. Hybrid models, which blend quantum and classical algorithms, rely on classical optimization techniques to refine quantum operations, highlighting the importance of robust integration between these two paradigms.
- Quantum-Classical Hybrid Workflow: QML thrives on hybrid workflows that combine quantum and classical computations, leveraging the strengths of each domain. Effective integration ensures seamless quantum-classical data flow, involving the encoding of classical data into quantum states, processing it through quantum feature mapping, and interpreting the results using classical post-processing techniques. Hybrid algorithms, such as VQAs and quantum-enhanced classical methods like quantum kernel-based SVMs, exemplify this synergy by utilizing quantum circuits for computation and classical optimization techniques for refinement. Given the susceptibility of quantum systems to noise and errors caused by decoherence and gate imperfections, strategies like QEC and noise-aware training are essential to maintaining the accuracy and reliability of these algorithms.
- QML Algorithms: QML algorithms harness the power of QC to enhance classical ML models, introducing novel approaches to data processing and learning. QNN extend classical neural networks by utilizing quantum gates for non-linear transformations of input data, showing potential to outperform classical models in tasks such as classification, regression, and clustering. Quantum Kernels enable quantum-enhanced SVMs by mapping classical data into higher-dimensional quantum spaces, often yielding superior performance compared to their classical counterparts. Additionally, Quantum Feature Maps encode classical data into quantum states, leveraging quantum parallelism to efficiently learn from complex data structures, paving the way for advancements in ML.
- Scalability and Deployment: Scalability is a critical aspect of QML frameworks, enabling models to efficiently tackle complex, large-scale problems. Distributed Computing frameworks enhance scalability by parallelizing quantum operations across multiple quantum processors or hybrid quantum-classical systems, accelerating computations and making extensive quantum experiments feasible. Cloud-Based QC platforms, such as IBM Q Experience, Microsoft Azure Quantum, and Amazon Braket, provide remote access to quantum hardware and simulators, allowing researchers to scale their quantum models without needing on-premises quantum devices. Additionally, Deployment Tools are essential for integrating quantum models into practical applications, ensuring seamless compatibility with existing software environments, enabling remote execution on cloud systems, and supporting model monitoring and maintenance for production-ready quantum solutions.

4.1. Major Quantum Machine Learning Frameworks

In the rapidly advancing field of QML, various frameworks have been developed to facilitate the integration of QC with ML techniques as shown in Figure 29. These frameworks provide tools and platforms that enable researchers and developers to design, simulate, and optimize quantum algorithms, making it easier to harness the power of quantum hardware for ML applications. This subsection introduces some of the leading QML frameworks as shown in Figure.29 , each offering unique features and capabilities. From versatile platforms like Amazon Braket and Azure Quantum, to specialized tools such as D-Wave for quantum annealing and Strawberry Fields for photonic computing, this section explores how these frameworks support the development and deployment of quantum-enhanced ML models. By examining these tools, we highlight the diverse approaches to QC and ML, enabling a deeper understanding of their potential and applications in the field.

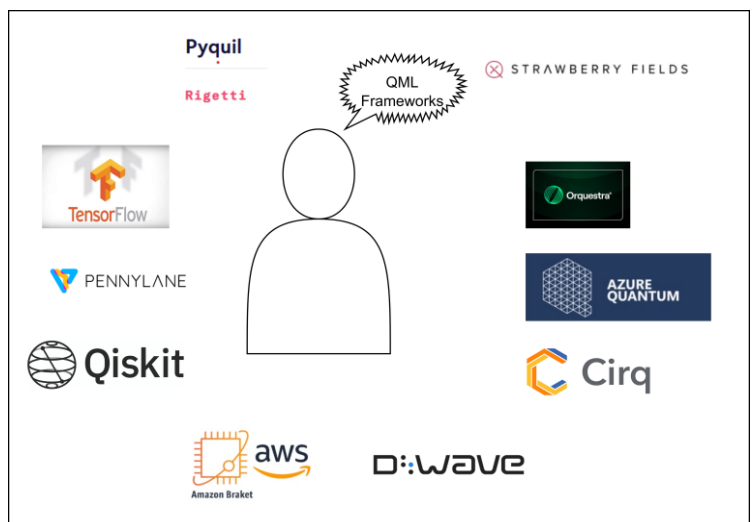

*Figure 29: Overview of major Quantum Machine Learning frameworks and platforms*

4.1.1. TensorFlow Quantum: A Powerful Platform for Quantum Machine Learning

TFQ, developed by Google AI Quantum and collaborators, is an open-source software framework designed specifically to advance the integration of QC and ML. TFQ is built on TensorFlow, one of the most popular ML libraries, and focuses on providing tools for quantum data processing, hybrid quantum-classical models, and the development of QML algorithms.

TensorFlow Quantum (TFQ) is a cutting-edge framework that seamlessly integrates quantum computing (QC) into classical machine learning (ML) workflows, enabling researchers to construct hybrid models that combine quantum circuits with TensorFlow's robust classical tools. This integration is particularly useful for studying variational quantum circuits (VQCs), essential for hybrid quantum-classical algorithms where quantum computers handle complex operations and classical computers optimize parameters. TFQ accelerates the exploration of hybrid workflows for solving real-world challenges in optimization, ML, and quantum chemistry. Additionally, TFQ provides efficient quantum circuit simulation, allowing scalable testing in classical environments before deployment on quantum hardware. Researchers can also define quantum datasets to encode classical data into quantum states, enabling quantum-enhanced pattern recognition and data processing. The framework's deep integration with TensorFlow's ecosystem ensures compatibility with tools like TensorBoard for visualization and Keras for model development, while pre-built quantum layers simplify quantum operations, making it accessible to users without deep quantum mechanics expertise. TFQ supports diverse applications, such as quantum neural networks (QNNs) for image recognition, NLP, and drug discovery; quantum generative models for realistic data generation in tasks like fraud detection; and variational quantum algorithms (VQAs) like VQE and QAOA for tackling complex optimization problems in quantum chemistry and materials science. Furthermore, TFQ facilitates quantum model benchmarking to assess the quantum advantage over classical methods. With its scalable performance, GPU/TPU acceleration, intuitive interface, and extensive ecosystem support, TFQ empowers researchers to experiment with advanced quantum algorithms and hybrid models, driving progress in quantum machine learning (QML).

TFQ bridges the gap between QC and classical ML, empowering researchers to explore the potential of quantum-enhanced models. TFQ's hybrid approach [218] is pivotal for advancing VQAs and integrating quantum data processing into classical workflows. The pre-built quantum layers and compatibility with TensorFlow's ecosystem, as discussed in How (2022), make TFQ an indispensable tool for QML research and applications. Its ability to scale simulations and support quantum-classical experimentation positions TFQ as a leader in the burgeoning field of QML.

### 4.1.2. Pennylane: A Versatile Framework for Quantum Machine Learning

Pennylane, developed by Xanadu, is a powerful open-source software framework designed to bridge the gap between quantum hardware and classical ML libraries. By seamlessly integrating quantum circuits with classical computation, Pennylane empowers researchers and developers to explore the potential of quantum-enhanced algorithms and applications.

PennyLane is a versatile framework that empowers the development of hybrid quantum-classical models by seamlessly integrating the strengths of quantum and classical computing. This integration is particularly valuable for tasks such as parameter optimization in variational quantum algorithms (VQAs), where quantum circuits are optimized using classical gradient-based methods. A standout feature of PennyLane is its automatic differentiation capability, which efficiently computes gradients of quantum circuits, enabling the optimization of both quantum and classical parameters. PennyLane also offers device agnosticism, supporting a wide range of quantum hardware platforms, including IBM Q, Rigetti, Google's quantum processors, and Xanadu's photonic devices, allowing users to experiment with diverse hardware and choose the best fit for their applications. Further simplifying development, PennyLane provides a library of pre-built quantum circuit templates and functions tailored for common machine learning tasks like classification, clustering, and regression. Its capabilities extend to quantum reinforcement learning (QRL), where quantum models are used to optimize complex decision-making processes. PennyLane excels in quantum-enhanced machine learning (QML) tasks, quantum chemistry simulations (e.g., calculating molecular ground-state energies using VQE), and solving combinatorial optimization problems through algorithms like the QAOA. With its seamless integration with classical ML tools such as TensorFlow, PyTorch, and JAX, PennyLane ensures accessibility for ML researchers and developers. Moreover, its hardware versatility and vibrant community, coupled with extensive documentation and tutorials, make it an invaluable resource for researchers. By combining automatic differentiation, hardware flexibility, and an active support ecosystem, PennyLane empowers users to tackle complex challenges in optimization, quantum chemistry, and quantum machine learning.

Pennylane exemplifies the intersection of QC and ML, with its hybrid architecture making it indispensable for quantum research and practical applications. Its automatic differentiation capabilities [219] streamline the training of VQAs, while its use in QRL [220] highlights its potential in cutting-edge AI research. By supporting multiple hardware platforms and integrating seamlessly with classical tools, Pennylane democratizes QC, paving the way for widespread adoption in academia and industry.

### 4.1.3. Qiskit: A Powerful Framework for Quantum Computing

Qiskit, an open-source QC framework developed by IBM, has become a cornerstone for researchers and developers in the QC field. It provides a comprehensive set of tools for designing, simulating, and executing quantum circuits, making it a versatile platform for exploring the potential of QC.

Qiskit is a comprehensive framework for quantum computing that supports the entire quantum computing (QC) workflow, from quantum circuit design to deployment on real quantum hardware. With its intuitive Python-based interface, Qiskit enables users to build and simulate quantum circuits, making it an essential tool for experimenting with quantum algorithms and understanding their behavior. Its direct integration with IBM Quantum's hardware provides access to real quantum processors, allowing researchers to execute quantum circuits and explore the practical limitations and capabilities of quantum devices. Qiskit's powerful transpiler optimizes quantum circuits to meet hardware-specific constraints, ensuring efficient execution and maximizing algorithm performance. Additionally, Qiskit includes a comprehensive library of pre-built quantum algorithms, such as the Variational Quantum Eigensolver (VQE) and the QAOA, simplifying the exploration of quantum chemistry, optimization, and machine learning applications. The framework also excels as an educational tool, offering extensive resources like tutorials, textbooks, and an active community to support learners and practitioners at all levels. Its strengths in quantum algorithm development, hardware-specific research, and the deployment of quantum applications on real devices make Qiskit a versatile tool for researchers and educators alike. Furthermore, Qiskit's advanced optimization techniques and hardware-agnostic design empower users to experiment with simulators and real quantum hardware, making it suitable for both prototyping and practical quantum computing solutions. With its extensive documentation, vibrant community, and robust set of tools, Qiskit bridges the gap between theoretical exploration and real-world quantum computing applications.

Qiskit's combination of powerful features, an active community, and integration with real quantum hardware makes it a leading platform for QC research and development. As quantum hardware continues to advance, Qiskit is poised to play a pivotal role in shaping the future of quantum technology. The flexibility of the framework in adapting to new quantum technologies is emphasized, with its contributions to advancing quantum applications in fields like chemistry, optimization, and ML highlighted in reference [221]. Additionally, its role in bridging the gap between theoretical QC and practical hardware implementation is underscored in reference [222], making it an indispensable tool for researchers. The application of Qiskit in quantum circuit development is further demonstrated in reference [223], reinforcing its importance as a cornerstone of the QC ecosystem.

### 4.1.4. Amazon Braket: A Gateway to Quantum Computing

Amazon Braket is a comprehensive QC service offered by Amazon Web Services (AWS). It provides a user-friendly platform for researchers, developers, and businesses to explore the potential of QC.

Amazon Braket is a versatile quantum computing service that provides access to diverse quantum hardware platforms, including gate-based quantum computers and quantum annealers. This hardware diversity enables users to experiment with different quantum technologies and tailor their approach to specific needs. Braket also offers high-performance managed quantum simulators, allowing researchers to test and debug quantum algorithms without relying on physical quantum devices—a valuable feature for scaling complex simulations. Its integration with Jupyter Notebooks provides a familiar and user-friendly development environment, streamlining the process of designing and running quantum algorithms. By enabling hybrid quantum-classical workflows, Braket empowers users to leverage the strengths of both quantum and classical computing resources to tackle real-world challenges. The platform's flexible, pay-as-you-go pricing model makes quantum computing more accessible to a broader audience. Amazon Braket supports a range of applications, including solving complex optimization problems like supply chain and financial portfolio optimization, advancing quantum machine learning (QML) in areas such as drug discovery and materials science, and providing a platform for prototyping and testing new quantum algorithms. Its user-friendly interface and educational resources make it an excellent tool for quantum education and training. Strengthened by its seamless integration with AWS services, Braket facilitates the development of advanced quantum applications while offering scalable simulation capabilities to explore complex quantum systems. With its combination of diverse hardware access, hybrid workflows, and a robust ecosystem, Amazon Braket serves as a powerful platform for both researchers and educators in quantum computing.

Amazon Braket is a versatile and accessible platform for QC, offering unified access to diverse hardware and robust simulation tools. Its seamless integration with AWS cloud services supports the development of hybrid quantum-classical workflows, which are critical for solving complex real-world problems. The platform has been recognized for its role in democratizing access to quantum resources, allowing users to experiment with different hardware paradigms [224]. Additionally, its utility in practical applications, such as combinatorial optimization and ML [225]. With its user-friendly interface and pay-as-you-go pricing model, Amazon Braket stands out as a powerful tool for advancing QC research and applications in academia and industry. Amazon Braket empowers researchers, developers, and businesses to explore the potential of QC. By providing access to diverse quantum hardware, advanced simulation capabilities, and a user-friendly development environment, Amazon Braket is accelerating the development of quantum technologies and driving innovation [224][225].

### 4.1.5. Azure Quantum: A Powerful Platform for Quantum Computing

Azure Quantum, Microsoft's cloud-based QC platform, offers a comprehensive suite of tools for building, simulating, and deploying quantum applications. It seamlessly integrates quantum hardware, classical cloud computing, and hybrid workflows, making it a valuable tool for researchers and developers exploring the potential of QC.

Azure Quantum is a comprehensive platform that provides a robust environment for developing and deploying quantum solutions. At its core is the Quantum Development Kit (QDK), which features the Q# programming language, designed for creating quantum algorithms with ease. Azure Quantum emphasizes hybrid quantum-classical workflows, enabling users to harness the strengths of both quantum and classical computing for powerful algorithm development. The platform provides access to multiple quantum hardware providers, such as IonQ, Quantinuum, and Rigetti, offering flexibility in exploring diverse quantum technologies. Leveraging Microsoft's scalable cloud infrastructure, Azure Quantum supports large-scale quantum simulations and hybrid workloads, making it well-suited for complex applications. The platform also includes pre-built quantum machine learning (QML) libraries, simplifying the creation of quantum-enhanced ML models. A unified ecosystem integrates seamlessly with Microsoft's Azure cloud platform, ensuring a cohesive development environment for both quantum and classical applications. Azure Quantum's hardware-agnostic design allows users to explore and select from multiple quantum technologies, while its focus on hybrid models reflects the platform's commitment to enabling near-term quantum advantage. With developer-friendly tools like the QDK and its scalable cloud infrastructure, Azure Quantum is accessible to a broad range of users, from individual researchers to large enterprises. The platform's applications span various fields, including solving complex optimization problems in supply chains, finance, and traffic management; advancing drug discovery and material science through molecular simulations; developing quantum-enhanced financial models for risk assessment and portfolio optimization; and improving classical ML performance with quantum-assisted feature engineering. Azure Quantum's versatility and scalability position it as a powerful tool for accelerating innovation in quantum computing.

Azure QML exemplifies Microsoft's commitment to bridging classical and QC through a unified cloud-based ecosystem. The combination of QDK, hybrid model support, and enterprise-grade infrastructure positions it as a leader in accessible QC platforms [226]. By supporting diverse quantum backends, scalable simulations, and seamless integration with the Microsoft ecosystem, Azure Quantum is ideal for tackling optimization problems, quantum-enhanced ML, and industry-specific applications such as finance and healthcare. These strengths make Azure Quantum not just a tool for researchers but a strategic enabler for enterprises preparing for the quantum era [226].

### 4.1.6. D-Wave: A Pioneer in Quantum Annealing

D-Wave Systems is a leading company in the field of QC, specializing in quantum annealing technology. Unlike gate-based quantum computers, which operate on qubits using quantum gates, D-Wave's quantum annealers are designed to solve optimization problems efficiently.

D-Wave's quantum annealers leverage the principles of quantum mechanics to solve complex optimization problems by finding the minimum energy state of a system, corresponding to the optimal solution. These quantum annealers are composed of thousands of qubits, allowing them to handle large-scale quantum systems and tackle highly complex problems. By integrating quantum annealing with classical computing, D-Wave creates powerful hybrid systems capable of solving real-world challenges across a variety of fields. The qubits in D-Wave's systems possess relatively long coherence times, ensuring stable and reliable quantum operations for solving optimization problems efficiently. D-Wave's quantum annealers are especially adept at solving optimization problems such as vehicle routing, protein folding, financial portfolio optimization, and supply chain management. Additionally, they can accelerate machine learning tasks, including training support vector machines (SVMs), and can simulate material properties to aid in discovering new materials with desirable characteristics. D-Wave's systems are specialized for optimization, offering scalability to handle large-scale problems with high efficiency. The hybrid quantum-classical computing approach enables seamless integration with classical resources, enhancing the development of practical quantum applications. Real-world success stories across industries further underscore the practical value of D-Wave's quantum annealers in addressing complex challenges and demonstrating the potential of quantum computing for tangible solutions.

D-Wave Systems offers a unique QC framework that excels in optimization tasks, making it particularly suitable for industries that require fast solutions to complex, large-scale problems. The quantum annealing approach is especially advantageous for optimization in ML, combinatorics, and operations research [227]. The company's use of quantum annealers for solving practical problems in finance, logistics, and drug discovery provides clear evidence of its real-world impact. The integration of classical and QC in D-Wave's hybrid models allows users to take advantage of quantum processing power while maintaining the familiarity and flexibility of classical computing systems. The ability to scale to large problems, coupled with a specialized focus on optimization, makes D-Wave an invaluable tool in the field of QML. D-Wave's quantum annealing systems represent a pioneering approach to QC, offering practical solutions to optimization problems that classical systems struggle to address efficiently. By focusing on real-world applications like financial modeling, ML, and logistics, D-Wave's technology promises to unlock new possibilities across a variety of fields [227][228][229].

### 4.1.7. Google Cirq: A Powerful Tool for Quantum Computing

Google Cirq is a versatile and powerful framework for QC, designed to bridge the gap between theoretical quantum algorithms and practical quantum hardware implementations. It provides a user-friendly interface for creating, simulating, and executing quantum circuits, making it a valuable tool for both researchers and developers.

Google Cirq is a powerful quantum computing framework that enables users to design, simulate, and implement quantum circuits. It provides a user-friendly interface, allowing the creation of quantum operations with a visual representation of their execution. Cirq seamlessly integrates with Google's quantum hardware, giving users access to real quantum devices for executing their circuits. The framework includes advanced features such as noise modeling and error mitigation tools, which are essential for improving the accuracy and reliability of quantum computations. Cirq also supports quantum machine learning (QML) algorithms, including quantum neural networks (QNN) and variational quantum algorithms (VQAs), making it ideal for developing hybrid quantum-classical algorithms. As an open-source and community-driven project, Cirq benefits from continuous development and innovation, with an active developer community providing resources, tutorials, and support. Google Cirq is well-suited for a range of applications, including quantum machine learning for tasks like classification, regression, and clustering; simulating molecular properties in quantum chemistry to aid in drug discovery and materials science; solving optimization problems such as vehicle routing and financial portfolio optimization; and developing quantum cryptography protocols like quantum key distribution (QKD). Its hybrid quantum-classical computing capabilities further position Cirq as a tool for practical and impactful quantum applications in various industries.

Google Cirq is a powerful QC framework that is well-suited for near-term QML applications. Cirq provides a flexible and developer-friendly environment for building quantum circuits, making it an excellent tool for both researchers and companies exploring the potential of QC in ML [230]. Cirq's compatibility with Google's quantum processors, as well as its support for hybrid quantum-classical models, positions it as a key framework for advancing QML. Moreover, its error mitigation techniques, quantum circuit simulation capabilities, and focus on quantum optimization provide significant advantages for real-world applications such as drug discovery, optimization, and ML. Thus, Cirq is instrumental in enabling the exploration of quantum algorithms that could offer superior performance in specific ML tasks when compared to classical counterparts [231].

### 4.1.8. Orquestra: A Quantum Software Development Platform

Orquestra is a powerful quantum software development platform designed to bridge the gap between theoretical quantum algorithms and practical quantum hardware. It provides a user-friendly interface and a comprehensive set of tools for developing, testing, and deploying quantum applications.

Orquestra is a unified development platform that integrates quantum computing (QC) tools, cloud services, and classical computing resources, enabling seamless quantum software development. The platform supports the orchestration of quantum algorithms through its Quantum Path, simplifying the development and execution of complex quantum workflows. Orquestra is designed to work with NISQ devices, which are the current state-of-the-art in quantum hardware, and enables the integration of quantum and classical computing for hybrid quantum-classical algorithms. It provides robust tools for quantum machine learning (QML) applications, including quantum neural networks (QNN), and supports cloud-based quantum development, offering convenient access to quantum hardware for experimentation. Orquestra's strengths include its user-friendly interface, which makes quantum algorithm development accessible to a wide range of users. The platform's emphasis on hybrid quantum-classical workflows further enhances its ability to create practical quantum applications. Additionally, Orquestra includes tools for error mitigation and noise handling, ensuring reliable quantum computations. It also offers cross-platform compatibility, supporting a variety of quantum hardware platforms, which provides flexibility for users. As quantum hardware evolves, Orquestra is built to scale with increasing capabilities, ensuring long-term usability. Orquestra finds applications in diverse fields such as quantum machine learning (e.g., image recognition, NLP, and drug discovery), quantum chemistry (simulating molecular properties), quantum optimization (solving problems like vehicle routing and financial portfolio optimization), and quantum cryptography (developing secure communication protocols). Its broad applicability, ease of use, and scalable architecture position Orquestra as a powerful tool for advancing quantum technologies.

Orquestra is a powerful quantum software development platform that simplifies the creation, testing, and execution of QML algorithms. Its support for hybrid quantum-classical workflows, cloud-based quantum processing, and integration with NISQ systems makes it an invaluable tool for researchers and developers working in QC. Orquestra's flexible design [232] allows for seamless orchestration of quantum algorithms, enabling advancements in areas like QML, optimization, and simulation. With its error mitigation tools and hybrid approach, Orquestra is well-suited for addressing the challenges of current quantum hardware, offering significant potential for real-world applications in finance, logistics, and drug discovery, among others. The role of platforms like Orquestra is emphasized [233] in the development of QNN, further demonstrating its importance in advancing the field of QML.

### 4.1.9. Strawberry Fields: A Quantum Leap in Photonic Computing

Strawberry Fields is a powerful quantum software platform developed by Xanadu, designed to harness the potential of photonic QC. By leveraging the unique properties of light particles (photons), Strawberry Fields offers a novel approach to QC, particularly in the realm of QML.

Strawberry Fields is a specialized quantum computing platform focusing on continuous-variable (CV) quantum computing, using continuous variables like position and momentum to represent quantum states. This unique approach offers distinct advantages for certain quantum algorithms. The platform provides robust tools for simulating quantum circuits on photonic quantum hardware, allowing researchers to test and refine quantum algorithms before deploying them on physical devices. Strawberry Fields is well-suited for quantum machine learning (QML), particularly for tasks involving quantum neural networks (QNN) and support vector machines (SVMs). It can be integrated with other quantum computing platforms, enabling hybrid quantum-classical workflows. As an open-source platform, Strawberry Fields fosters community contributions and allows for customization, ensuring continuous improvement and adaptability. Its strengths lie in its specialization in photonic quantum computing, which leverages light particles for quantum information processing, offering advantages in scalability and high-performance quantum computing. This makes it an ideal platform for developing QML algorithms that have the potential to revolutionize fields like artificial intelligence and drug discovery. Furthermore, its support for hybrid quantum-classical computing enhances its versatility, while the vibrant open-source community accelerates innovation. Strawberry Fields finds applications in diverse fields such as QML (image recognition, natural language processing, drug discovery), quantum cryptography (developing quantum key distribution protocols for secure communication), quantum sensing (creating high-precision sensors), and quantum simulation (simulating complex quantum systems to explore material and molecular properties). Its unique combination of features, scalability, and performance makes Strawberry Fields a powerful tool for advancing quantum technologies.

Strawberry Fields is a powerful tool that is driving innovation in the field of QC. By leveraging the unique advantages of photonic QC, Strawberry Fields has the potential to revolutionize a wide range of industries, from healthcare to finance to materials science. The platform's capabilities in photonic QC enable advancements in QML and optimization, positioning it as a key player in the development of quantum technologies. Its flexibility and scalability make it an invaluable resource for researchers working to unlock the full potential of QC. The effectiveness of Strawberry Fields in QML is further demonstrated by its benchmarking results, showing its ability to tackle complex tasks in various domains [234][235].

### 4.1.10. PyQuil: A Framework for Quantum Computing

PyQuil is a robust and flexible quantum programming framework developed by Rigetti Computing. It provides a user-friendly interface for building, simulating, and executing quantum circuits on Rigetti's quantum hardware.

PyQuil is a quantum computing framework that leverages the Quantum Instruction Language (Quil) for writing quantum programs, offering a high-level and intuitive way to define quantum circuits. It supports hybrid quantum-classical computing, enabling the development of powerful algorithms that integrate quantum and classical computing resources. PyQuil is well-suited for quantum machine learning (QML) tasks, including the implementation of quantum neural networks (QNN) and support vector machines (SVMs). The platform provides seamless integration with Rigetti's quantum hardware, allowing users to run quantum algorithms on real quantum devices and experiment with actual hardware. Additionally, PyQuil includes classical simulators, making it possible to test and debug quantum algorithms before execution on physical hardware. One of its key strengths is its user-friendly interface, which simplifies the learning curve for those new to quantum computing. The hybrid quantum-classical capabilities of PyQuil empower the development of practical quantum applications, and its support for QML allows researchers to explore new possibilities in fields like AI and drug discovery. With direct access to Rigetti's hardware, PyQuil offers an opportunity to work with real quantum devices, and its growing community ensures that users have the support and resources needed to make the most of the platform. PyQuil's applications include quantum chemistry (simulating molecular systems and predicting chemical properties), quantum machine learning (classification, regression, and clustering), quantum optimization (solving problems like vehicle routing and financial portfolio optimization), and quantum cryptography (developing secure communication protocols). PyQuil's combination of accessibility, integration with hardware, and versatility in various domains makes it a valuable tool for advancing quantum research and applications.

PyQuil is a powerful tool that is driving innovation in the field of QC. By providing a user-friendly interface and access to cutting-edge quantum hardware, PyQuil is empowering researchers and developers to explore the potential of QC and develop groundbreaking applications. It is particularly well-suited for QML, quantum chemistry, and optimization problems. Its integration with Rigetti's quantum hardware allows users to run quantum algorithms on actual quantum processors, while its support for hybrid quantum-classical models makes it an ideal tool for combining classical and quantum computational resources. The framework's focus on quantum chemistry, optimization, and ML demonstrates its potential across multiple industries [236] [237]. The open-source nature of PyQuil, along with its scalability and ability to integrate with classical systems, positions it as a critical tool in the ongoing development of practical quantum algorithms and applications.

*Table 4: Comparison of various Quantum Machine Learning frameworks*

| Framework | Key Focus | Strengths | Typical Use Cases | Limitations |
|---|---|---|---|---|
| **TFQ (TFQ)** | Hybrid quantum-classical algorithms, QML | Integration with TensorFlow, Supports deep quantum circuits, Scalable | QML, optimization, chemistry simulations | Limited quantum hardware support, High learning curve |
| **Pennylane** | Hybrid quantum-classical computation | Supports multiple quantum devices, Strong in QNN | QML, optimization, quantum chemistry, RL | May have integration challenges with some quantum devices |
| **Qiskit** | Quantum circuit design, quantum algorithm execution | Wide device compatibility, Open source, large community support | QC, quantum simulation, optimization, QML | Limited scalability for larger systems |
| **Amazon Braket** | Quantum algorithms on cloud, Hybrid quantum-classical | Cloud-based, multi-platform support (IONQ, D-Wave, Rigetti) | Quantum optimization, chemistry simulations, ML | Limited quantum hardware variety for some algorithms |
| **Azure QML** | Quantum programming, Hybrid quantum-classical | Supports various quantum processors, Cloud-based, Integration with ML | Quantum optimization, quantum chemistry, QML, finance | Can be complex to integrate with existing systems |
| **D-Wave** | Quantum annealing, Optimization algorithms | Best suited for quantum annealing, Fast optimization solutions | Optimization problems, ML, QML, logistics | Only supports quantum annealing, not suitable for general QC |
| **Google Cirq** | Quantum circuit design, QC with Python | Well-integrated with Google hardware, Focus on near-term quantum devices | Quantum simulations, optimization, QML | Limited to specific hardware, Lack of flexibility for large-scale systems |
| **Orquestra** | Quantum software development platform | Integrates with multiple quantum frameworks, Focus on software lifecycle | Hybrid quantum-classical algorithms, neural networks, QML | Complex setup, Requires strong understanding of quantum systems |
| **Strawberry Fields** | Photonic QC, Quantum circuits | Strong in photonic quantum systems, High- | Quantum photonic computing, optimization, QML | Limited to photonic quantum systems, not widely compatible |

|  |  | level abstraction for photonic QML |  |  |
| --- | --- | --- | --- | --- |
| **PyQuil** | Quantum circuit design, Hybrid quantum-classical | Direct integration with Rigetti hardware, Classical simulation support | QML, optimization, chemistry simulations | Limited to Rigetti hardware, smaller community support |

The various QML frameworks reviewed in the paper cater to different needs in the evolving landscape of QC shown in Table 4. TFQ and Google Cirq are key players in hybrid quantum-classical models, offering integration with classical ML models for optimization and quantum simulations. Pennylane provides flexibility for QNN and RL, supporting multiple quantum devices, while Qiskit is a versatile and well-established framework ideal for quantum simulation and circuit design. Cloud-based platforms like Amazon Braket and Azure Quantum make QC accessible with hybrid models and multi-platform support, especially for quantum optimization and ML. On the other hand, D-Wave's focus on quantum annealing makes it an excellent tool for optimization-heavy industries. Strawberry Fields specializes in photonic QC, targeting quantum optics and photonic ML, while Orquestra and PyQuil facilitate the integration of quantum and classical systems for a wide range of applications in QML and optimization. Each of these frameworks brings a unique set of tools and strengths, making them suitable for specific use cases within the broad domain of QML.

4.2. Interoperability with Classical Machine Learning Libraries

Interoperability between quantum and classical ML libraries is one of the key aspects of a successful QML framework. The ability to seamlessly integrate quantum algorithms with classical ML tools enables researchers to leverage the best of both worlds: classical ML's established techniques and QC's potential for solving complex problems. This integration allows QML systems to be more practical, adaptable, and scalable, accelerating the development of hybrid quantum-classical models. Here's a deeper look into how QML frameworks can ensure smooth interoperability with classical ML libraries:

4.2.1. Hybrid Quantum-Classical Workflows

In QML, many algorithms employ a hybrid approach, combining quantum and classical components to solve problems more efficiently. Frameworks like TFQ and PennyLane are crucial for enabling smooth integration between classical and quantum models, facilitating the development of hybrid workflows that leverage the strengths of both.

- Quantum Data Encoding: Classical ML models typically process data in standard formats, such as vectors or matrices. In QML, this data needs to be encoded into quantum states for quantum processing. Libraries such as PennyLane [219] and TFQ [217] provide tools for data encoding and decoding between quantum and classical systems, enabling a seamless flow of data between the two paradigms.
- Classical Preprocessing: Classical ML libraries, such as Scikit-learn, offer a wide array of data preprocessing tools, including normalization, PCA, and feature extraction. These tools are essential in preparing classical data for use in quantum algorithms, especially in applications such as quantum-enhanced clustering or classification.
- Quantum Feature Maps: Quantum feature maps transform classical data into quantum states, allowing quantum algorithms to leverage quantum parallelism and entanglement for more efficient processing. This feature mapping is integral to quantum-enhanced versions of classical algorithms, such as SVMs or k-means clustering. Frameworks like Qiskit [221] and PennyLane [219] support these techniques, enabling classical data to be mapped onto quantum systems for enhanced performance.

4.2.2. Integration with Classical ML Libraries

QML frameworks are designed to work seamlessly with classical ML libraries, making it easier to adopt quantum techniques without abandoning classical workflows.

- TensorFlow and PyTorch: These widely used DL libraries can be extended with quantum functionality through frameworks like PennyLane, which integrates quantum operations into these ecosystems. This enables the combination of classical neural networks with quantum-enhanced models for improved performance in tasks like classification and regression.
- Scikit-learn: For traditional ML tasks, Scikit-learn is a go-to tool. QML frameworks allow the integration of quantum kernels or feature maps to augment classical models, enhancing performance in tasks like classification, clustering, and regression.
- XGBoost: Known for its power in classification and regression tasks, XGBoost can be augmented by quantum-enhanced kernels, potentially improving its accuracy on complex datasets. Quantum feature maps can be integrated to further boost the capabilities of classical algorithms.

This interoperability allows researchers to train and evaluate quantum-enhanced models alongside classical models, thus offering the best of both worlds and ensuring a smoother transition into QML.

4.2.3. Quantum-Enhanced Classical Algorithms

Many classical ML algorithms can be enhanced by QC, particularly in optimization tasks, which is one of the key benefits of QML frameworks. Quantum kernel methods compute high-dimensional feature mappings of classical data, enhancing classical models like SVMs. Quantum-enhanced SVMs, using quantum kernels, can better capture complex patterns that are difficult for classical methods to detect. Libraries like Qiskit ML and PennyLane [219] provide tools for integrating quantum kernels into classical ML algorithms, enabling quantum-enhanced performance without fully transitioning to quantum models.

Quantum algorithms like the QAOA have the potential to revolutionize optimization tasks, including portfolio optimization, combinatorial optimization, and hyperparameter tuning for classical ML models. Quantum-enhanced optimization can help classical algorithms converge faster or find better solutions, especially in complex, high-dimensional optimization landscapes.

#### 4.2.4. Quantum Machine Learning and Classical Machine Learning Models

QML represents an exciting frontier in the field, integrating quantum mechanics with ML architectures to enhance performance. QML algorithms leverage quantum properties such as superposition and entanglement to improve the efficiency of ML tasks, providing a potential edge over classical methods.

- Frameworks like PennyLane [219] facilitate the integration of quantum operations within classical ML models, enabling quantum layers to be added to existing classical networks implemented in TensorFlow or PyTorch. This hybrid approach allows quantum algorithms to process data in parallel and exploit quantum entanglement, improving performance in tasks like image recognition, NLP, and more. The synergy of quantum and classical techniques has the potential to significantly enhance the capabilities of ML models, especially in scenarios where classical methods face limitations.
- Hyperparameter optimization is a crucial aspect of ML, and quantum techniques can substantially enhance this process. Quantum algorithms can explore hyperparameter spaces more efficiently than classical methods, offering faster convergence and better solutions.
- Algorithms like Quantum Gradient Descent and QAOA enable quantum systems to navigate hyperparameter spaces more effectively, allowing for quicker optimization and potentially better outcomes, especially in high-dimensional or complex optimization landscapes. QML frameworks such as Qiskit [221] and TFQ [217] integrate quantum optimization algorithms for hyperparameter tuning, improving the performance of classical ML models. By leveraging the power of QC, researchers can fine-tune models with greater efficiency and precision, especially for challenging tasks like model selection or hyperparameter search in large, multi-dimensional spaces.

#### 4.2.5. Cloud-Based Platforms for Classical-Quantum Interoperability

Cloud platforms have become a vital tool for accessing QC power and integrating it with classical systems, making QML more accessible.

- IBM Quantum Experience: IBM provides a cloud platform that integrates QC with classical workflows through Qiskit. Researchers can access IBM's quantum computers and simulators to run hybrid quantum-classical algorithms.
- Microsoft Azure Quantum: This platform allows for hybrid workflows, integrating quantum operations with classical models such as TensorFlow and PyTorch, and provides access to various quantum hardware backends.
- Amazon Braket: Amazon's cloud platform offers APIs to integrate quantum algorithms with classical ML models. It provides access to both quantum hardware and simulators for hybrid workflows, making QML more accessible to researchers.

These cloud platforms simplify the development of QML algorithms by offering remote access to both quantum and classical computing resources.

#### 4.2.6. Error Mitigation and Scalability in Hybrid Systems

One of the main challenges in QML is dealing with quantum noise and ensuring scalability. Classical ML models are already mature and robust, and ensuring the seamless operation of both quantum and classical components is crucial for successful hybrid systems. QML frameworks provide various techniques for error mitigation and noise reduction, such as:

- QEC: This technique addresses errors that occur during quantum computations, ensuring more reliable results.
- Noise-Aware Training: This method adjusts classical models or quantum circuits to minimize the impact of quantum noise, ensuring the model remains accurate and reliable.
- Hybrid Training Approaches: Combining quantum and classical error correction methods can enhance model accuracy, especially in noisy quantum environments.

These error mitigation techniques are essential for ensuring that hybrid systems are both scalable and reliable, paving the way for the practical use of QML. The ability to seamlessly integrate quantum algorithms with classical ML tools is central to the adoption of QML. By combining quantum advantages such as quantum kernels, feature maps, and optimization techniques with classical models, QML frameworks enable:

- The enhancement of classical models, improving performance in complex tasks.
- Seamless hybrid workflows that combine the strengths of both quantum and classical systems.
- Tools for data preprocessing, model training, and hyperparameter optimization across both domains.

As quantum hardware continues to evolve, the integration of quantum and classical workflows will become even more essential for advancing real-world QML applications in various industries. Frameworks like TFQ [217], PennyLane [219], Qiskit [221], and cloud platforms such as Amazon Braket [224], Microsoft Azure Quantum [226], and IBM Quantum Experience [222] will play a pivotal role in this transformation.

4.3. Benchmarking and Comparison between Quantum Machine Learning and Classical Machine Learning

This expanded section provides detailed insights into the performance metrics used for benchmarking QML against CML. Each subsection delves into how these metrics are evaluated, the comparative results, and examples of applications supported by research studies. The evaluation of QML and CML can be categorized into five critical performance metrics: accuracy, training time, scalability, robustness, and resource utilization. These metrics define the utility and feasibility of these paradigms in various applications and guide their future development. Each metric is explored below in Table 5, focusing on its essence, how it can be achieved, why it is essential, and its distinct applications in QML and CML.

*Table 5: Performance metric for classical Machine Learning and Quantum Machine Learning*

| Metric | Quantum Machine Learning Performance | classical Machine Learning Performance | Relevant Studies |
|---|---|---|---|
| **Accuracy** | Quantum models excel in structured and combinatorial tasks using entanglement and superposition. | Dominates in large-scale, noisy datasets due to advanced optimization and pre-trained architectures. | [238], [240], [241] |
| **Training Time** | Faster for small, well-structured datasets using quantum parallelism; suffers from barren plateaus in larger variational models. | Efficient for large datasets using stochastic gradient descent and GPU/TPU optimization. | [239], [240] |
| **Scalability** | Limited by hardware constraints such as qubit count and inefficiencies in data encoding. | Scales effectively with modern distributed computing frameworks like TensorFlow and PyTorch. | [238], [241] |
| **Robustness** | Vulnerable to noise and decoherence in quantum states; error correction remains a bottleneck. | Highly resilient due to decades of refinement in algorithms and hardware. | [238], [240], [241] |
| **Resource Utilization** | High operational costs due to specialized hardware, including cryogenic cooling and superconducting qubits. | Cost-effective with widespread use of GPUs and CPUs, optimized for production-scale deployments. | [239], [241] |

- Accuracy: Accuracy reflects the model's capability to make correct predictions or classifications, a cornerstone for evaluating the reliability of any machine learning system. Higher accuracy ensures dependable decision-making in critical areas such as healthcare, finance, and autonomous systems. In QML, accuracy can be enhanced through the exploitation of quantum-specific phenomena like entanglement and superposition. Algorithms such as QAOA leverage these features to solve structured and combinatorial problems with precision. For CML, accuracy is achieved through well-established optimization techniques, robust feature extraction mechanisms, and pre-trained models that excel in large-scale datasets. High accuracy is vital to establish trust and ensure performance in practical applications, especially in fields where errors can lead to significant consequences, such as diagnostic tools or financial risk assessment. Quantum systems demonstrate superior accuracy in structured tasks like combinatorial optimization and molecular property prediction under low-noise environments. For example, QML has been used in portfolio optimization to identify optimal asset allocations [238], [240]. Classical models dominate tasks involving large-scale and noisy datasets, such as NLP (e.g., language translation) and computer vision (e.g., CIFAR-10 and MNIST image recognition). These applications benefit from decades of architectural refinements and algorithmic optimization [241].
- Training Time: Training time refers to the duration required to optimize the parameters of a model and prepare it for accurate predictions. It is a crucial factor for time-sensitive applications like autonomous vehicles and real-time analytics. QML benefits from quantum parallelism, allowing certain problems to be solved faster by simultaneously evaluating multiple states. However, issues like barren plateaus in variational quantum circuits—where gradients vanish during optimization—pose challenges. In CML, techniques like stochastic gradient descent and parallel processing with GPUs/TPUs significantly reduce training time for large-scale datasets. Reduced training time ensures timely deployment of models, which is critical for applications requiring frequent updates or real-time processing, such as recommendation systems or dynamic pricing algorithms. Quantum systems are advantageous in small dataset scenarios, such as optimization tasks in financial forecasting, where rapid convergence can be achieved under ideal conditions [238], [240]. Classical models excel in large-scale applications like training generative models (e.g., GANs and transformers) used for image synthesis or language generation, owing to their consistent and efficient optimization processes [241].

- Scalability: Scalability measures a model's ability to handle increasing data volumes or problem complexity without significant degradation in performance. It is essential for adapting models to real-world applications where data grows over time. In QML, advancements in qubit fidelity, error correction, and hybrid approaches are necessary to overcome limitations in scalability. For CML, distributed computing frameworks like TensorFlow and PyTorch enable efficient scaling by dividing tasks across clusters of GPUs or TPUs. Scalability is crucial for applications requiring the processing of massive datasets, such as big data analytics or climate modeling. Without it, models would become impractical for use in dynamic environments. Quantum systems have shown promise in high-dimensional clustering tasks, where the complexity of classical solutions becomes prohibitive. However, these applications remain constrained by hardware limitations [240]. Classical models excel in tasks like ImageNet classification, where distributed frameworks enable them to scale effectively across massive datasets and complex architectures [241].
- Robustness: Robustness defines a model's ability to maintain performance under noise, perturbations, or other adverse conditions. It determines the reliability of a system in real-world environments. QML systems require enhanced error mitigation and fault-tolerant quantum computing to address their sensitivity to noise and decoherence. For CML, robustness is achieved through years of algorithmic and architectural refinement, making models resilient even in challenging conditions. Robustness ensures the reliability of machine learning systems, especially in safety-critical applications like autonomous driving, where noisy or incomplete data can otherwise lead to catastrophic outcomes. Quantum models perform well in controlled environments with minimal noise, such as specific forecasting tasks in finance or physics simulations [239]. Classical models are used in anomaly detection for fraud prevention and cybersecurity, where they consistently handle noisy and incomplete datasets without significant performance loss [240].
- Resource Utilization: Resource utilization examines the computational and physical resources required to train and deploy machine learning models effectively. It includes considerations of energy efficiency, hardware availability, and cost. QML systems currently rely on specialized hardware, such as superconducting qubits and cryogenic cooling systems, which are expensive and energy intensive. Research into scalable quantum architectures and hybrid systems aims to reduce these costs. In CML, resource efficiency is achieved through commodity hardware (e.g., GPUs and CPUs) and optimized algorithms that balance performance and energy consumption. Efficient resource utilization is essential for cost management, environmental sustainability, and scaling technologies for broader adoption in industry and research. Quantum chemistry simulations, where classical resources fall short, highlight the value of QML despite its high operational costs [241]. Applications like recommendation systems in e-commerce benefit from the widespread availability of affordable and energy-efficient hardware for large-scale deployment [239].

Quantum Machine Learning finds niche applications in problems where quantum advantages can be leveraged effectively. Portfolio optimization in finance, molecular property prediction in drug discovery, and high-dimensional clustering are areas where QML demonstrates superiority under low-noise conditions [238], [240]. On the other hand, Classical Machine Learning dominates in diverse applications requiring robustness and scalability, such as large-scale image recognition (e.g., ImageNet) and NLP tasks like machine translation using transformer architectures [241]. Recent developments have enhanced the applicability of QML and CML across multiple domains. Frameworks such as Qiskit Runtime and PennyLane provide standardized tools for benchmarking quantum models, focusing on accuracy and time-to-solution metrics [238], [239]. Hybrid quantum-classical approaches have emerged as a significant innovation, combining classical preprocessing for data reduction with quantum backends for solving computationally intensive tasks. Public datasets like MNIST and CIFAR-10, alongside quantum-specific datasets like MoleculeNet, offer platforms for consistent benchmarking, facilitating comparisons between quantum and classical approaches [240], [241]. Several challenges persist in benchmarking and deploying QML. Hardware limitations, such as the limited availability of high-fidelity qubits and high noise levels, restrict scalability and robustness. The lack of standardized benchmarking protocols complicates direct comparisons between QML and CML models. Additionally, debates regarding the origins of quantum advantages—whether stemming from quantum effects or innovative model designs—continue to fuel research and innovation [239].

The benchmarking of QML and CML across these performance metrics underscores their complementary strengths. QML excels in specialized tasks requiring quantum-specific advantages, such as structured optimization and molecular simulations, while CML dominates in large-scale, noisy environments where its robustness, scalability, and resource efficiency shine. Despite QML's current limitations, advances in quantum hardware, hybrid approaches, and error mitigation strategies promise to expand its applicability. The ongoing convergence of quantum and classical paradigms may redefine the future of machine learning, enabling more versatile and powerful systems to address complex real-world challenges.

## 5. APPLICATIONS OF QUNATUM MACHINE LEARNING

This section provides an in-depth exploration of the diverse applications of QML across various domains shown in Figure 30, showcasing its potential to address complex computational challenges. It systematically examines fields such as Quantum Image Processing (QIP) and quantum finance, outlining the role of ML in these areas and the enhancements brought by QML. Each application is analyzed through a structured framework, including the significance of the domain, the advantages of QML over classical ML methods, and practical implementations. Comparative analyses, detailed in tables, highlight key performance metrics, while case studies provide insights into datasets, hardware, and outcomes from real-world applications. This section aims to bridge theoretical advancements with practical relevance, illustrating the transformative impact of QML across industries.

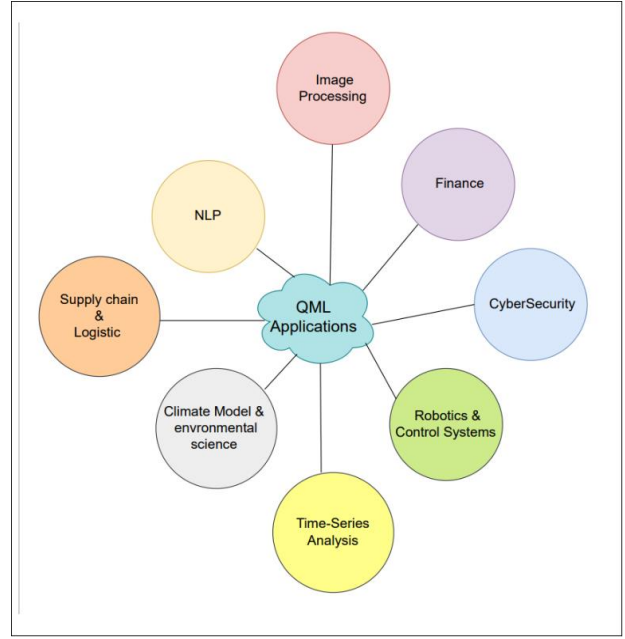

*Figure 30: Applications of Quantum Machine Learning across domains*

5.1. Finance

QC has emerged as a transformative force in financial technology, promising to address computational bottlenecks in high-dimensional optimization, risk assessment, and asset pricing. Traditional methods such as Monte Carlo simulations and matrix-based optimization are often resource-intensive, limiting their scalability to complex, multi-asset financial systems. Quantum algorithms, leveraging principles like superposition and entanglement, offer significant computational advantages, enabling faster and more accurate solutions to these problems.

- Portfolio Optimization: Portfolio optimization is a cornerstone of financial management, involving the allocation of assets to balance risk and returns effectively. Classical methods rely on quadratic programming, which becomes infeasible for large datasets due to exponential growth in computational complexity. Quantum approaches, particularly the QAOA, exploit the parallelism of quantum states to encode and solve optimization problems efficiently. By utilizing quantum amplitude encoding, QAOA directly accesses expected returns and covariance matrices, reducing computational overheads associated with classical models [242][243]. Furthermore, qRAM-based data structures enhance the ability to handle real-time financial data updates, providing a scalable framework for dynamic portfolio management [244].
- Risk Analysis and Pricing: QC significantly enhances risk analysis by accelerating dimensionality reduction techniques such as PCA. QPCA provides exponential speedups by operating on high-dimensional datasets in a compressed quantum state, enabling financial institutions to evaluate market volatilities more accurately [245]. In pricing derivatives, quantum Monte Carlo simulations reduce the number of iterations required for accurate price estimation, achieving quadratic speedups compared to classical approaches. These advantages are particularly impactful in multi-asset scenarios, where classical methods are constrained by the curse of dimensionality [242] [246].
- Forecasting and Fraud Detection: QML models, including quantum-enhanced neural networks and kernel methods, have demonstrated superior performance in financial forecasting by capturing non-linear relationships in large datasets. These models leverage quantum feature spaces to identify patterns and correlations that classical systems often overlook. For fraud detection, quantum classifiers offer enhanced precision in anomaly detection, a critical requirement for minimizing financial crime and operational risks [247][248].

QC outperforms classical methods in speed, scalability, and accuracy for specific financial applications. However, practical deployment faces limitations such as noise, decoherence, and the restricted qubit count in current NISQ devices. Additionally, the integration of quantum systems into existing financial infrastructure requires robust hybrid frameworks to bridge classical and quantum computational paradigms [249], [250]. Comparison of classical Finance and Quantum Finance is shown in Table 6.

*Table 6: Comparison of Classical finance and Quantum finance*

| Aspect | Classical Finance | Quantum Finance |
|---|---|---|
| Speed | Limited by sequential computation, especially for Monte Carlo simulations. | Achieves quadratic or exponential speedups through quantum parallelism [242][244]. |
| Scalability | Struggles with high-dimensional datasets and matrix operations. | Scales efficiently using quantum states and amplitude encoding [242][251]. |
| Accuracy | Accuracy decreases as data complexity increases. | Maintains accuracy with reduced computational resources by leveraging quantum kernel methods [246]. |
| Practical Implementation | Mature but resource-intensive and slower in dynamic systems. | Promising but constrained by current quantum hardware (NISQ devices) [250]. |

Despite its potential to revolutionize financial modelling and decision-making, quantum finance faces several significant challenges that hinder its widespread adoption.

- Hardware Constraints: Current NISQ devices are prone to noise and decoherence, which disrupt the stability and reliability of quantum computations. These issues stem from the inherent fragility of quantum states and the limited coherence times of qubits. Efforts to mitigate these limitations are ongoing, with research focused on developing error-correction techniques such as surface codes and fault-tolerant architectures. Additionally, the advancement of quantum hardware, including superconducting qubits and ion-trap systems, aims to improve coherence and gate fidelity [249][250].
- Data Encoding Overheads: Mapping classical financial data into quantum states requires sophisticated techniques like quantum Random Access Memory, which remains a significant technical challenge due to its scalability and error resilience. Current methods often demand substantial quantum resources, which are scarce. To address these issues, researchers are exploring hybrid encoding strategies that combine classical preprocessing with efficient quantum data representation methods. Innovations in quantum hardware, such as integrated photonic circuits, are also being investigated to streamline the encoding process [242][255].
- Algorithm Development: Many quantum algorithms for finance remain in the theoretical or early experimental stages, limiting their immediate applicability. Moreover, their advantages over classical methods have not been consistently demonstrated across all financial use cases. To overcome this, research communities are prioritizing algorithm refinement and benchmarking studies. Collaborative initiatives between academia and industry are accelerating the experimental validation of quantum algorithms, aiming to demonstrate their real-world feasibility and superiority [251][247].
- Integration with Classical Systems: The hybrid nature of quantum and classical computational models poses challenges in integrating quantum algorithms into existing financial infrastructures. Compatibility issues arise from differences in data formats, computational paradigms, and performance expectations. Solutions involve developing robust middleware and quantum-classical interfaces, as well as designing scalable hybrid workflows that leverage the strengths of both paradigms. Industry-driven collaborations, such as IBM Q Network partnerships, are facilitating this integration process [249][250].
- Ethical and Regulatory Issues: QC introduces ethical and regulatory challenges, particularly in ensuring data privacy and preventing potential misuse, such as market manipulation. Regulatory frameworks for quantum applications in finance are still underdeveloped, leading to uncertainties for stakeholders. Policymakers and industry leaders are engaging in discussions to establish guidelines and best practices. Efforts are also directed toward designing secure quantum algorithms that prioritize data confidentiality, aligning with ethical standards and legal compliance [250].

These concerted efforts underscore the determination of the quantum finance community to overcome current limitations. While challenges remain, ongoing advancements in hardware, algorithm development, and regulatory frameworks pave the way for a more robust and scalable quantum finance ecosystem.

Quantum finance presents transformative opportunities by enhancing computational efficiency, scalability, and accuracy. However, addressing its current limitations will require advancements in quantum hardware, robust hybrid frameworks, and interdisciplinary collaboration to align quantum methods with practical financial applications. As research progresses, quantum finance is poised to redefine the landscape of financial technology, offering novel solutions to longstanding challenges. Table 7 highlights some of the practical studies done across this application.

*Table 7: Practical case study in quantum finance*

| Reference | Challenge Tackled | Quantum Advantage Shown | Dataset Used (Public/Not) | Framework & Simulator Used | Type of Quantum Hardware |
|---|---|---|---|---|---|
| [243] | Simulation of financial assets and quantum state preparation for risk analysis. | Enhanced efficiency in Monte Carlo simulations for option pricing compared to classical techniques. | Synthetic financial datasets. | IBM Qiskit, Cirq. | NISQ (IBM QPU). |
| [244] | Review of QML applications in fraud detection and risk modelling. | Demonstrates potential for faster fraud detection and optimization tasks in financial services. | Multiple (some public, e.g., S&P 500). | PennyLane, TFQ. | Not specified. |
| [245] | Forecasting stock trends using quantum-enhanced neural networks. | Improved predictive accuracy compared to classical RNN. | NASDAQ stock price datasets (public). | PennyLane, D-Wave Ocean SDK. | D-Wave annealer (hybrid systems). |
| [246] | Quantum impact on pricing derivatives and portfolio optimization. | Achieved faster convergence in derivative pricing through quantum-assisted Monte Carlo methods. | Proprietary financial data. | IBM Qiskit. | IBM Quantum System One. |
| [248] | Exploring state-of-the-art quantum algorithms in portfolio optimization and risk modelling. | Demonstrates quantum speedup for solving optimization subroutines in portfolio management. | Not specified. | D-Wave Ocean SDK, Rigetti Forest. | D-Wave and Rigetti processors. |
| [249] | Opportunities for integrating quantum systems into enterprise financial workflows. | Focuses on quantum-classical hybrid approaches for real-time financial data processing. | Synthetic financial data. | AWS Braket. | AWS IonQ systems (hybrid approaches). |
| [250] | Overview of QC applications in asset management and pricing. | Highlights the potential for quadratic speedups in financial optimization tasks. | Simulated data. | ProjectQ, Qiskit. | Simulated quantum processors. |
| [251] | Carbon emissions financial modelling and investment decision-making. | Faster option pricing for carbon rights trading using quantum annealing. | Proprietary carbon finance datasets. | D-Wave Ocean SDK. | D-Wave quantum annealers. |

5.2. Image Processing

QIP leverages QC to address the challenges of classical image processing, such as scalability, high-dimensional data management, and computational intensity. This emerging field integrates quantum principles into image encoding, enhancement, classification, encryption, and reconstruction tasks, promising advancements in fields like medicine, security, and pattern recognition. Image processing is a critical domain encompassing tasks like feature extraction, object detection, image classification, and enhancement. Classical techniques often struggle with large-scale datasets and real-time processing demands due to the limitations of sequential computations and high storage requirements. Applications in medical imaging, satellite image analysis, and automated visual recognition systems require advanced solutions for improving efficiency and accuracy [252][253][254]. QIP introduces quantum algorithms to represent and manipulate image data using qubits. Methods such as Quantum Fourier Transforms, Quantum Arithmetic Operations, and QNN enable efficient encoding, processing, and transformation of image data [259][262][263]. Notable areas include quantum image encryption for secure communication and quantum-enhanced CNN for medical diagnostics and real-time object recognition [255][256][258]. QC transforms image processing by introducing novel capabilities in the following areas:

- Image Representation: Quantum states encode image information efficiently, allowing for the representation of exponentially large datasets in a compressed form. Methods like Flexible Representation of Quantum Images and Quantum Haar Transform enhance storage and retrieval [260][263].
- Classification and Detection: Accurate classification and detection are essential for various domains, from medical diagnostics to satellite imaging. As data complexity and volume grow, traditional methods face limitations in processing efficiency and precision. QC offers a paradigm shift by introducing novel approaches that harness quantum mechanics to enhance these tasks. Quantum-based models excel at analyzing high-dimensional data, extracting features, and recognizing patterns, often outperforming classical techniques. These advancements are driving breakthroughs in fields that rely on rapid and reliable classification, enabling applications that range from identifying diseases to monitoring Earth's surface changes.
    a. QCNN utilize quantum circuits to parallelize operations like feature extraction, enabling faster and more accurate image classification. QCNNs have demonstrated superior performance in tasks such as medical image diagnosis and satellite imaging [253][256][257].
    b. Quantum adaptive learning techniques accelerate pattern recognition by operating on high-dimensional data directly [254][255].
- Image Encryption and Security: Quantum algorithms, including chaotic maps and amplitude encoding, offer robust encryption schemes resistant to classical decryption attacks. These methods secure sensitive data in medical imaging and other critical domains [258][289].
- Image Enhancement and Reconstruction: Quantum-based enhancement techniques, such as quantum histogram equalization, improve visual quality for degraded images, while quantum image reconstruction algorithms restore images with higher fidelity [260][261][262].

Comparison of Classical image processing and Quantum Image Processing is shown in Table 8.

*Table 8: Comparison of Classical image processing and Quantum Image Processing*

| Aspect | Classical Image Processing | Quantum Image Processing |
| --- | --- | --- |
| Data Representation | Relies on binary encoding, leading to storage bottlenecks. | Efficiently encodes large datasets using qubits [260][262]. |
| Processing Speed | Sequential; slow for high-resolution data. | Parallel; exponential speedups in operations [253][256]. |
| Scalability | Limited by memory and computational resources. | Scales efficiently with quantum states [258][263]. |
| Encryption Security | Vulnerable to quantum decryption methods. | Offers robust encryption through quantum chaotic maps [259][261]. |
| Accuracy and Quality | Often constrained by hardware limitations. | Enhanced precision and fidelity via quantum transformations [254][256]. |

QIP offers transformative potential for visual data analysis, but several hurdles impede its practical implementation.

- Hardware Limitations: Current quantum processors, particularly NISQ devices, face constraints in qubit count, coherence times, and error rates. These limitations make it challenging to process large-scale and high-resolution images effectively. To address this, researchers are focusing on hardware advancements, such as developing qubit architectures with higher coherence and exploring scalable quantum error-correction techniques. Emerging technologies like quantum photonics and superconducting circuits are also being explored to enable larger and more reliable quantum systems [257][259].
- Complex Data Mapping: The process of translating classical images into quantum states requires sophisticated quantum circuits and incurs high computational overhead. Methods like amplitude encoding, which efficiently represent pixel intensities in quantum states, are computationally expensive and demand precise qubit manipulation. Current efforts aim to simplify quantum encoding techniques, integrating classical preprocessing to reduce resource demands. Additionally, advancements in quantum compilers and optimized circuit designs are helping to streamline data mapping processes [260][262].

- Algorithm Maturity: Many quantum algorithms for image processing remain experimental, with limited validation on real-world datasets. The theoretical nature of most QIP algorithms raises questions about their scalability and practical advantages over classical methods. To overcome this, researchers are emphasizing the development of robust algorithms that can handle noise and imperfections in quantum hardware. Experimental benchmarking on hybrid quantum-classical platforms is also being pursued to evaluate and refine these algorithms for practical use [261][263].
- Integration Challenges: The hybridization of quantum and classical systems in image processing pipelines introduces significant compatibility issues. Seamlessly integrating quantum algorithms with classical preprocessing and postprocessing stages requires the development of efficient interfaces and middleware. Current efforts include designing hybrid frameworks that distribute computational tasks between quantum and classical systems based on their respective strengths. Initiatives like quantum cloud platforms are facilitating this integration by providing accessible hybrid computing environments [258][260].

The field of QIP is actively evolving to address these limitations. By advancing hardware capabilities, optimizing algorithms, and enhancing hybrid workflows, researchers aim to unlock the full potential of quantum technologies in visual data analysis. These efforts promise to transform image processing, paving the way for innovative applications across industries such as healthcare, security, and entertainment.

As quantum hardware evolves and error correction techniques improve, QIP is expected to revolutionize fields like medical imaging, autonomous driving, and cybersecurity. The integration of QCNNs into diagnostic workflows and the adoption of quantum encryption in secure communications highlight its transformative potential. Continued research into efficient encoding schemes, algorithm optimization, and scalable hardware is essential to overcome current limitations and unlock QIP's full capabilities. Table 9 highlights some of the practical studies done across this application.

*Table 9: Practical case study in Quantum Image Processing*

| Reference | Challenge Tackled | Quantum Advantage Shown | Dataset Used (Public/Not) | Framework & Simulator Used | Type of Quantum Hardware |
|---|---|---|---|---|---|
| [252] | QML for image classification, focusing on enhancing accuracy. | Achieved faster and more accurate classification compared to classical CNN. | Public image datasets like MNIST. | Qiskit, TFQ. | IBM QPU. |
| [253] | QCNN for image recognition. | Demonstrated improved feature extraction and classification performance in visual datasets. | Proprietary synthetic image datasets. | Variational Quantum Circuit (VQC) simulators. | Simulated quantum processors. |
| [254] | Medical image diagnostics using quantum adaptive techniques. | Enhanced diagnostic precision in detecting medical anomalies. | Proprietary medical image data. | PennyLane. | NISQ hardware (D-Wave). |
| [255] | QCNN for efficient image processing tasks in pattern analysis. | Reduced computational time for feature extraction in large-scale image datasets. | CIFAR-10 (public dataset). | TFQ. | Not specified (simulated results). |
| [256] | Survey of QML techniques in medical image analysis. | Highlights advantages in scalability and speed for medical imaging tasks. | Public and proprietary datasets. | Various platforms including Cirq and IBM Qiskit. | Multiple quantum simulators. |

| [257] | Early developments in QIP (QUIP). | Introduced foundational methods for encoding and processing images using quantum algorithms. | Simulated datasets. | Custom frameworks for experimental purposes. | Simulated hardware. |
|---|---|---|---|---|---|
| [258] | Quantum image encryption for secure communication of image data. | Demonstrated enhanced cryptographic security using quantum chaotic maps. | Publicly available encrypted datasets. | Custom quantum encryption simulators. | Simulated quantum processors. |
| [259] | Review of security advancements in QIP. | Identified potential quantum speedups in secure image transmission and processing tasks. | Not specified. | Various experimental frameworks. | Simulated platforms. |
| [262] | Development of efficient quantum arithmetic operations for QIP. | Improved computational efficiency for image transformations and manipulations. | Not specified. | IBM Qiskit. | IBM QPU (simulated results). |
| [263] | Quantum mesh neural networks for precise diagnostics in image analysis. | Enhanced precision in pattern recognition tasks in medical datasets. | Private medical datasets. | D-Wave Ocean SDK. | D-Wave annealers (hybrid systems). |

5.3. Natural Language Processing

Quantum Natural Language Processing (QNLP) is an emerging discipline that combines QC with linguistic models to revolutionize how machines understand and process human language. Traditional NLP systems rely heavily on classical computing techniques, which face challenges in capturing contextual nuances, semantics, and complex dependencies in language efficiently. QNLP leverages quantum mechanics to address these limitations, offering novel solutions for tasks like sentiment analysis, text classification, and grammar-aware modeling. NLP focuses on enabling machines to process and analyze large volumes of text, speech, or language data. Applications include machine translation, chatbots, text summarization, and sentiment analysis. Classical NLP approaches often depend on resource-intensive algorithms, such as RNN or transformers, to extract features and understand context. These methods struggle with scalability, especially when analyzing multi-language datasets or long-form texts [264][269][271].

QNLP uses quantum principles to encode linguistic information into quantum states, which can then be processed with quantum algorithms. This encoding enables exponential parallelism, allowing efficient manipulation of sentence structures and context-sensitive language models. Key applications of QNLP include sentiment classification, low-resource language processing, and compositional models of meaning [268][271][273]. Quantum is a good aid in this field in various forms like:

- Quantum Representation of Text: QNLP encodes text into quantum states, such as tensor-product representations, which inherently preserve syntactic and semantic relationships. Techniques like Quantum Contextual Compositionality allow efficient modeling of sentence structures and meaning [270][272].
- Enhanced Text Classification: NLP has undergone significant advancements in recent years, yet challenges remain in processing complex linguistic structures, addressing low-resource languages, and improving the efficiency of text classification. QC introduces innovative approaches that leverage its fundamental principles, such as superposition and entanglement, to overcome these barriers. By integrating quantum mechanics with ML architectures, researchers are pioneering new methods to enhance the speed, accuracy, and scalability of NLP applications. These advancements are not only reshaping how we process and understand text but also opening new avenues for tackling linguistic challenges across diverse languages and datasets.
  a. Quantum self-attention mechanisms in transformers enable faster and more precise text classification by leveraging quantum superposition to analyze multiple sequences simultaneously [265][267].
  b. QRNN address low-resource language challenges by reducing the computational overhead of processing smaller datasets [268].
  c. Grammar-Aware Language Models: QNLP frameworks integrate grammatical structures into quantum models, achieving better sentence classification and syntactic parsing than traditional NLP algorithms [269][272].

- Quantum Sentiment Analysis and Transfer Learning: Quantum models for sentiment analysis, especially in social media data, have shown improved accuracy due to their ability to capture non-linear patterns in textual data. QTL enhances pre-trained models by adapting them to new linguistic tasks efficiently [267][270].

*Table 10: Comparison of Classical Natural Language Processing and Quantum Natural Language Processing*

| Aspect | Classical Natural Language Processing | Quantum Natural Language Processing |
|---|---|---|
| **Contextual Understanding** | Relies on complex embeddings like word2vec. | Encodes contextual meaning directly into quantum states [270][273]. |
| **Computation Speed** | Limited by sequential processing of large datasets. | Exploits quantum parallelism for faster results [264][269]. |
| **Scalability** | Struggles with long sequences and resource-intensive datasets. | Efficiently handles high-dimensional linguistic representations [265][272]. |
| **Grammar Sensitivity** | Grammar-awareness requires separate modules. | Directly integrates grammatical structure into quantum models [269][270]. |

Table 10 shows the Comparison of Classical NLP and QNLP. QNLP holds the potential to revolutionize how machines understand and process human language. However, several significant challenges must be addressed before its widespread adoption becomes feasible.

- Hardware Challenges: Current quantum computers lack the qubit stability and coherence needed to support the complex operations required for large-scale QNLP applications. The limitations of NISQ devices, including high error rates and short coherence times, restrict their ability to process intricate language models. Research is focused on enhancing qubit stability through error-correction codes and improving coherence times with advancements in materials and architecture. Efforts to transition from NISQ devices to fault-tolerant quantum systems are also a priority for scaling QNLP [268][272].
- Data Encoding Overhead: Translating classical text data into quantum-compatible formats is a computationally intensive process. Techniques such as quantum word embeddings and amplitude encoding require significant quantum resources and precise control of qubits. To mitigate these overheads, researchers are developing hybrid encoding methods that preprocess data classically before mapping it to quantum states. Additionally, the exploration of quantum-native language representations aims to streamline the encoding process, reducing the computational load [270][273].
- Algorithm Development: Many QNLP algorithms are still in their infancy, with most existing as theoretical constructs or experimental prototypes. Their practical deployment remains limited due to the lack of robust validation and benchmarking on real-world datasets. Current efforts are directed toward refining these algorithms to improve their scalability and noise resilience. Experimental collaborations between academia and industry are accelerating the transition from theory to application by testing QNLP models on hybrid quantum-classical platforms [269][271].
- Integration and Accessibility: Combining QNLP with classical NLP pipelines presents significant integration challenges. Building robust hybrid systems requires the development of middleware that can efficiently coordinate quantum and classical computations. Additionally, the accessibility of quantum cloud platforms is being expanded to allow NLP practitioners to experiment with QNLP without requiring in-depth quantum expertise. These initiatives aim to bridge the gap between classical NLP techniques and emerging quantum capabilities, fostering broader adoption [267][272].

While the field of QNLP is in its early stages, ongoing advancements in hardware, algorithms, and integration techniques hold the promise of overcoming these limitations. As these developments progress, QNLP is poised to complement and eventually transform traditional NLP methods, enabling breakthroughs in areas such as machine translation, sentiment analysis, and conversational AI.

QNLP is set to redefine language understanding by integrating quantum mechanics into linguistic models. With advancements in quantum hardware, error correction, and algorithm development, applications such as multilingual translation, adaptive chatbots, and sentiment analysis in low-resource languages will become more efficient and accurate. Continued research into scalable QNLP frameworks will ensure broader applicability across diverse domains, including healthcare, education, and social media analysis. Table 11 highlights some of the practical studies done across this application.

*Table 11: Practical case study in Quantum Natural Language Processing*

| Reference | Challenge Tackled | Quantum Advantage Shown | Dataset Used (Public/Not) | Framework & Simulator Used | Type of Quantum Hardware |
|---|---|---|---|---|---|
| [264] | Compositional models for meaning representation in QNLP. | Demonstrated efficient parsing and contextual analysis using quantum compositional approaches. | Publicly available linguistic datasets. | PennyLane, Cirq. | IBM QPU. |
| [265] | Quantum self-attention mechanisms for text classification. | Enhanced feature extraction and attention mechanisms in quantum recurrent networks. | Proprietary datasets. | TFQ. | Simulated quantum processors. |
| [266] | Exploration of QC applications in NLP. | Highlights speedups in semantic text analysis and translation tasks using quantum-enhanced transformers. | Multiple public datasets (e.g., SST-2). | Qiskit, TFQ. | IBM QPU. |
| [267] | QTL for acceptability judgments in language models. | Achieved faster convergence and improved accuracy in language acceptability tasks. | Proprietary text datasets. | Variational Quantum Circuits (VQC). | Simulated hardware. |
| [268] | QRN for low-resource language classification. | Improved performance in resource-constrained language classification compared to classical RNNs. | Low-resource language datasets. | Qiskit. | IBM QPU and classical hybrid simulators. |
| [269] | Practical implementations of compositional meaning models in QNLP. | Demonstrated scalable implementations of compositional models for sentence processing. | Synthetic text datasets. | D-Wave Ocean SDK. | D-Wave annealers. |
| [270] | Grammar-aware quantum-enhanced sentence classification. | Enhanced syntactic parsing and classification of text with grammar awareness. | Linguistic grammar datasets (public). | PennyLane. | Simulated quantum devices. |
| [271] | Sentiment classification for Arabic social media using QML. | Improved accuracy in sentiment analysis compared to classical models. | Arabic sentiment datasets (public). | TFQ, custom frameworks. | Not specified. |
| [272] | Early-stage QNLP models for syntax-based sentence representation. | Explored potential quantum speedups in grammatical parsing. | Synthetic grammar-based datasets. | ProjectQ. | Simulated processors. |
| [273] | Foundational work introducing QNLP and its applications in acoustics and signal processing. | Theoretical advantages for audio and textual NLP tasks using quantum-enhanced models. | Not specified. | PennyLane, Qiskit. | IBM quantum systems. |

## 5.4. Optimization for Supply Chain and Logistics Improvement

QC offers revolutionary solutions to the challenges of modern supply chain and logistics management. Traditional approaches to supply chain optimization involve solving combinatorial problems such as route planning, inventory management, and resource allocation. These methods are computationally expensive, especially as the complexity of global logistics networks increases. Quantum optimization provides an efficient way to tackle these challenges by leveraging quantum principles like superposition, entanglement, and tunneling.

- Supply chain and logistics optimization involves ensuring the efficient movement of goods, minimizing costs, and maintaining service quality. Key tasks include:
- Route Optimization: Finding optimal delivery routes for fleets to reduce travel time and fuel consumption.
- Inventory Management: Balancing stock levels across warehouses to minimize holding costs and meet demand efficiently.
- Demand Forecasting: Predicting future demand to optimize production and transportation schedules.
- Network Design: Structuring supply chain networks to ensure reliability and resilience against disruptions.

Classical approaches often rely on linear programming, genetic algorithms, or heuristic methods, which struggle to scale efficiently with large datasets and complex constraints [274][278]. QC enhances supply chain optimization through advanced algorithms and quantum-inspired approaches:

- Quantum Annealing for Optimization: Quantum annealing algorithms, such as those implemented on D-Wave systems, excel in solving combinatorial optimization problems. These methods identify near-optimal solutions for complex routing and inventory problems faster than classical techniques [278][279].
- Hybrid Quantum-Classical Approaches: Quantum-inspired algorithms integrate classical methods with quantum principles, improving scalability and accuracy for large supply chain networks [276][279].
- Hybrid models use QC for computationally intensive parts of the problem, such as constraint satisfaction in network design, while classical systems handle data preprocessing and integration [275][277].
- AI-Infused Quantum Forecasting: QML models, including QNN, enhance demand forecasting by analyzing historical data with improved precision. QNNs utilize quantum feature spaces to model non-linear patterns and correlations [277][280].
- Sustainable Supply Chains: Quantum optimization supports sustainability by reducing resource wastage through precise inventory management and optimal routing, helping organizations achieve environmental goals [275][276].

*Table 12: Comparison of Classical optimization and Quantum optimization in logistic & supply chain*

| Aspect | Classical Optimization in logistic & supply chain | Quantum Optimization in logistic & supply chain |
|---|---|---|
| Computation Speed | Slows significantly with complex, large-scale problems. | Offers exponential speedups for combinatorial tasks [274][278]. |
| Scalability | Struggles with multi-variable constraints. | Efficiently scales using quantum annealing techniques [278][279]. |
| Accuracy | Relies on approximations or heuristics for large datasets. | Provides near-optimal solutions with higher precision [274][277]. |
| Energy Efficiency | Energy-intensive due to prolonged computation times. | Operates more efficiently, reducing computational energy requirements [276][280]. |

Table 12 shows the Comparison of Classical optimization and Quantum optimization in logistic & supply chain. Quantum optimization offers transformative potential for addressing the complexities of supply chain and logistics management. However, several limitations currently hinder its widespread adoption.

- Hardware Constraints: Quantum annealers and gate-based quantum computers are limited by the constraints of the NISQ era, including limited qubit counts, high error rates, and short coherence times. These factors restrict their ability to handle the scale and complexity of real-world supply chain problems. Efforts to overcome these constraints include advancements in qubit technologies, such as superconducting qubits and trapped ions, as well as the development of error-corrected quantum systems. Researchers are also exploring hybrid models that offload certain tasks to classical systems while leveraging quantum capabilities for optimization [278][280].

- Integration Issues: Integrating quantum solutions into existing classical supply chain systems presents significant challenges. Differences in computational paradigms and data formats create compatibility issues. Current efforts to address these include developing robust hybrid architectures that facilitate seamless communication between quantum and classical systems. Middleware solutions and API standardization are being designed to ensure smooth integration and interoperability, enabling organizations to experiment with quantum optimization without overhauling their entire infrastructure [275][279].
- Algorithm Maturity: While quantum algorithms for logistics optimization, such as QAOA and VQE solver (VQE), show promise, they remain largely experimental. These algorithms require refinement to scale effectively and demonstrate consistent performance across diverse supply chain scenarios. Research initiatives are focusing on validating these algorithms with real-world datasets and creating benchmark studies to highlight their advantages and limitations. Collaborative efforts between academia and industry are accelerating the adaptation of these algorithms for practical use [276][278].
- Cost Barriers: The high cost of quantum hardware and the specialized expertise required to implement quantum solutions pose significant barriers, particularly for small and medium enterprises (SMEs). To address these issues, quantum cloud platforms are being developed to provide cost-effective access to QC resources. Additionally, training programs and open-source tools are being introduced to democratize quantum expertise, making the technology more accessible to organizations of all sizes [274][275].

Despite these limitations, quantum optimization holds immense promise for revolutionizing supply chain and logistics operations. As hardware matures, algorithms advance, and integration frameworks improve, quantum optimization is expected to become a cornerstone of efficient, scalable, and cost-effective supply chain solutions.

As quantum hardware evolves and algorithms mature, quantum optimization will play a transformative role in supply chain management. Enhanced forecasting, route planning, and inventory optimization will enable organizations to meet global logistical demands while reducing costs and environmental impact. By integrating AI with quantum systems, supply chain networks can become more adaptive, sustainable, and efficient. Future research should focus on scaling algorithms, improving hybrid architectures, and ensuring accessible quantum technologies for broader industry adoption. Table 13 highlights some of the practical studies done across this application.

*Table 13: Practical case study in Quantum logistic & supply chain*

| Reference | Challenge Tackled | Quantum Advantage Shown | Dataset Used (Public/Not) | Framework & Simulator Used | Type of Quantum Hardware |
|---|---|---|---|---|---|
| [274] | Overview of QC applications in logistics and supply chain management. | Highlights potential speedups in solving combinatorial optimization problems. | Simulated supply chain datasets. | D-Wave Ocean SDK, IBM Qiskit. | D-Wave annealers, IBM QPU. |
| [275] | Integrating AI and quantum technologies for sustainable supply chain optimization. | Demonstrated faster route optimization and inventory management solutions. | Proprietary logistics datasets. | Cirq, TFQ. | Simulated processors and D-Wave annealers. |
| [276] | Data-driven decision-making using quantum-inspired algorithms in supply chain logistics. | Showed improvements in supply chain resilience and cost efficiency. | Public and synthetic datasets. | IBM Qiskit. | IBM QPU (simulated results). |
| [277] | AI-infused QML for enhanced forecasting in supply chains. | Improved demand forecasting accuracy and efficiency using QML techniques. | Proprietary supply chain datasets. | PennyLane. | NISQ systems. |

| [278] | Comparison of quantum and classical annealing for logistics optimization. | Achieved reduced computation times for route and resource optimization problems using quantum annealing. | Not specified. | D-Wave Ocean SDK, AWS Braket. | D-Wave annealers. |
|---|---|---|---|---|---|
| [279] | Quantum-inspired computing for logistics operations and supply chain management. | Demonstrated reduced complexity in warehouse layout design and transportation planning. | Mixed datasets. | Rigetti Forest, IBM Qiskit. | Rigetti processors. |
| [280] | QNN for logistics applications. | Showed potential improvements in dynamic inventory and scheduling tasks. | Synthetic and public datasets. | TFQ. | Simulated hardware. |

5.5. Climate Modeling and Environmental Science

QC presents transformative potential for advancing climate modeling and environmental science. Traditional methods for analyzing climate data and ecosystem dynamics often face scalability and computational bottlenecks due to the complexity and size of the datasets involved. Quantum algorithms, with their ability to process vast data spaces efficiently, offer novel solutions for simulating, modeling, and mitigating the impacts of climate change. Climate modeling involves simulating the Earth's atmospheric, oceanic, and terrestrial processes to predict climate patterns, analyze the impact of human activities, and guide policy decisions. Applications include:

- Earth Observation and Remote Sensing: Processing satellite data to monitor deforestation, polar ice melting, and weather patterns.
- Sustainability and Resource Management: Optimizing energy consumption, reducing carbon footprints, and enhancing crop yields.
- Environmental Risk Assessment: Identifying and mitigating the risks of natural disasters like hurricanes, droughts, and floods.

Classical approaches to climate modeling depend heavily on high-performance computing resources. These methods, however, often struggle with the stochastic and nonlinear nature of environmental systems, leading to approximations that can compromise accuracy [283][286]. QC aids environmental science and climate modeling in several key areas:

- Enhanced Climate Simulations: Quantum simulations can model complex systems like atmospheric dynamics and ocean circulation with greater precision, leveraging quantum parallelism to process nonlinear interactions more effectively. For example, quantum annealing algorithms are used to optimize satellite mission planning for Earth observation, reducing computational overheads [288][289].
- Data Processing and Analysis: As the world generates unprecedented volumes of data from satellites, sensors, and other digital sources, the need for efficient and accurate processing techniques has become critical. QC, with its unique capabilities, offers transformative solutions for tackling these challenges. By leveraging quantum algorithms and ML models, researchers can process and analyze massive datasets at speeds far beyond what classical computers can achieve. This has profound implications for fields like environmental monitoring, urban planning, and climate science, where timely and precise insights are essential for informed decision-making.
  a. Quantum algorithms enable faster processing of vast satellite and sensor datasets, improving the accuracy of deforestation monitoring and urbanization trends [284][288].
  b. QML techniques enhance the analysis of climate trends by identifying subtle patterns and correlations that classical methods might miss [281][286].
- Sustainable Resource Management: Quantum optimization algorithms support sustainable agriculture by optimizing crop management, irrigation systems, and yield prediction [284][285]. Additionally, they help design more efficient energy systems to reduce emissions and enhance sustainability [281][286].
- Advanced Photonic Applications: Quantum-enhanced photonic technologies improve environmental monitoring through better imaging, sensing, and illumination, offering unprecedented precision in detecting atmospheric pollutants and greenhouse gases [285][286].

*Table 14: Comparison of Classical climate modelling and Quantum climate modelling*

| Aspect | Classical Climate Modelling | Quantum Climate Modelling |
|---|---|---|

| Data Processing Speed | Limited by High Performance Computing (HPC) scalability for large datasets. | Processes complex data efficiently using quantum states [283][288]. |
|---|---|---|
| Accuracy | Subject to approximations due to nonlinear complexities. | Achieves higher precision through quantum simulations of interactions [289]. |
| Resource Efficiency | High energy and computational resource requirements. | Operates with lower energy requirements for equivalent tasks [285][286]. |
| Integration of Models | Often sequential and modular. | Simultaneously models interconnected systems like atmosphere and oceans [283][288]. |

Comparison between of Classical climate modelling and Quantum climate modelling is represented in Table 14. Quantum climate modeling has the potential to revolutionize our understanding of complex environmental systems. However, significant limitations hinder its current applicability and scalability.

- Hardware Constraints: Current quantum devices are limited by insufficient qubits, short coherence times, and high error rates, which constrain their ability to process large-scale and high-resolution climate simulations. These limitations are being addressed through advancements in QEC techniques and the development of next-generation quantum hardware, such as superconducting and ion-trap qubits with improved stability. Research into fault-tolerant quantum systems is also paving the way for more reliable and scalable climate modeling capabilities [283][286].
- Algorithmic Challenges: Many quantum algorithms designed for climate modeling remain theoretical or are in the proof-of-concept stage. They lack the robustness and scalability required for real-world applications. Efforts are underway to refine these algorithms, focusing on their noise resilience and adaptability to diverse climate scenarios. Collaborative initiatives between climate scientists and QC researchers aim to test and validate these algorithms using hybrid quantum-classical models, bridging the gap between theory and practice [289].
- Integration with Existing Systems: Transitioning from classical high-performance computing (HPC) systems to quantum-enhanced systems presents technical and financial barriers. Existing climate models and datasets are deeply integrated into classical systems, making a seamless transition challenging. To mitigate these issues, hybrid architectures are being developed to complement classical HPC with QC, enabling gradual integration. Middleware solutions and interoperability frameworks are being explored to facilitate compatibility and optimize workflow efficiency [283][288].
- Environmental Data Encoding: Mapping large and complex environmental datasets onto quantum systems is computationally demanding. Techniques such as amplitude encoding and tensor network representations, while promising, require significant quantum resources and precise qubit manipulation. Current research is focused on optimizing data encoding methods and leveraging classical preprocessing to reduce the computational load. Advanced quantum compilers and data compression algorithms are also being developed to streamline the encoding process and improve efficiency [284][289].

Quantum climate modeling is an evolving field with immense potential to address the complexities of environmental science. As quantum hardware improves, algorithms mature, and hybrid systems become more accessible, quantum technologies are expected to play a transformative role in advancing climate research, helping to model and mitigate the impacts of climate change more effectively.

QC holds significant promise for revolutionizing climate modeling and environmental science. Future advancements in quantum hardware, coupled with the development of scalable quantum algorithms, will enable researchers to model environmental systems with unprecedented accuracy. This progress is vital for understanding and mitigating the effects of climate change, optimizing resource management, and guiding global sustainability initiatives. Table 15 highlights some of the practical studies done across this application.

*Table 15: Practical case study in Quantum climate modelling*

| Reference | Challenge Tackled | Quantum Advantage Shown | Dataset Used (Public/Not) | Framework & Simulator Used |
|---|---|---|---|---|
| [281] | Developing economic models for Earth conservation using quantum theories. | Introduced quantum models to better account for economic interdependencies in environmental decision-making. | Not specified. | Custom frameworks for quantum economics. |
| [282] | Strategic integration of quantum AI for environmental modelling. | Improved accuracy in climate prediction models and resource optimization tasks. | Proprietary climate datasets. | TFQ. |

| [283] | Leveraging quantum resources in high-performance computing for environmental science. | Enhanced computational efficiency for simulating atmospheric processes. | Public weather datasets (e.g., NOAA). | IBM Qiskit, QAOA simulators. |
| [284] | Sustainable agriculture through quantum-assisted crop management and yield prediction. | Improved precision in optimizing crop yield using quantum-enhanced predictive models. | Agricultural datasets (not specified). | PennyLane, TFQ. |
| [285] | Application of quantum-enhanced photonics for environmental monitoring and resource efficiency. | Achieved greater sensitivity in detecting environmental changes using quantum photonic systems. | Sensor-generated environmental data. | Custom frameworks for quantum photonics. |
| [287] | Broad review of QC and AI applications in climate science. | Highlighted potential for faster climate model simulations and improved anomaly detection in atmospheric data. | Global climate datasets. | IBM Qiskit, Cirq. |
| [288] | Quantum algorithms for satellite mission planning and Earth observation. | Demonstrated faster task scheduling and resource optimization for satellite-based climate monitoring missions. | Earth observation datasets (public). | D-Wave Ocean SDK. |
| [289] | Applications of quantum AI for understanding climate change dynamics. | Showed potential speedups in processing large-scale climate data and performing high-resolution modelling tasks. | Climate science datasets (proprietary). | TFQ, custom hybrid frameworks. |

5.6. Cybersecurity

QC is set to revolutionize the field of cybersecurity by addressing existing cryptographic challenges and introducing new, more secure encryption methods. The increasing computational power of quantum computers poses both opportunities and threats to traditional cryptographic techniques. While QC can enhance cybersecurity through QKD and PQC, it also presents risks, such as the potential to break widely used encryption schemes like RSA and ECC. As such, the integration of quantum technologies into cybersecurity is essential for creating resilient, future-proof systems. Quantum cybersecurity focuses on harnessing QC to develop novel encryption methods, improve data privacy, and mitigate potential threats from quantum-enabled attackers. Applications in this domain include:

- QKD: Using quantum mechanics to securely distribute cryptographic keys between parties.
- PQC: Developing cryptographic algorithms that are resistant to quantum-based attacks, ensuring long-term data protection.
- Botnet Detection and Security Analytics: Leveraging quantum algorithms for efficient detection of cybersecurity threats and attacks such as botnet activities.
- Quantum-Enhanced SIEM (Security Information and Event Management): Accelerating data analysis in cybersecurity analytics through quantum-enhanced SIEM systems to detect vulnerabilities and threats more effectively [294][295].

Classical cryptographic methods, while effective, are becoming increasingly vulnerable to quantum-enabled attacks. For example, Shor's algorithm can factor large numbers exponentially faster than classical algorithms, rendering RSA encryption insecure against quantum threats [293][296]. Quantum can be proved to be a strong ally in this field in the following ways:

- QKD: QKD uses quantum mechanics to secure the communication channel, making it impossible for eavesdroppers to intercept encrypted data without being detected. This is achieved by exploiting the no-cloning theorem of quantum mechanics, which ensures that any attempt to measure quantum states disturbs them and is easily detected [290][291]. Recent developments in QKD include improvements in efficiency and practical implementation over long distances [294][296].
- PQC: As quantum computers could potentially break classical encryption systems, PQC aims to develop new algorithms that are resistant to quantum attacks. Algorithms like lattice-based cryptography, code-based cryptography, and hash-based cryptography are being designed to withstand quantum computation [292][294]. These cryptographic systems focus on using mathematical problems that are hard for quantum computers to solve efficiently, offering security even in the face of quantum threats [291][295].
- Botnet Detection and Security Analytics: Quantum algorithms can enhance cybersecurity analytics by improving the speed and accuracy of threat detection. Quantum-enhanced SIEM systems, combined with ML techniques, enable the identification of botnet activities and anomalies in network traffic more efficiently than classical methods [294][295]. Quantum-based ML algorithms, such as quantum-enhanced Hoeffding trees, improve the detection rate of malicious behaviors in real-time [294].

- Quantum-Enabled Network Security: Quantum encryption algorithms provide stronger network security, particularly in communications between distributed systems. Quantum-safe protocols ensure that data transmitted over quantum networks is secure from potential eavesdropping or interception [293][295].

Table 16: Comparison of Classical cybersecurity and Quantum cybersecurity

| Aspect | Classical Cybersecurity | Quantum Cybersecurity |
|---|---|---|
| Encryption Strength | Dependent on the computational difficulty of factoring large numbers. | Quantum-resistant encryption methods, such as lattice-based algorithms, offer security even against quantum attacks [290][294]. |
| Key Distribution | Relies on trusted third parties and secure channels. | QKD ensures secure, eavesdrop-resistant key exchanges [291][294]. |
| Threat Detection | Uses classical ML models for anomaly detection. | Quantum-enhanced SIEM and ML accelerate detection, reducing response times [294][295]. |
| Scalability | Struggles with increasing data and growing attack vectors. | Scales effectively for large, decentralized systems using quantum-enhanced techniques [292][295]. |

Comparison between Classical cybersecurity and Quantum Cybersecurity is depicted in Table 16. Quantum cybersecurity represents a critical frontier in safeguarding digital infrastructure against emerging threats. However, its practical implementation faces several challenges that require immediate attention.

- Quantum Hardware Limitations: The current generation of QC technology is constrained by limited qubits, short coherence times, and high error rates, characteristic of the NISQ era. These limitations restrict the scale and reliability of quantum-based cybersecurity applications. Ongoing advancements in quantum hardware, including error-corrected quantum systems and improved qubit stability, aim to address these constraints. Research into fault-tolerant quantum architectures is expected to enhance the feasibility of large-scale quantum cryptographic systems [295].
- Transition from Classical to Quantum Systems: Integrating quantum-resistant cryptographic methods into existing infrastructures poses significant challenges. Compatibility issues with legacy systems and the resource demands of implementing quantum-safe protocols create barriers for widespread adoption. Efforts to overcome these challenges include the development of standardized frameworks for PQC and hybrid cryptographic models that enable gradual transitions. Collaborative initiatives among governments, industry, and academia are driving the adoption of quantum-safe technologies, ensuring compatibility while minimizing disruption to existing systems [294][296].
- Algorithm Development: While there has been significant progress in quantum encryption algorithms, many remain in theoretical or experimental stages. Their effectiveness and resilience against real-world threats are yet to be thoroughly validated. To address this, researchers are conducting extensive testing and benchmarking of quantum encryption protocols, such as QKD and lattice-based cryptography. Pilot projects and field trials are being implemented to accelerate the transition of these algorithms from theory to practical deployment [292][293].
- Quantum Attack Vectors: While quantum technologies enhance cybersecurity, they also introduce new vulnerabilities. Potential threats include quantum eavesdropping, where adversaries exploit quantum properties to intercept secure communications, and quantum-enabled attacks that can break classical encryption systems. Addressing these risks involves developing quantum-resistant protocols and leveraging advanced detection methods, such as quantum intrusion detection systems. Additionally, the establishment of quantum security standards and continuous monitoring of emerging quantum threats are critical to mitigating these risks [295][296].

Despite these limitations, quantum cybersecurity is poised to redefine the security landscape, offering robust protection against evolving threats. As quantum technologies mature and collaborative efforts expand, the implementation of quantum-safe systems will become increasingly viable, ensuring a secure digital future in the quantum era.

As QC technology matures, it will play a pivotal role in enhancing cybersecurity defenses, particularly in securing communications and protecting sensitive data. However, widespread adoption of quantum cybersecurity will require significant advancements in quantum hardware, algorithmic development, and integration strategies. Research into quantum-safe cryptographic methods, QKD, and quantum-enhanced ML for threat detection will be essential to creating resilient, future-proof cybersecurity systems. The next decade will be crucial in ensuring that quantum cybersecurity can meet the evolving challenges of the digital age. Table 17 highlights some of the practical studies done across this application.

*Table 17: Comparison of Classical cybersecurity and Quantum cybersecurity*

| Reference | Challenge Tackled | Quantum Advantage Shown | Dataset Used (Public/Not) | Framework & Simulator Used | Type of Quantum Hardware |
|---|---|---|---|---|---|
| [290] | Quantum-enhanced encryption methods for secure communication. | Demonstrated improved resilience of cryptographic protocols against quantum attacks. | Not specified. | Custom quantum encryption simulators. | Simulated quantum processors. |
| [292] | Integrating AI and quantum techniques for enhancing cybersecurity measures. | Enhanced threat detection and response times in real-time using quantum-enhanced SIEM architectures. | Public cybersecurity datasets. | TFQ. | Simulated quantum platforms. |
| [293] | QC applications for encryption, key distribution, and secure networks. | Demonstrated potential improvements in secure data transmission using quantum cryptography techniques. | Mixed datasets (some public, e.g., KDDCup). | Qiskit, IBM QKD simulators. | IBM QPU. |
| [295] | Quantum-enhanced SIEM architecture for botnet detection. | Demonstrated speed-ups in botnet detection using quantum-enabled Hoeffding tree algorithms. | Proprietary botnet datasets. | PennyLane, TFQ. | Hybrid quantum-classical systems. |

5.7. Robotics and Control Systems

Quantum robotics represents an exciting frontier in the integration of QC with autonomous systems. By leveraging QC's potential, quantum robotics aims to improve task planning, decision-making, control, and coordination of robots in environments that require precision and adaptability. QC can dramatically enhance the performance of robots by improving their ability to process complex decision-making algorithms and real-time control systems, enabling tasks that were previously beyond the capabilities of classical robotics. Quantum robotics combines QC with robotics to enhance the capabilities of autonomous machines in complex, dynamic environments. Key applications include:

- QRL: Used to optimize robot behavior through trial-and-error learning in uncertain environments. This is particularly useful in tasks like landing computation-limited reusable rockets, where real-time decision-making is crucial [296][297].
- Swarm Robotics: Quantum algorithms are applied to the coordination and planning of multiple robots working in tandem, optimizing performance across a range of collective tasks [298][299].
- Quantum Control Systems: These systems enable highly efficient, adaptive control over robotic systems, particularly for nonlinear or complex systems, such as PID controllers for real-time adjustments in dynamic environments [300].
- Quantum-enhanced Navigation: Quantum-controlled mobile robots can explore complex environments with enhanced precision, handling tasks such as mapping, object recognition, and pathfinding [301].

Classical robotics often relies on algorithms like RL (RL) or PID (proportional-integral-derivative) controllers, but these approaches can struggle with computational limitations, especially when high-dimensional decision spaces or real-time adaptive control is required [296][302]. Quantum can be proved to be a useful aid in this field in following ways:

- QRL: QRL provides exponential speedups in solving decision-making tasks in robotics, especially in dynamic and uncertain environments. This method allows robots to learn optimal policies for tasks like landing reusable rockets or navigating complex terrains [296]. The quantum version of RL can handle larger state spaces and provide faster convergence compared to classical RL algorithms [296][300].
- Quantum Control Systems: Quantum-enhanced PID controllers are used to adaptively manage robotic movements and operations in nonlinear systems. QC enables more efficient calculations for real-time feedback, improving system stability and reducing the need for human intervention [300][302]. Quantum algorithms improve the ability to control quantum states in robotics systems, facilitating the precise manipulation of physical processes such as nitrogen pressure regulation in industrial settings [301].
- Swarm Robotics and Quantum Planning: Quantum algorithms are applied to the coordination of large swarms of robots. Quantum planning optimizes how robots work together to perform tasks more efficiently by solving coordination problems that would be difficult for classical systems to handle. Quantum swarm robotics can perform complex collective tasks like environmental monitoring or automated search and rescue [298][299].
- Navigation and Control of Mobile Robots: Quantum algorithms allow for more efficient navigation and decision-making in robots operating in unknown or dynamic environments. Quantum-controlled robots can more precisely navigate complex environments, such as through enhanced mapping or pathfinding, ensuring more reliable autonomous operations [301].

*Table 18: Comparison of Classical robotics and Quantum robotics*

| Aspect | Classical Robotics | Quantum Robotics |
| --- | --- | --- |
| Decision-Making | Relies on classical RL or rule-based systems. | QRL provides faster learning and better decision-making capabilities in dynamic environments [296][297]. |
| Control Systems | Classical PID controllers are often limited by computational constraints. | Quantum PID controllers enable real-time adaptive control, handling complex nonlinear systems more efficiently [300][302]. |
| Swarm Coordination | Classical swarm robotics struggles with coordination in highly dynamic environments. | Quantum swarm robotics optimize coordination, enabling large-scale collaborative tasks [298][299]. |
| Navigation Precision | Depends on classical algorithms that may struggle with precision and efficiency in complex environments. | Quantum-enhanced navigation allows for more precise pathfinding and decision-making [301]. |

Table 18 shows the comparison between Classical robotics and Quantum robotics. QC has the potential to revolutionize robotics and control systems by enhancing computational efficiency and decision-making capabilities. However, several limitations must be addressed to realize this potential.

- Hardware Limitations: Quantum systems in the NISQ era are prone to high error rates, short coherence times, and limited qubit counts, which restrict their ability to handle complex and large-scale robotic tasks. These constraints make it challenging to implement real-time quantum control systems. To address these issues, advancements in fault-tolerant QC and error-correction techniques are underway. Efforts to improve quantum hardware, such as exploring topological qubits and superconducting circuits, are also being pursued to increase reliability and scalability [300][301].
- Algorithmic Maturity: Many quantum algorithms designed for robotics and control systems remain theoretical or in early experimental stages. Their practical application is hindered by a lack of robust, real-world implementations and benchmarking studies. Current research is focused on refining these algorithms for noise resilience and optimizing them for hybrid quantum-classical platforms. Collaborations between roboticists and QC experts are being encouraged to accelerate the development and validation of quantum algorithms tailored to robotic applications [298][300].
- Integration Challenges: The integration of QC with existing classical robotic systems presents significant technical barriers. These include compatibility issues between quantum algorithms and classical hardware, as well as the need for efficient quantum-classical interfaces. To address these challenges, hybrid architectures are being developed to allow quantum processors to perform specific tasks, such as optimization and ML, while classical systems manage real-time robotic operations. Middleware solutions and standardized protocols are also being designed to facilitate seamless integration [302].
- Scalability: Scaling quantum robotics to handle large, real-time operations remains a significant challenge. The computational demands of robotic systems require quantum hardware capable of managing high-dimensional data and rapid processing, which current NISQ devices cannot yet achieve. To overcome this, research is focusing on modular quantum systems and distributed QC, where multiple quantum processors can work together to tackle complex tasks. These approaches aim to extend the scalability of quantum robotics and enable more practical applications [300][301].

While quantum robotics is still in its infancy, ongoing advancements in hardware, algorithms, and integration techniques are paving the way for its future success. By addressing current limitations, QC is expected to play a transformative role in robotics and control systems, enabling innovative solutions for complex, dynamic environments.

Quantum robotics has the potential to revolutionize the way robots interact with their environment and perform tasks autonomously. By enhancing decision-making, control systems, and swarm coordination, QC offers solutions to the challenges classical systems cannot efficiently address. However, for quantum robotics to realize its full potential, significant advancements in quantum hardware and algorithm development are needed. The coming years will likely see continued progress, particularly in real-time applications like landing rockets, swarm management, and precision control systems. Table 19 highlights some of the practical studies done across this application.

*Table 19: Comparison of Classical robotics and Quantum robotics*

| Reference | Challenge Tackled | Quantum Advantage Shown | Dataset Used (Public/Not) | Framework & Simulator Used | Type of Quantum Hardware |
|---|---|---|---|---|---|
| [296] | Stabilizing rocket landing for reusable systems using QRL. | Demonstrated faster convergence in learning optimal control policies under computational constraints. | Simulated rocket dynamics datasets. | TFQ. | IBM QPU. |
| [298] | Quantum planning methods for coordinating robotic swarms. | Improved efficiency in collaborative task execution among multiple robots. | Simulated swarm datasets. | D-Wave Ocean SDK. | D-Wave quantum annealers. |
| [299] | Designing and modeling robotic swarms using quantum algorithms. | Achieved enhanced coordination and scalability in swarm robotics. | Synthetic datasets. | PennyLane, Qiskit. | Simulated quantum platforms. |
| [300] | Real-time adaptive PID control for nonlinear systems using QNN. | Improved precision and adaptability in dynamic control systems. | Experimental nonlinear systems data. | TFQ, custom quantum frameworks. | IBM QPU. |
| [301] | Intelligent quantum control of nitrogen pressure in cryogenic facilities. | Enhanced response times and control precision in industrial systems. | Proprietary cryogenic data. | Custom quantum control systems simulators. | Simulated quantum hardware. |

5.8. Time Series Analysis

Time-series analysis plays a crucial role in a wide range of applications, from financial forecasting to climate modeling and energy prediction. Classical time-series models like Long Short-Term Memory (LSTM) networks are widely used, but they face limitations when handling high-dimensional, complex, or non-linear data. Quantum Time-Series Analysis (QTSA) harnesses the power of QC to overcome these limitations, offering faster processing and improved accuracy, particularly for large-scale datasets with intricate dependencies.

QTSA focuses on using QC techniques to analyze sequential data over time. Classical time-series models, such as ARIMA (Autoregressive Integrated Moving Average), and DL models like LSTM, can handle time-dependent data but often struggle with high-dimensional, noisy, or non-linear data patterns. QTSA uses quantum algorithms and quantum-inspired methods to improve forecasting accuracy, pattern recognition, and anomaly detection across multiple domains, including:

- Energy Forecasting: Energy forecasting involves predicting energy production and consumption patterns to optimize resource allocation and grid management. For example, forecasting solar power generation requires analyzing vast datasets of historical weather conditions, solar radiation levels, and seasonal patterns. QTSA can enhance these predictions by leveraging quantum algorithms to process high-dimensional data more efficiently, improving the accuracy of models for intermittent and renewable energy sources. This enables better integration of renewable energy into the grid, reducing reliance on fossil fuels and enhancing sustainability efforts.

- Financial Markets: The analysis of stock prices, trading volumes, and market trends is critical for decision-making in financial markets. Classical methods often struggle with the computational complexity of high-frequency trading data and the nonlinear relationships between market variables. QTSA can accelerate these processes by using quantum algorithms to identify patterns and correlations within massive datasets, enabling more accurate predictions and real-time market insights. This can improve risk management, portfolio optimization, and trading strategies, giving financial institutions a competitive edge.
- Climate and Environmental Monitoring: Predicting weather patterns and analyzing long-term climate data are essential for understanding and mitigating the effects of climate change. Classical models require extensive computational resources to simulate complex interactions among atmospheric, oceanic, and terrestrial systems. QTSA can process these multidimensional datasets more efficiently, facilitating the generation of precise and timely climate models. This can aid in forecasting extreme weather events, improving disaster preparedness, and supporting environmental policy decisions.
- Anomaly Detection: Anomaly detection involves identifying irregularities in time-series data, which is crucial for applications like fraud detection, system health monitoring, and predictive maintenance. Traditional methods often struggle with detecting subtle anomalies in noisy or high-dimensional datasets. Quantum algorithms can enhance anomaly detection by exploring multiple potential irregularities simultaneously, increasing detection speed and accuracy. This capability is particularly valuable in industries like cybersecurity, where rapid identification of threats can prevent significant damage, and in manufacturing, where it ensures the reliability of critical systems [303][305][309].

QC has emerged as a powerful tool in time-series analysis, addressing some of the critical limitations of classical approaches, such as handling high-dimensional datasets, managing computational complexity, and improving prediction accuracy. By leveraging quantum principles like superposition, entanglement, and quantum parallelism, quantum techniques provide innovative solutions for analyzing sequential data. Below are specific contributions QC has made in this field:

- Quantum Long Short-Term Memory (QLSTM): QLSTM networks leverage QC to improve the performance of classical LSTMs by enhancing their ability to store and retrieve long-range dependencies in time-series data. Quantum circuits enable faster state updates, allowing QLSTMs to handle more complex time-series forecasting tasks, such as predicting solar power production, with increased accuracy and speed compared to classical methods [303][304].
- Quantum Dynamic Mode Decomposition: This algorithm extends classical dynamic mode decomposition methods by using quantum principles to decompose high-dimensional time-series data more efficiently. Quantum-enhanced DMD allows for faster identification of dominant modes and patterns in time-series data, which is valuable in fields like fluid dynamics, stock market analysis, and climate modeling [305][308].
- Quantum Reservoir Computing: Quantum reservoir computing models, including those using weak and projective measurements, enhance the ability to process time-series data by leveraging quantum entanglement and superposition. These models improve prediction accuracy for non-linear time-series data, making them suitable for applications such as anomaly detection and system health monitoring [306][309].
- Quantum-Inspired Methods: Quantum-inspired algorithms like online spiking neural networks and random oscillator networks are being developed to predict future values in time-series data. These quantum-inspired models offer improvements in processing speed and adaptability, especially when dealing with large-scale, real-time data feeds [307][308].

QC is revolutionizing the field of time-series analysis by addressing some of the limitations faced by classical methods, such as computational inefficiency, scalability issues, and the ability to process complex, high-dimensional data. Quantum systems leverage phenomena like superposition, entanglement, and quantum parallelism to accelerate data processing, enhance pattern recognition, and improve prediction accuracy in various time-series applications. Table 20 shows the comparison of Classical time-series and Quantum time-series.

*Table 20: Comparison of Classical Time-series model and Quantum Time-series model*

| Aspect | Classical Time-Series Models | Quantum Time-series model |
| --- | --- | --- |
| Data Handling | Struggles with high-dimensional, noisy, or non-linear data. | Efficiently handles high-dimensional, complex, and non-linear data [303][305]. |
| Computation Speed | Slower processing, especially for large datasets. | Offers exponential speedups through quantum algorithms [307][308]. |

| Forecasting Accuracy | Limited by the capacity of classical models (e.g., LSTM). | Enhanced prediction accuracy due to quantum-enhanced learning capabilities [303][304]. |
|---|---|---|
| Scalability | Computationally expensive as data size grows. | Scales more efficiently, handling vast amounts of time-series data [306][309]. |
| Pattern Recognition | Uses linear or non-linear techniques that may miss subtle relationships. | Quantum algorithms capture complex relationships and hidden patterns more effectively [308]. |

QTSA offers the potential to revolutionize the processing of temporal data by enhancing computational speed and accuracy. However, its practical implementation faces several challenges.

- Hardware Constraints: Current quantum hardware, characterized by the limitations of the NISQ era, lacks the capacity to handle large-scale time-series analysis with high precision. Issues such as limited qubit counts, short coherence times, and high error rates significantly hinder performance. Efforts to overcome these constraints include the development of fault-tolerant quantum systems and scalable quantum hardware architectures. Research into next-generation qubit designs and error mitigation techniques is also underway to enable more reliable quantum computations for time-series applications [306][307].
- Algorithmic Development: Quantum algorithms for time-series analysis, while promising, remain largely experimental and lack extensive validation on real-world datasets. The theoretical nature of many algorithms limits their deployment in practical scenarios. Current initiatives are focused on refining these algorithms to improve their noise resilience, scalability, and integration capabilities. Collaboration between academia and industry is accelerating the adaptation of quantum time-series algorithms for real-world use cases, ensuring their robustness and reliability [304][307].
- Integration with Classical Systems: Integrating quantum-enhanced models with existing classical infrastructure presents compatibility challenges, particularly in resource management and data format standardization. Hybrid quantum-classical frameworks are being developed to address these issues, allowing quantum systems to handle computationally intensive tasks while classical systems manage real-time processing and storage. Middleware solutions and APIs are also being designed to ensure seamless interoperability between quantum and classical systems [305][309].
- Data Encoding and Measurement: Encoding classical time-series data into quantum systems, particularly for high-dimensional or real-time data streams, remains a significant technical hurdle. Techniques like amplitude encoding and phase encoding, while effective, are resource-intensive and require high precision in qubit manipulation. Researchers are exploring optimized encoding schemes and preprocessing techniques to reduce the computational overhead. Additionally, advancements in quantum measurement protocols are being pursued to improve data retrieval accuracy and efficiency [307][308].

Despite these limitations, QTSA continues to evolve, driven by advancements in hardware, algorithms, and integration strategies. As these challenges are addressed, QC is expected to significantly enhance time-series analysis, enabling breakthroughs in fields such as finance, healthcare, and environmental monitoring.

QTSA is an exciting field with the potential to transform how we forecast and analyze time-dependent data. With continued advancements in quantum hardware and algorithm development, QTSA will enable faster, more accurate predictions and pattern recognition, offering a competitive edge in industries like energy, finance, and healthcare. However, significant challenges remain in scaling quantum systems, optimizing quantum algorithms, and integrating them with classical systems. As these challenges are addressed, QTSA will play an increasingly central role in data-driven decision-making. Table 21 highlights some of the practical studies done across this application.

*Table 21: Practical case study in Quantum Time-series model*

| Reference | Challenge Tackled | Quantum Advantage Shown | Dataset Used (Public/Not) | Framework & Simulator Used | Type of Quantum Hardware |
|---|---|---|---|---|---|
| [303] | Comparison of QLSTM and classical LSTM for solar power forecasting. | Achieved higher prediction accuracy with fewer parameters compared to classical LSTM models. | Solar energy datasets (public). | TFQ. | IBM QPU. |

| [304] | Clustering high-dimensional time-series images using quantum-inspired digital annealers. | Demonstrated faster clustering for large-scale data compared to classical approaches. | Public image datasets. | Quantum-inspired digital annealer (Fujitsu). | Quantum-inspired annealer. |
|---|---|---|---|---|---|
| [305] | Dynamic Mode Decomposition (DMD) for analyzing high-dimensional time-series data. | Improved scalability and efficiency in extracting dominant modes from complex systems. | Climate and financial datasets. | Qiskit, PennyLane. | IBM QPU. |
| [306] | Reservoir computing for time-series analysis using quantum measurements. | Enhanced performance in pattern recognition for non-linear time-series data. | Synthetic time-series data. | TFQ. | Simulated quantum platforms. |
| [307] | Quantum-inspired spiking neural networks for real-time time-series predictions. | Achieved faster and more adaptive predictions in streaming data environments. | Streaming financial data (public). | PennyLane, D-Wave Ocean SDK. | D-Wave quantum annealers. |
| [308] | Online processing of time-series using quantum random oscillator networks. | Demonstrated potential for real-time data processing with higher efficiency than classical methods. | Public sensor data streams. | IBM Qiskit, custom quantum network simulators. | Simulated quantum processors. |
| [310] | Fuzzy-Quantum Time Series Forecasting Model (FQTSFM) for financial predictions. | Improved forecasting accuracy for noisy financial datasets. | Proprietary financial datasets. | Qiskit. | IBM QPU. |

## 6. LIMITATIONS AND FUTURE RESEARCH DIRECTIONS

In this section, we provide a comprehensive exploration of the current challenges and promising future directions in Quantum Machine Learning (QML). Despite its transformative potential at the intersection of quantum computing and artificial intelligence, QML faces numerous hurdles that must be addressed to unlock its full capabilities. The section is divided into two subsections: Limitations, which examines the technical, theoretical, and ethical barriers impeding progress in QML, and Future Research Directions, which highlights emerging opportunities and innovations shaping the field. Together, these subsections aim to provide a holistic view of the ongoing efforts and critical paths forward for advancing QML.

6.1. Limitations

This subsection delves into the key limitations currently hindering the advancement of Quantum Machine Learning (QML). Despite its promise, QML is confronted with various challenges, including issues related to noise, error correction, scalability, interpretability, and resource requirements. Additionally, the complexity of algorithm development, hybridization concerns, and the limitations of quantum data encoding further complicate the practical implementation of QML. By addressing these hurdles, researchers aim to improve the reliability, efficiency, and accessibility of QML, paving the way for more widespread adoption. QML stands at the intersection of QC and AI, promising transformative advantages in data processing, optimization, and pattern recognition. However, realizing its full potential requires addressing numerous technical, theoretical, and ethical challenges. Below is a detailed exploration of the limitations and open problems in QML, supported by references from the provided and additional sources.

- Noise and Error Correction: Quantum systems are inherently prone to errors caused by decoherence, gate inaccuracies, and environmental noise. While QEC techniques like surface codes and real-time decoding show promise, they require significant hardware overhead and are computationally expensive. Current systems require thousands of physical qubits to maintain a single logical qubit, making scalable error correction a long-term challenge [312] [319][320]. Progress in fault-tolerant computing is critical to ensure reliable QML operations [9]. Recent breakthroughs in QEC by Microsoft and Google involve real-time decoding algorithms and surface code-based logical qubits, reducing error rates significantly [48]. Hybrid error mitigation strategies, such as tensor network-based methods, are being explored to bridge NISQ-era hardware with fault-tolerant systems [313].

- Scalability: Scaling quantum systems to handle large datasets and complex computations remains a bottleneck. Current quantum hardware is limited in the number of qubits and gate operations it can reliably perform. Achieving scalability involves addressing both physical hardware constraints and algorithmic efficiency to manage resource requirements effectively [9][48][314]. Advances in modular quantum architectures, like those by IBM and Atom Computing, aim to scale systems by interconnecting smaller, error-corrected quantum modules [48][312]. Quantum-inspired algorithms also provide interim solutions, leveraging classical platforms to simulate larger problem spaces [315][318].
- Interpretability: Quantum models often operate in high-dimensional Hilbert spaces, making their operations less interpretable compared to classical ML models. Developing tools and frameworks to explain QML decisions is crucial for trust and usability. This limitation affects adoption in critical domains like healthcare and finance, where explainability is paramount [9][54][313]. Research into explainable quantum AI (XQAI) is emerging, where techniques like quantum state tomography and hybrid classical quantum explainability frameworks are being developed to provide insights into quantum models' behavior [314][318].
- Resource Requirements: Quantum hardware demands significant physical resources, including advanced cryogenic systems for maintaining qubit coherence and high-fidelity gates. The cost and complexity of such systems are barriers to widespread adoption. Additionally, the energy requirements of quantum machines, although lower for certain tasks, remain non-trivial compared to classical supercomputers [7][313][314]. Efforts to optimize quantum resources include the development of energy-efficient qubits and integration with classical HPC systems to reduce overhead [48][314]. Quantum cloud platforms (e.g., AWS Braket, Azure Quantum) are democratizing access, lowering resource barriers [317][318].
- Algorithm Development: Most quantum algorithms for ML remain theoretical or experimental. Many lack clear demonstrations of quantum advantage over classical counterparts. Addressing issues such as optimization, convergence guarantees, and robustness against noise is essential to advance QML algorithms from theory to practice [9][48][317]. Advances in Quantum Inspired Machine Learning (QIML) are bridging the gap by developing hybrid models that incorporate quantum principles into classical algorithms [313][315]. Variational algorithms, such as QAOA, are gaining traction for near-term applications [318].
- Hybridization Issues: Hybrid quantum-classical models are a practical approach in the NISQ era but pose integration challenges. Efficiently partitioning tasks between classical and quantum processors, minimizing communication overhead, and ensuring compatibility across frameworks are critical to unlocking the potential of hybrid QML [312][314]. Projects like TFQ and IBM Qiskit Runtime are streamlining hybrid workflows, enabling faster task partitioning and optimized quantum-classical communication [312][314].
- Data Encoding and Representation: Mapping classical data into quantum states is a significant hurdle. Methods like amplitude encoding and tensor-product representation are resource-intensive and scale poorly with dataset size. Developing more efficient encoding techniques is a pressing area of research [48][314][7]. New encoding methods, such as tensor-product networks and hybrid classical-quantum embeddings, are reducing overheads while preserving data structure [315][318]. Classical ML preprocessing remains a critical enabler for efficient quantum encoding [313][314].
- Understanding Quantum Dynamics: The dynamics of quantum systems in the context of ML are not fully understood. Theoretical advancements are needed to explain how quantum entanglement, superposition, and interference can be effectively leveraged for learning tasks. This gap hinders both algorithm design and hardware optimization [313][315][318]. Interdisciplinary studies in quantum mechanics and ML are advancing theoretical models to better understand phenomena like quantum generalization and entanglement's role in learning [315][318].
- Benchmarking and Evaluation Metrics: Unlike classical ML, where standardized datasets and benchmarks exist, QML lacks universal evaluation metrics. Establishing such benchmarks is essential for comparing QML algorithms, hardware, and performance against classical systems. This will also facilitate reproducibility in QML research [48][315][316]. Initiatives are emerging to establish benchmarks tailored to QML, including quantum-specific datasets and evaluation criteria that account for quantum noise and hardware constraints [314][318].
- Theoretical Foundations: The theoretical underpinnings of QML, such as generalization bounds and the relationship between quantum and classical complexity, remain underexplored. Addressing these gaps will provide clearer insights into the advantages and limitations of QML [54][312][315]. Studies are exploring connections between quantum information theory and classical learning theory to define quantum complexity classes and generalization bounds [313][316].
- Training Complexity: Training quantum models is computationally challenging due to problems like barren plateaus, where gradients vanish during optimization, hindering convergence. Developing optimization methods that avoid these issues is critical for the practical training of QNN and other QML models [314][315][317]. Alternatives like quantum natural gradient optimization and parameter initialization heuristics are mitigating barren plateau effects in variational quantum circuits [314][315].
- Ethical Considerations: As with classical ML, ethical concerns surrounding QML include biases in data and algorithms, misuse of advanced QML systems, and challenges to data privacy. Quantum technologies could exacerbate existing inequalities in AI access and capability, raising concerns about their responsible use [9][54][312]. Discussions on responsible AI are expanding to include quantum technologies, with guidelines focusing on fairness, accountability, and transparency in QML applications [318].

- Long-Term Viability of Quantum Technologies: The future of QML depends on sustainable advancements in quantum technologies. This includes reducing the cost of quantum systems, developing robust error correction techniques, and scaling hardware to meet practical needs. Ensuring the environmental and economic sustainability of quantum systems is critical for their long-term viability [312][313][316]. Quantum hardware developers are focusing on scalable, energy-efficient designs, and cloud-based quantum solutions are reducing the need for resource-intensive setups [313][314].

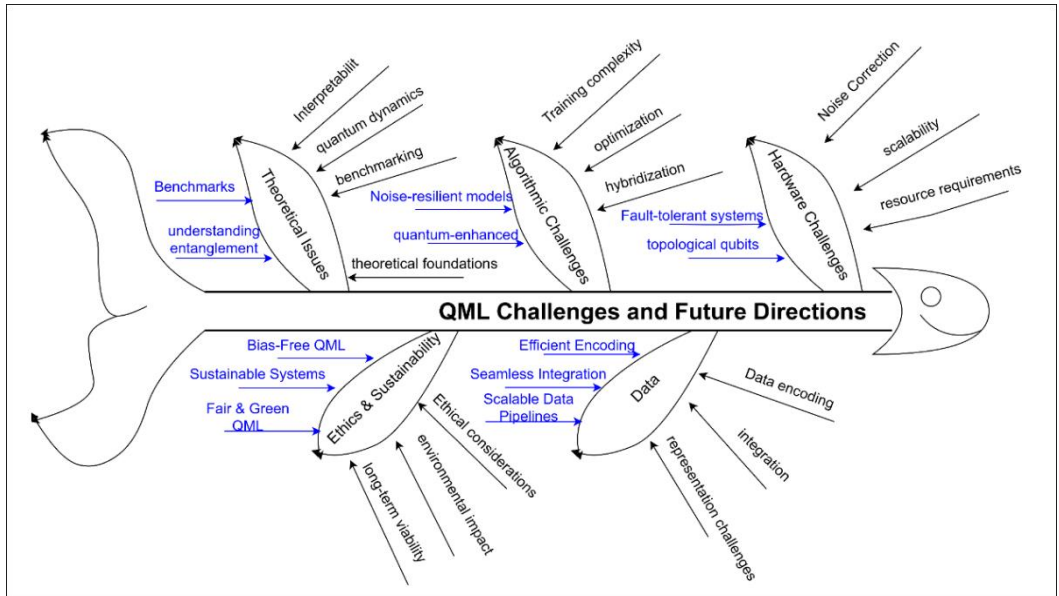

*Figure 31 : Fishbone diagram depicting major challenges and future research directions in Quantum Machine Learning, including technical, ethical, and practical considerations for advancement in the field. (**Black arrow**- Challenges, **Blue arrow**- Future Research directions)*

6.2. Future Research direction

In this subsection, we explore the exciting future directions in Quantum Machine Learning (QML), focusing on the ongoing research efforts and innovations that could drive the field forward as shown in Figure 31. Key areas of development include advancements in quantum hardware, quantum-classical hybrid systems, and novel algorithmic techniques. Additionally, new data representation methods and interdisciplinary collaborations are set to play crucial roles in the evolution of QML. By addressing current challenges and exploring these emerging avenues, future research holds the potential to unlock the full power of QML, enabling it to tackle complex real-world problems more efficiently and effectively. QML is advancing rapidly, driven by a combination of theoretical research, hardware improvements, and algorithmic innovation. Below is a comprehensive exploration of research horizons and the ongoing efforts propelling the field. Hardware Advances: As QML continues to evolve, one of the most critical challenges is improving quantum hardware. This involves developing fault-tolerant quantum systems with higher qubit counts, enhanced gate fidelity, and longer coherence times. The latest advancements in quantum technologies such as topological qubits, superconducting circuits, and quantum dots aim to overcome these challenges. These innovations will enable more complex QML models while reducing quantum errors. For example, IBM's latest quantum processors, capable of executing circuits involving thousands of gates, have made significant progress in improving coherence times and error correction techniques. Similarly, companies like Microsoft and Atom Computing have made strides toward creating logical qubits with higher fidelity, which is essential for scalable quantum systems [54][322]. As these hardware improvements continue, the scope of QML applications will expand, allowing for more efficient and accurate ML models that can process vast and complex datasets.

- Quantum Software Ecosystems: The development of robust quantum software ecosystems is critical to advancing QML. These ecosystems must include tools for algorithm testing, noise management, and integrating quantum-classical workflows to maximize computational efficiency. As quantum systems mature, these software platforms will incorporate automatic optimizations that enable seamless partitioning of tasks between quantum and classical systems. Tools like Qiskit and TFQ are evolving, with libraries that support noise-resilient algorithms and hybrid quantum-classical workflows. IBM's introduction of tensor network-based error mitigation, for example, enhances the reliability of quantum circuits by addressing common issues such as noise and decoherence [322][324][325]. As these ecosystems continue to grow, they will allow researchers and developers to more easily create, test, and deploy advanced QML algorithms, accelerating the transition from theory to practical, real-world applications.
- Integration with Classical ML: One of the most promising future directions in QML is the integration of quantum and classical ML workflows. Quantum-classical hybrid systems will enable computational efficiency by offloading resource-intensive tasks to quantum processors while utilizing classical systems for preprocessing, visualization, and

- less demanding computations. Microsoft's Azure Quantum platform exemplifies this approach, providing developers with a unified framework to combine high-performance classical computing with quantum resources. This integration is crucial for real-world applications in fields such as material science simulations and financial portfolio optimization, where quantum processors can significantly speed up certain calculations while classical systems handle other tasks [54][322][323]. The future of QML will likely involve the seamless coexistence of classical and quantum systems, optimizing resources to tackle complex problems more efficiently.
- New Potentials of Algorithms: The development of novel quantum algorithms has the potential to revolutionize AI. Quantum-enhanced transformers, generative models, and QNN are at the forefront of this research, promising to expand the capabilities of AI. These algorithms exploit quantum phenomena like entanglement and superposition to enhance tasks such as unsupervised learning, anomaly detection, and feature engineering. Quantum dynamic mode decomposition, for instance, is being explored as a method for analyzing high-dimensional datasets in domains such as fluid dynamics and financial forecasting [324][325]. In the future, these algorithms will redefine the scope of AI, enabling the analysis of increasingly complex datasets and improving decision-making processes in various industries.
- Enhanced Quantum Data Representation: Efficient data representation is a key challenge for QML. Innovations in data encoding techniques, such as amplitude encoding and hybrid tensor networks, aim to reduce computational overhead while maintaining essential features of classical datasets. The current inefficiencies in data encoding are being addressed through new methods, including tensor-product-based encodings, which have shown promise in handling large-scale datasets in fields like finance and healthcare. As these encoding techniques are optimized, they will enable quantum systems to process larger and more complex datasets, leading to more accurate ML models and predictions [323][324][325]. This will play a pivotal role in the future of QML, allowing quantum systems to handle the vast quantities of data generated in many modern applications.
- Interdisciplinary Collaborations: The integration of QML into real-world applications will accelerate through interdisciplinary collaborations between quantum scientists and experts in fields like healthcare, climate science, and finance. QC has already shown its potential in optimizing drug discovery pipelines and improving precision in satellite-based Earth observations. These interdisciplinary efforts highlight the wide-reaching impact of QML and its potential to solve critical challenges in various industries [323][325]. As these collaborations continue to grow, we can expect to see more practical QML applications that address global issues, from climate change to disease prevention, driving the widespread adoption of quantum technologies.
- QFL: QFL is an exciting area of research that combines QC with privacy-preserving distributed AI. This framework enables secure model training across decentralized nodes, making it suitable for sectors such as healthcare, where data privacy is a critical concern. Early implementations of QFL are being explored in projects like secure diagnostics in medical imaging, where QC can enhance the accuracy and privacy of AI systems. The development of QFL will make it possible to securely train models without needing to share sensitive data, allowing for advancements in privacy-preserving AI technologies [323][324][325]. As the field progresses, QFL could become a key tool for industries that rely on distributed data and require robust security measures.
- Improved Error Mitigation Techniques: One of the main obstacles to realizing practical QML systems on NISQ devices is the error rates inherent in QC. As quantum systems become more advanced, better error mitigation strategies are being developed to improve the reliability of quantum circuits. Techniques such as adaptive error correction and noise-resilient algorithms are crucial for bridging the gap between NISQ systems and fault-tolerant quantum computers. IBM's tensor network-based error mitigation tools, for instance, have shown promise in reducing the impact of decoherence, making quantum circuits more reliable and less susceptible to errors [324][325]. As error mitigation techniques continue to improve, we will see more practical and stable QML applications emerge.
- Quantum Simulation for Real-World Applications: Quantum simulations are poised to revolutionize fields like climate modelling, drug discovery, and materials science by solving problems that are currently computationally infeasible for classical systems. Companies like Microsoft and IBM have demonstrated the use of quantum simulations for molecular modelling, contributing to breakthroughs in materials design and pharmaceuticals. These quantum simulations have the potential to unlock new insights in areas like drug development and renewable energy, providing solutions to problems that were once thought to be beyond reach [321][325]. In the future, quantum simulations will be a critical tool for solving complex, real-world problems that require significant computational resources.
- Scalable Quantum Learning Models: A key challenge for the future of QML is the development of scalable models that can efficiently process vast datasets. Innovations in distributed QC frameworks and hierarchical learning models are expected to address these challenges. By enabling quantum systems to scale, researchers will be able to apply QML to larger datasets, such as those found in genomics and financial analysis. These scalable systems will be essential for broader adoption of QML, enabling more industries to harness the power of QC for data analysis and decision-making [196][321]. As scalability becomes more achievable, QML will open new doors for industries that rely on big data.
- Ethical Frameworks for QML: As QML systems become more integrated into sensitive domains such as healthcare, finance, and surveillance, it will be critical to address ethical concerns surrounding data privacy, fairness, and accountability. Establishing standards for responsible AI development will ensure that QML is used in a way that benefits society while minimizing risks such as bias and privacy violations. Ongoing academic discussions are exploring the ethical implications of QML, with guidelines being proposed to mitigate potential risks in automated decision-making and surveillance systems [196][321]. As quantum technologies continue to advance, creating a robust ethical framework will be essential for ensuring that QML benefits all sectors of society.

The future of QML is poised for transformative growth, driven by advancements in both hardware and software, as well as innovative algorithms. Key areas of development include the continued improvement of quantum hardware, particularly through fault-tolerant systems, higher qubit counts, and enhanced error correction techniques, which will enable more complex QML models. Software ecosystems tailored for QML are evolving to support noise-resilient algorithms and hybrid quantum-classical workflows, making it easier to integrate quantum accelerators with classical systems for improved computational efficiency. Future QML applications will also benefit from the development of novel algorithms, such as quantum-enhanced transformers and generative models, which can tackle tasks like unsupervised learning and anomaly detection with greater efficiency. Enhanced quantum data representation techniques will further optimize the handling of large-scale datasets, making it possible to apply QML in industries like finance and healthcare. Interdisciplinary collaborations, particularly between quantum scientists and domain experts in healthcare, climate science, and finance, will accelerate the practical deployment of QML. QFL and improved error mitigation strategies will play a pivotal role in addressing privacy concerns and stabilizing quantum circuits for real-world applications. Quantum simulations are also set to revolutionize fields such as climate modelling, drug discovery, and materials science, providing solutions to previously intractable problems. Additionally, scalable quantum learning models will pave the way for handling vast datasets in genomics and financial analysis. As QML continues to evolve, it will be essential to address ethical considerations related to privacy, fairness, and accountability, ensuring that the technology benefits society while minimizing risks. Overall, the future of QML holds immense promise, with its potential to solve complex problems and redefine AI capabilities in ways that were once thought impossible.

## 7. CONCLUSION

QML stands at the intersection of two transformative domains: QC and ML. This survey has explored the potential of QML to address computational bottlenecks in handling high-dimensional datasets, complex optimization problems, and large-scale data analysis, which challenge classical systems. By leveraging quantum principles like superposition, entanglement, and quantum parallelism, QML offers promising solutions that can significantly outperform classical approaches in specific scenarios. Through a detailed examination of various algorithms—such as SVMs, QSVM, QPCA, and QNN—it is evident that quantum approaches hold the potential to revolutionize areas like pattern recognition, optimization, and simulation. Applications in fields as diverse as healthcare, finance, quantum chemistry, and NLP highlight QML's versatility and its capacity to unlock new capabilities in data analysis and AI.

However, the field is not without challenges. The limitations of current quantum hardware, including noise, limited coherence times, and scalability issues, impose significant constraints on the practical implementation of QML algorithms. Moreover, the integration of classical and quantum systems in hybrid approaches, while promising, introduces additional complexities in terms of workflow design and interoperability. These factors underscore the need for advancements in error correction, hardware reliability, and algorithm optimization. Research Horizons in QML research should focus on addressing these challenges by developing noise-resilient quantum algorithms, improving qubit quality, and exploring innovative hybrid models. Benchmarking quantum algorithms against classical counterparts will be crucial for identifying practical use cases where quantum advantages can be realized. Additionally, interdisciplinary collaborations between quantum physicists, computer scientists, and domain experts will play a pivotal role in accelerating the adoption of QML across industries. As quantum technologies mature, QML is poised to become a cornerstone of the computational landscape, offering groundbreaking solutions to problems that were previously considered intractable. While achieving full-scale quantum supremacy remains a long-term goal, the incremental advancements in QML methodologies and hardware capabilities provide a clear trajectory for its evolution. This convergence of QC and ML heralds a new era of innovation, with the potential to redefine the boundaries of computation and its applications.